\documentclass[reprint,prl,aps,nobibnotes,nofootinbib]{revtex4-2}
\usepackage{enumitem}
\usepackage[utf8]{inputenc}
\usepackage{times}
\usepackage{graphicx}
\usepackage{amsmath, braket, amsfonts}
\usepackage{amssymb}
\usepackage{bm, color, ulem}
\usepackage{ragged2e}
\usepackage{wrapfig,lipsum,booktabs}
\usepackage{lineno}
\usepackage{color}
\usepackage{url}
\usepackage{comment}
\usepackage{hyperref}
\usepackage{epstopdf}
\usepackage[version=4]{mhchem}
\usepackage{xcolor}
\usepackage{xcolor}
\definecolor{green}{RGB}{0,115,70}
\usepackage{placeins}
\usepackage[T1]{fontenc}
\usepackage[a4paper, left=0.8in, right=0.8in, top=0.9in, bottom=0.9in]{geometry}

\makeatletter
\newcommand{\SIcontents}{%
    \begin{center}
        {\large\bfseries CONTENTS}
    \end{center}
    \vspace{0.5em}

    \begingroup
    \large
    \renewcommand{\baselinestretch}{1.25}\selectfont
    \@starttoc{sitoc}
    \endgroup
}

\makeatother
\usepackage{soul}
\modulolinenumbers[1]        
\hypersetup{
    colorlinks=true,
    linkcolor=blue,
    filecolor=blue,      
    urlcolor=blue,
    citecolor=blue,
    pdftitle={Overleaf Example},
    pdfpagemode=FullScreen,
    }
\modulolinenumbers[5]
\definecolor{green}{RGB}{0,0,0}
\definecolor{magenta}{RGB}{0,0,0}
\let\OriginalSIcontents\SIcontents
\renewcommand{\SIcontents}{%
    \begingroup
    \color{black}%
    \hypersetup{linkcolor=blue}%
    \colorlet{green}{blue}%
    \colorlet{magenta}{blue}%
    \OriginalSIcontents
    \endgroup
}

\makeatletter
\renewcommand{\frontmatter@above@affiliation}{\vspace{6pt}}
\makeatother
\begin{document} 

\title{Strain-Engineering Giant Pseudo-Electric Fields in Multilayer Graphene}

\author{Yiwen Liu$^{1,*,\dagger}$}
\author{Ambikesh Gupta$^{1,\dagger,\text{\P}}$}
\author{Adrián Ceferino$^{2,\dagger}$}
\author{Youngjoon Choi$^{3,\ddagger}$}
\author{Pierre A. Pantaleón$^{2,\S}$}
\author{Zhen Zhan$^{2}$}
\author{Suman Sanki$^{1}$}
\author{Hari Stoyanov$^{3}$}
\author{Takashi Taniguchi$^{4}$}
\author{Kenji Watanabe$^{5}$}
\author{Francisco Guinea$^{2,6}$}
\author{Andrea Young$^{3}$}
\author{Haim Beidenkopf$^{1,*}$}
\author{Nurit Avraham$^{1,*}$}
\affiliation{$^{1}$Department of Condensed Matter Physics, Weizmann Institute of Science, Rehovot 7610001, Israel}
\affiliation{$^{2}$IMDEA Nanoscience, C/ Faraday 9, 28049 Madrid, Spain}
\affiliation{$^{3}$Department of Physics, University of California Santa Barbara, Santa Barbara, California 93106, USA}
\affiliation{$^{4}$Research Center for Electronic and Optical Materials, National Institute for Materials Science, Tsukuba, Japan.}
\affiliation{$^{5}$Research Center for Materials Nanoarchitectonics, National Institute for Materials Science, Tsukuba, Japan}
\affiliation{$^{6}$Donostia International Physics Center, Paseo Manuel de Lardizábal 4, 20018 San Sebastián, Spain}
\maketitle 

\section{Abstract} 
\textbf{Electromagnetic scalar and vector potentials can be engineered in solids through lattice distortions, providing a versatile means of tuning electronic band structures. The pseudovector component, which generates pseudo-magnetic fields, has received much attention for its ability to realize a zero-field quantum Hall effect. However, the pseudoscalar component, giving rise to pseudo-electric fields to which many electronic systems are intrinsically susceptible, has not yet been experimentally explored. Here, we demonstrate the emergence and profound consequences of the pseudoscalar potential in a multilayer graphene system. Atomic relaxation reconstructs the system into large Bernal-stacked domains separated by one-dimensional rhombohedral channels. Within this inter-dimensional system, we achieve the first direct atomic-scale imaging of distinct lattice deformations in graphene under strain. In rhombohedral channels, we observe that the tensile lattice deformation induces giant pseudo-electric fields, producing ultra-flat dispersion with divergent density of states and a large energy gap. 
This results in enhanced electronic correlations, leading to flat band instabilities and striped modulation in the density of states that would otherwise require strong external displacement fields. Moreover, the gap opened by the pseudoscalar potential protects topological boundary states that persist in the absence of applied fields. Our results provide a comprehensive atomic-scale characterization of graphene elasticity and direct spectroscopic evidence for the nontrivial pseudo-electric field arising from lattice distortion. More broadly, it provides an intrinsic and tunable pathway to new quantum states of matter devoid of external electromagnetic fields. }

\begin{figure*}
    \centering
    \includegraphics[scale = 0.28]{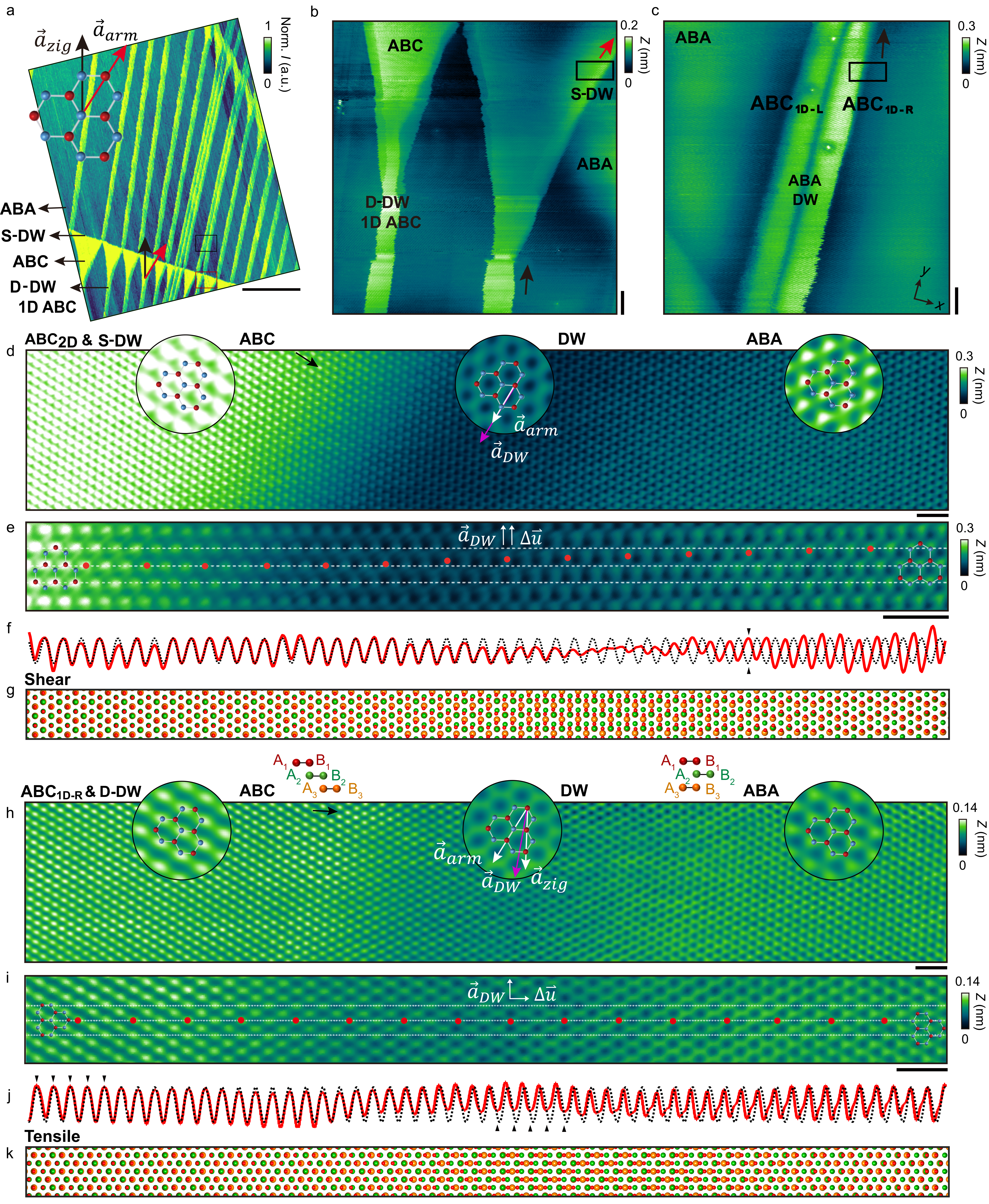}
    \caption{\textbf{Structural transition from 2D-ABC to 1D-ABC.} 
    \textbf{a}, cAFM image of the ABA- and ABC-stacked domains.
    \textbf{b}, STM topography showing triangular ABC domains transitioning into 1D channels. A shear single DW (S-DW) lies between the triangular ABC and ABA domains along the armchair direction (top red arrow). 
    \textbf{c}, STM topography of conjoined 1D-ABC channels with DWs oriented close to the zigzag direction. DWs and a narrow ABA-channel separate two 1D-ABC channels. 
    \textbf{d}, Sublattice-resolved STM topography of an ABC–DW–ABA region, acquired in a different triangular domain with the same DW orientation as the black box in \textbf{b}, showing lattice displacements as the stacking changes across a shear DW. Insets: zoomed-in image overlaid with the graphene lattice. 
    \textbf{e}, Cropped slice of \textbf{d}, with red dots tracing the displacement of A-sublattice shifted along the armchair direction in the S-DW. The image is a zoomed-in area perpendicular to the DW, showing lattice movement in the sheared direction. $\Delta u$ is the lattice displacement at each position.
    \textbf{f}, Line profile across the interface (red, along the black arrow in \textbf{d}). The dotted black curve represents a standard sinusoidal fit on the ABC side, highlighting the phase shift induced by shear atomic displacement.
    \textbf{g}, Simulation of lattice displacements in a shear DW. Top-, middle-, and bottom-layer atoms are denoted by red, green, and orange dots.  
    \textbf{h}, Sublattice-resolved STM topography of a 1D-ABC–DW–ABA region (black box in panel \textbf{c}), showing lattice distortions around a tensile DW along the nearly zigzag direction. 
    \textbf{i}, Cropped slice of panel \textbf{h}, with red dots tracing the displacement of atoms compressed along the zigzag direction due to tensile strain. 
    \textbf{j}, Line profile across the interface (red, along the black arrow in \textbf{h}). The black curve is a fit to the ABC side.
    \textbf{k}, Simulation of lattice displacements of tensile ABC-DW-ABA stacking. Scale bars: \textbf{a}, 500 nm; \textbf{b,c}, 20 nm; \textbf{d,e,h,i}, 1 nm. 
    }
    \label{fig1}
\end{figure*}

\section{Introduction}

Electromagnetic fields are among the most powerful tools for controlling quantum matter \cite{Nagaosa-2012,xiao-2010}. Electric fields polarize charge carriers and open bandgaps, whereas magnetic fields quantize electronic motion and reconstruct energy spectra \cite{xiao-2010}. Crystals provide an alternative route by which the lattice itself can generate analogous electromagnetic fields. Because electronic motion is intimately tied to atomic and bond arrangement, strain can act as an emergent electromagnetic environment for charge carriers. This concept of emergent electromagnetism has become a powerful framework for engineering electronic structure in quantum materials \cite{zeljkovic2015strain,liu2014tuning,kamboj2019generation,cheng2010landau}. 
So far, however, the main focus has been on the magnetic vector-potential component of this emergent electromagnetism. Non-uniform shear strain couples the electronic states with the pseudovector potential \cite{guinea-energy-2010,guinea-generating-2010, mann-tunable-2020, vozmediano-gauge-2010,ceferino-pseudomagnetic-2024},
enabling the experimental realization of Landau quantization in the absence of an applied magnetic field \cite{levy-strain-induced-2010,zhou-imaging-2023,nigge-room-2019,georgi-tuning-2017,liu-realizing-2022,liu-tailoring-2018,ho2021hall,hsu-nanoscale-2020,kang-pseudo-magnetic-2021,shi-large-area-2020}. Such emergent magnetic fields can reach magnitudes of tens to hundreds of tesla and have been harnessed as a platform for correlated states through flat bands \cite{levy-strain-induced-2010,mao2020evidence}. 

The pseudoscalar potential \cite{suzuura2002phonons, aharonov-significance-1959, ilan-pseudo-electromagnetic-2020,choi-controlling-2010} has received far less attention than the pseudovector potential \cite{strain-choi-2010, kerjouan-quantum-2024, shioya-electron-2015,wang-global-2021,ilan-pseudo-electromagnetic-2020,layek2025modulation}. A key limitation is that in single-layer systems the scalar potential acts as a deformation potential that shifts the Dirac point energy, resulting primarily in modest variations in the local carrier density \cite{strain-choi-2010, kerjouan-quantum-2024, shioya-electron-2015,wang-global-2021,ilan-pseudo-electromagnetic-2020,zeljkovic2015strain}. However, displacement-susceptible two-dimensional (2D) quantum materials have recently taken center stage. These include rhombohedral multi-layer graphene \cite{zhou-half-and-2021,chen-correlated-2022,liu2024spontaneous,lu2024fractional,xie2025tunable,zhou2024layer, kumar2025superconductivity} and transition metal dichalcogenides (TMDCs) \cite{redekop-direct-2024}.
In such systems, interlayer potential differences can significantly modify the low-energy band structure and give rise to exotic phases including unconventional superconductivity \cite{zhou-superconductivity-2021, kumar2025superconductivity,han2025signatures}, fractional quantum anomalous Hall states \cite{cai2023signatures,lu2024fractional}. Hydrostatic or tensile strain is predicted to induce a layer-asymmetric pseudoscalar potential and open up bandgaps in graphene \cite{choi-controlling-2010}. The large density of states in the resulting van Hove singularities (vHs) serves as a promising platform for promoting symmetry-broken phases \cite{zhou-half-and-2021,chen-correlated-2022,liu2024spontaneous,lu2024fractional,xie2025tunable,zhou2024layer, kumar2025superconductivity}, without relying on externally applied displacement fields.  

Such effects have not yet been directly resolved experimentally, and the extent of their consequences remains unknown \cite{song2025artificial}. The challenge lies in connecting strain textures and electronic response at the nanoscale level. 
Although transport measurements have demonstrated that strain in 2D van der Waals heterostructures can serve as a global tuning knob for the ground state, for example, in the control of the anomalous Hall effect \cite{ma2025}, it primarily acts as a source of disorder in realistic devices \cite{couto2014random}. 
Scanning probe techniques \cite{hoke2026imaging, zhou-imaging-2023} have revealed these local strain inhomogeneities, including modifications to moiré superlattices and electronic band structures at the hundred-nanometer scale. However, a direct experimental link between atomic-scale lattice deformation, the full local strain tensor, and the resulting electronic band structure reconstruction has remained elusive. Establishing this connection in displacement-field–sensitive systems, where different strain textures can generate qualitatively different pseudo-electromagnetic responses, will enable the manipulation and engineering of electronic states and promote emergent instabilities.

Using scanning tunneling microscopy and spectroscopy (STM/STS), combined with elasticity modeling and tight-binding calculations, we resolve atomic-scale strains by directly imaging the consequent lattice deformation and unravel the impact of the pseudoscalar potential in a trilayer graphene (TLG) system. We present the first atomic-scale images of a strain-induced structural transition from 2D to one-dimensional (1D) ABC stripes \cite{li2020global,kerelsky-moireless-2021, halbertal-moire-2021, park2025unconventional}, concomitant with a change in domain wall (DW) character from shear to tensile \cite{alden-strain-2013}. By comparing the local density of states (LDOS) in the two phases, we reveal clear imprints of a pseudo-electric field in the tensile 1D channels in the form of divergent density of states with $\sim$5-10 meV full width at half maximum (FWHM) and a large energy gap $\sim$35 meV. The reconstructed band structure originating from this pseudoscalar potential enables us to directly observe the correlated anisotropic density of states and unconventional topological modes devoid of external electromagnetic fields. These findings establish elasticity-driven pseudo-electric fields as a powerful and previously underappreciated mechanism for engineering band structure and emergent quantum states in displacement-field–sensitive systems.

\section{Strain-driven dimensional crossover}
We begin by visualizing the strain landscape of a minimally twisted TLG: a monolayer rotated by \(0.05^\circ\) on top of a Bernal bilayer graphene. The conductive-AFM map in Fig.~\ref{fig1}a reveals the strain-driven relaxation into alternating ABA (green) and ABC (yellow) commensurate domains \cite{kazmierczak-strain-2021, halbertal-moire-2021}. In the lower part of the image highlighted by the red box, ABC-stacked regions form a series of triangular 2D domains. Yet, most of the image is dominated by regions where ABC domains relax into long 1D channels (approximately 2 $\mu$m long). These regions, spanning the middle and top of the image, exhibit a structural transition in which 2D-ABC collapses into 1D-ABC channels under a global elongation. In some devices, 1D-ABC channels extend across the entire sample length as well (see Supplementary Sec.~1 and Fig.~S1 for more details \cite{SM}). This structural transition is consistent with well-studied theoretical models, which suggest that increasing biaxial elongation with additional shear strain drives a transition of triangular domains into stripe configurations beyond a critical strain, particularly when the twisted top layer is a monolayer, which is more elastic than thicker top multilayers \cite{lebedeva-energetics-2019,lebedeva-two-2020,escudero-designing-2024,ouyang2025structural}. These models further suggest that this shape evolution is accompanied by a change in DW character, from shear-type, aligned to the armchair direction with lattice movement predominantly along the DW, to tensile-type, aligned in the zigzag direction with lattice movement predominantly perpendicular to the DW. 
 
While the overall morphology we observe accords with the theoretical picture, a decisive test is to directly measure and identify the strain character. To that end, we imaged both regions at the atomic scale. Figures ~\ref{fig1}b and \ref{fig1}c provide representative topographic images of ABC-ABA interfaces in the confined triangular region  \cite{halbertal-moire-2021} and in the vicinity of the extended 1D channels, respectively (black squares in Fig.~\ref{fig1}a). Zooming into the single DW (S-DW) in the triangular region (Fig.~\ref{fig1}b, rectangle), the atomically resolved topography in Fig.~\ref{fig1}d reveals that the lattice is triangular in the adjacent ABC and ABA domains but becomes more hexagonal within the DW, whose orientation tracks the armchair direction (magenta arrows). Lattice displacement extracted from the topographic images (red dots, Fig.~\ref{fig1}e) shows atoms sliding parallel to the wall, and a line profile taken perpendicular to it (Fig.~\ref{fig1}f) compares the measured corrugation (red) with an ideal sinusoidal corrugation (black), exhibiting the expected out-of-phase shift from ABC to ABA that hallmarks a shear-type DW (Fig.~\ref{fig1}g) \cite{koshino-electronic-2013,alden-strain-2013,jiang-soliton-dependent-2016}. Thus, we find that armchair-aligned DWs exhibit shear strain, where the lattice movement is along the DW. In contrast, in the 1D-ABC-DW region marked by a rectangle in Fig.~\ref{fig1}c, the DW orientation is closer to zigzag (magenta arrow in Fig.~\ref{fig1}h and zoomed-in lattice shown in Fig.~\ref{fig1}i). Here, atoms shift predominantly perpendicular to the DW along the bond direction. A phase analysis of the atomic corrugation (Fig.~\ref{fig1}j) shows progressive compression as the DW is approached, indicating tensile strain in that direction (Fig.~\ref{fig1}k) \cite{mesple-giant-2023}. 
 
Overall, these observations establish a clear atomic-scale picture of an elongation-driven structural crossover from nearly strain-free 2D triangles to strained 1D-ABC channels, accompanied by a transition from shear (armchair-aligned) to tensile-type (zigzag-aligned) DWs, as also visualized in Fig.~S2 and summarized in Table~S1 \cite{lebedeva-two-2020,jiang-manipulation-2018,hasdeo-long-lived-2017,SM}. This strain-induced evolution provides the structural foundation for the distinct electronic responses described below.
\begin{figure*}
    \centering
    \includegraphics[width=\textwidth]{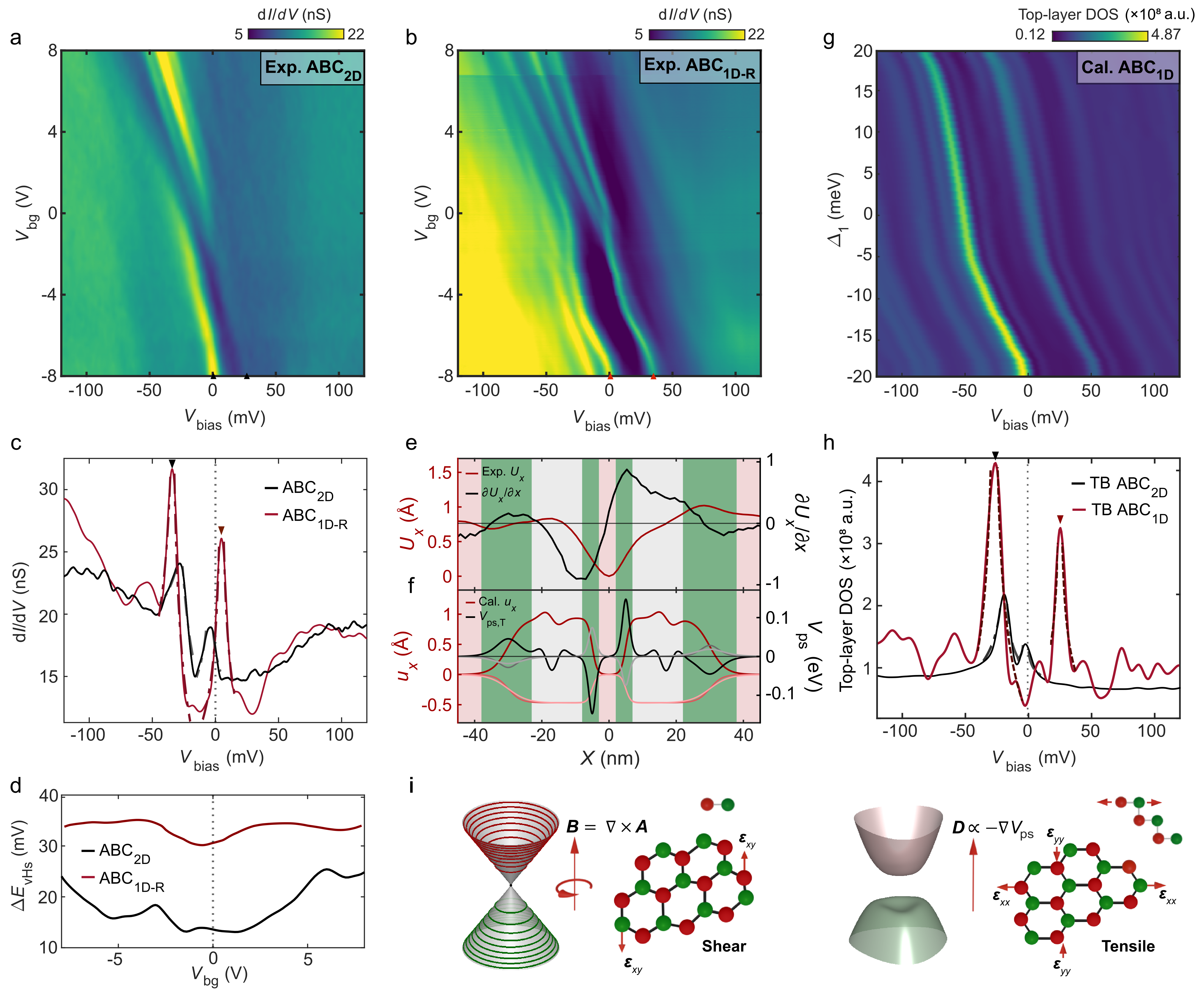}
    \caption{\textbf{Ultra-flat band and energy gap via tensile strain–induced pseudo-electric fields.} 
    \textbf{a}, $\mathrm{d}I/\mathrm{d}V$ ($V_{\rm bias}$, $V_{\rm bg}$) map measured
    at the center of a triangular ABC domain at $T=2\,\mathrm{K}$. $V_{\mathrm{set}}= -200$ mV, $I_{\mathrm{set}} = 100$ pA.
    \textbf{b}, Same measurement as in panel \textbf{a}, but taken in the middle of a ABC$_\mathrm{1D-R}$ channel at $T=4\,\mathrm{K}$. The vHs are indicated by the red arrows at $V_{bg}$= -8 V. $V_{\mathrm{set}}= 120$ mV, $I_{\mathrm{set}} = 800$ pA.
    \textbf{c}, $\mathrm{d}I/\mathrm{d}V$ spectra on ABC$_\mathrm{2D}$ (black) and ABC$_\mathrm{1D-R}$ channel (red) showing sharper CB and VB vHs at low energy for the 1D-ABC case. Both spectra are acquired at $V_{\mathrm{bg}} = 0~\mathrm{V}$.
   Red dashed line - 1D-ABC spectrum fit to a background-offset power-law model of the form $\mathrm{d}I/\mathrm{d}V(V) = a\,|V - V_{0}|^{d} + c$, where $a$ is an amplitude, $V_{0}$ the singular bias voltage, $d$ the power-law exponent with an exponent of $-1/4$, and $c$ a constant background. Black dashed line - 2D-ABC spectrum fit to a logarithmic divergence. It captures the broader, less divergent vHs features of 2D-ABC TLG at zero back-gate voltage.
    \textbf{d}, Energy spacing between the two vHs ($\Delta E_{\rm vHs}$) in the ABC$_\mathrm{2D}$ triangle and ABC$_\mathrm{1D-R}$ channel as a function of $V_{\rm bg}$.
    \textbf{e}, Left axis: Moiré displacement $U_x$ (solid red line) extracted from the profile in Fig.~S10 \cite{SM}. Right axis: $\partial U_x/\partial x$ in percentage, calculated from the profile (black). 
    \textbf{f}, Left axis: strain field $u_x$ in the theoretical model.
    Red curves show the lattice displacement $u_{x}$ for the top (dark red), middle (red), and bottom (light red) layers. 
    Right axis: Pseudoscalar potential derived from the strain gradient: $V_{ps} \propto \frac{\partial u_x}{\partial x}$ for the top (black), middle (dark gray), and bottom (light gray) layers. The potential varies spatially across the channel.
    \textbf{g}, Simulated top layer DOS map from the strain profile in \textbf{f}.
    \textbf{h}, Simulated top layer DOS for strained (red) and strain-free (black) domains at zero applied interlayer potential. The power-law fitting for the 1D-ABC and logarithmic fitting for the 2D-ABC are also applied to the theoretical DOS.
    \textbf{i}, A schematic of the electronic response to different types of elastic deformations. Left: the monolayer graphene band structure and its lattice. A non-uniform strain generates a pseudo-magnetic field in a Dirac band, manifested as Landau-level-like bands. Right: tensile strain applied to one layer of RTG modifies the electron density distribution and thus the electrostatic potential of layers. This produces an out-of-plane potential gradient and an effective pseudo-electric field.
    }  
    \label{fig2}
\end{figure*}
\section{Ultra-flat band and giant energy gap in 1D-ABC channels}

We next investigate how the strain-driven structural transition modifies the LDOS in the 1D-ABC channels. To this end, we perform gate-tunable STS measurements on the strained 1D channels and directly compare them with the same measurement on a nearly strain-free 2D-ABC domain. Figures~\ref{fig2}a and \ref{fig2}b present the corresponding  $\mathrm{d}I/\mathrm{d}V(V_{\mathrm{bias}}, V_{\mathrm{bg}})$ maps acquired in the 2D-ABC domain and in the middle of the 1D channel, respectively. In our single-gate STM geometry, sweeping $V_{\mathrm{bg}}$ simultaneously changes both the displacement field and the carrier density. Consistent with previous reports \cite{liu-visualizing-2024}, in the 2D-ABC domains, the application of a displacement field through the backgate voltage opens an energy gap between the conduction (CB) and valence (VB) bands that increases with the field and progressively flattens the band edges, such that the vHs (black arrows in Fig.~\ref{fig2}a at $V_{\mathrm{bg}} = -8$~V) become more pronounced with increasing field. This evolution is also accompanied by layer-polarization of the wavefunction, under which the VB vHs strengthens at negative gate voltages, while the CB vHs dominates at positive voltages, as measured on the top layer. 
In the 1D-ABC measurement (Fig.~\ref{fig2}b), the difference is conspicuous: a prominent gap and a sharp peak in LDOS can be clearly identified even at zero external $D$ ($V_{\mathrm{bg}} = 0$~V) and remain nearly unchanged with increasing external $D$ as shown in Fig.~\ref{fig2}d (see Fig.~S6 and Fig.~S15 for the spectral homogeneity \cite{SM}). Thus, tensile-strain dominated 1D spectra feature a larger gap, ultra-flat bands, and a redistribution of the wavefunction as if an intrinsic "pseudo-displacement field" were present in the 1D channels (see Fig.~S1 for the evolution from 2D to 1D and Fig. S3 for its robustness \cite{SM}). It is worth noting that we focus on the two-channel configuration, where ultra-squeezing of the inner DW–ABA–DW region generates exceptionally strong pseudoscalar potentials, as discussed below (see Fig. S1 \cite{SM}).

To quantify this enhanced band flattening, we compare the corresponding spectra in Fig.~\ref{fig2}c for 1D stripes and 2D triangles extracted at $V_{\mathrm{bg}}=0$~V. In the 1D channel, LDOS exhibits remarkably sharper vHs with a power-law divergence with an exponent of $\alpha=-0.25$ (see Figs.~S4 and ~S5 for vHs LDOS distributions \cite{SM}), consistent with a high-order vHs \cite{yuan-magic-2019}. This contrasts with a much weaker, nearly logarithmic divergence of vHs observed in the 2D-ABC domain. Indeed, as shown in Fig.~\ref{fig2}b, CB vHs in 1D case remains pinned at the Fermi energy ($E_F$, represented by $V_{\mathrm{bias}}=0$) over around $\Delta V_{bg} = 2~\mathrm{V}$ window, compared to only $\Delta V_{bg} = 0.4~\mathrm{V}$ in the 2D domain, indicative of ultra-flat bands with enhanced DOS. Such a long pinning range in the gate occurs in 2D triangular domains only under high-displacement fields (see the comparison in Fig. S16 \cite{SM}). We also extract in Fig.~\ref{fig2}d the separation between the two vHs peaks, $\Delta E_{\rm vHs}$ (hereafter referred to as the 'gap', see the discussion of Fig.~S5 \cite{SM} for the characterization of $\Delta E_{\rm vHs}$) as a function of $V_{\mathrm{bg}}$ for the two cases. The 1D spectrum displays a larger and rather constant apparent gap of 35 mV across the gate range, versus only 12 mV at the zero gate voltage, and increases on either side of the zero gate in 2D-ABC (see Table~S2 for the comparison \cite{SM}). Importantly, although the narrow width of the 1D-ABC channels naturally raises the possibility of finite-size confinement, our measurements can rule out this effect. The channel width varies from approximately $7$ to $15$~nm, yet $\Delta E_{\rm vHs}$ shows no systematic decrease with increasing width (Fig.~S8 \cite{SM}). This is inconsistent with the width dependence expected for finite-size quantization, which for the approximately quadratic band-edge dispersion scales as $E_{\rm conf}\propto W^{-2}$. We therefore attribute the enhanced single-particle gap primarily to pseudo–displacement fields-induced band-structure modification (see Fig.~S7 and Fig.~S8 for a detailed analysis of competing effects \cite{SM}).

A deeper understanding of the origin of the pseudo–displacement field emerges by comparing our experimental observations with elasticity modeling and tight-binding calculations (also see Fig. S12 \cite{SM}). To enable a better estimation of the band structure and LDOS profile in this system, we analyze the strain across the channels.
The strain profile across the dual 1D-ABC channel in Fig.~\ref{fig2}e is extracted from topographic measurements by analyzing the moiré pattern of unaligned TLG–hBN (also see Fig.~S10 \cite{SM}). The line profile along the direction of the moiré lattice shows the displacement $U_x(x)$ of the moiré pattern and its gradient $\partial U_x/\partial x$, providing a proxy for the graphene lattice movement perpendicular to the DW (see Fig.~S10 and Fig.~S11 for details of strain analysis and the magnitude of strain \cite{SM}). Note that this moiré analysis can provide a sensitive probe of the spatial variation of strain across the DW shown in  Figs.~\ref{fig1}h–k \cite{jiang-visualizing-2017,escudero-designing-2024,sequeira2024manipulating, kapfer2023programming}. However, because moiré contributions associated with different graphene layers cannot be independently resolved in our local Fourier analysis, the moiré pattern alone does not provide a unique layer-resolved strain reconstruction. We therefore use the experimentally obtained moiré distortion to constrain the spatial strain profile, while the layer-resolved strain distribution entering the electronic-structure calculation is obtained from the continuum relaxation model.

As shown in Fig.~\ref{fig2}f, a strain profile mimicking Fig.~\ref{fig2}e was used in the simulation. Additionally, a top-layer tensile strain in the 1D-ABC channel was included to account for the attractive forces between the inner and outer DWs surrounding the 1D-ABC channels. Here, the pseudopotentials of the middle and bottom layers are nearly identical because they have similar strains (see Table~S3 for the layer-resolved strain percentages \cite{SM}), while the top layer experiences the greatest shifts in energy derived from elastic deformations. In Fig. \ref{fig2}h, we show the resulting theoretical LDOS profile in a 1D channel as well as in a 2D sheet (also see Fig. S14 \cite{SM}). The calculated LDOS reproduces most of the features observed experimentally: a robust flat band (see Fig. \ref{fig2}h); a significant opening of the gap at zero applied $D$ with a weak dependence of the gap size on the interlayer potential, and a suppressed gate-modulation of the wavefunction layer polarization (see Fig. \ref{fig2}g).
Note that the 1D-ABC LDOS profile differs from the confined LDOS in TMDC channels \cite{li-imaging-2024}, where discrete spectral features and charging phenomena are observed. We attribute this difference to the gap closing in the DWs surrounding ABC channels, which results in transparent barriers around each 1D-ABC channel (also see Fig. S8 \cite{SM}). 

Our results demonstrate a fundamental distinction in how strain couples to the electronic degrees of freedom in different graphene systems. Strain generates a pseudo-electromagnetic potential $A=(V_{ps}, \mathbf{A}$) \cite{guinea-energy-2010,vozmediano-gauge-2010}, where $A_x\propto(\varepsilon_{xx}-\varepsilon_{yy})$, $A_y\propto (\varepsilon_{xy}+\varepsilon_{yx})$ and $V_{ps}\propto (\varepsilon_{xx}+\varepsilon_{yy})$, where $\varepsilon_{ij}$ is the strain tensor (see Supplementary Sec.~2.2(a) \cite{SM}
for details of the strain-induced pseudovector and pseudoscalar potentials). In a Dirac band system, such as monolayer graphene, non-uniform shear strain generates a pseudovector potential $\mathbf{A}$ that acts as a valley-contrasting magnetic field. Because the Dirac Hamiltonian couples linearly to momentum, the non-uniform vector potential enters as a gauge field, strongly modifying the electronic dispersion and producing Landau-level-like quantization (see Fig.~\ref{fig2}i, left). In contrast, tensile strain gives rise to a pseudoscalar potential $V_{ps}$ that primarily shifts the Dirac point energy without altering the overall band topology. Therefore, this effect has been neglected in most strain-related studies \cite{bi-designing-2019}. 

In multilayer systems, and specifically, in the rhombohedral graphene family, where the low-energy bands exhibit parabolic-like band touching and are highly sensitive to displacement fields, the role of the pseudoscalar potential becomes significant. Tensile strain in each layer alters the electron density and work function of that layer separately \cite{wang-global-2021}. As a result, asymmetric tensile strain across the layers can produce a potential gradient that effectively acts as a displacement field $D$ applied perpendicular to the layers (see Fig.~\ref{fig2}i, right). This pseudoscalar coupling directly opens a band gap and flattens the low-energy bands in the strained 1D-ABC channels, providing an elegant platform for exploring external-field-free correlated and topological phenomena. 

Note that the essential requirement is an imbalance between the layer potentials, which breaks inversion symmetry. This imbalance can be produced by different strain configurations, for example by preferentially straining the top, middle, or bottom layer. Specifically, in our system, the top layer experiences the most significant amount of strain (including extra elongation, lattice relaxation, and DW attraction \cite{halbertal-moire-2021}), due to the mono-bilayer twist configuration. This interlayer strain difference induces distinct pseudoscalar potentials in the top layer (Fig. \ref{fig2}f), thereby tuning the band structure and providing a direct visualization of the effect of tensile strain in the band structure of 1D-ABC channels.
Moreover, the coupling between multiple ABC channels and the ultra-squeezed DW–ABA–DW region also helps amplify the pseudoscalar potential, leading to substantial gap opening and flatter bands than those of a single ABC channel (see the robust flat band and its line profile in the dual-ABC and multi-ABC stripes in Fig. S3 \cite{SM}).
\section{Strain-Engineered Correlated States in 1D-ABC Flat Bands}
\begin{figure}
    \centering
    \includegraphics[scale=0.7]{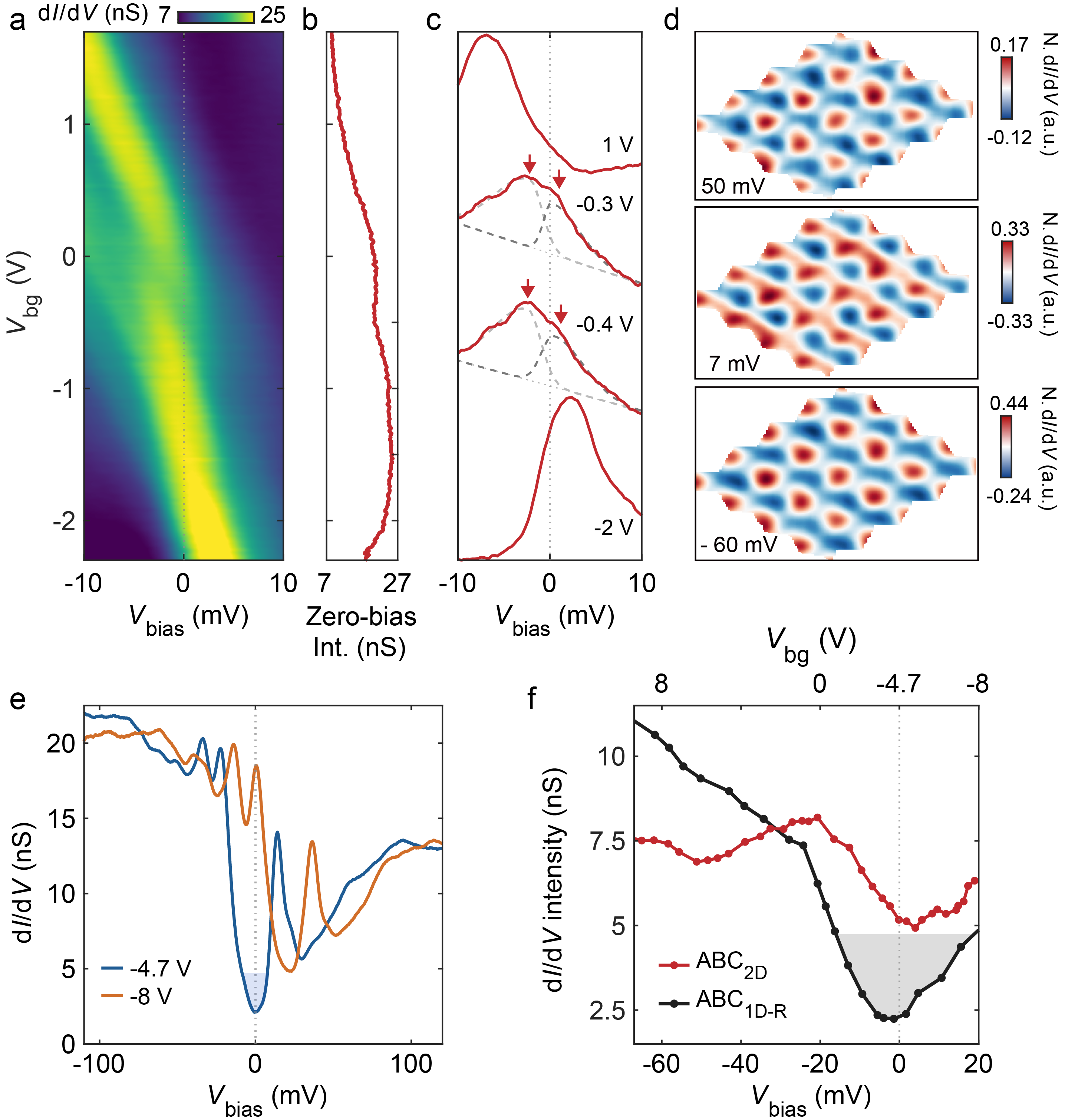}
    \caption{\textbf{Enhanced correlations and flat band instabilities in 1D-ABC channels engineered by pseudo-electric fields.}
    \textbf{a}, $\mathrm{d}I/\mathrm{d}V(V_{\rm bias}, V_{\rm bg})$ map showing vHs splitting near $E_F$ (right 1D-ABC channel). $T = 4$ K. See the comparison with 2D-ABC flat bands in Fig. S16 \cite{SM}. 
    \textbf{b}, Zero-bias intensity as a function of $V_{\rm bg}$.
    \textbf{c}, Representative $\mathrm{d}I/\mathrm{d}V$ spectra at different $V_{\rm bg}$ values. The dashed curves show the Gaussian components used to fit the split peaks.
    \textbf{d}, Super-resolved spatial normalized $\mathrm{d}I/\mathrm{d}V$ maps (N. $\mathrm{d}I/\mathrm{d}V$) at $V_{\rm bg}=0\,\mathrm{V}$ showing the LDOS at $V_{\rm bias}=-60\,\mathrm{mV}$ (bottom), $7\,\mathrm{mV}$ (middle), and $50\,\mathrm{mV}$ (top). Here, N. $\mathrm{d}I/\mathrm{d}V$ is defined as N. $\mathrm{d}I/\mathrm{d}V = \mathrm{d}I/\mathrm{d}V - \langle\mathrm{d}I/\mathrm{d}V \rangle_{\rm line}$. The LDOS maps are acquired at $V_{\mathrm{bg}} = 0~\mathrm{V}$. Note that the maps were acquired at a slightly different position from the STS data in panel a. Like in panel \textbf{a}, the band is partially filled but the right split vHs lies near $5$--$7$~mV, see details in Fig. S17 \cite{SM}.
    \textbf{e}, $\mathrm{d}I/\mathrm{d}V(V_{\rm bias}, V_{\rm bg})$ spectra measured in the middle of the 1D-ABC channel at $T=4\,\mathrm{K}$. The blue region highlights the strong suppression of the tunneling DOS near $E_F$. See Fig. S18 for further analysis \cite{SM}.
    \textbf{f}, $\mathrm{d}I/\mathrm{d}V$ intensity corresponding to the dip feature in the spectra in 1D-ABC (black) vs in 2D-ABC (red) extracted from Figs. \ref{fig2}b and \ref{fig2}a by following the dip position across the gate range (see Fig. S18 for the tracking trajectory in the maps \cite{SM}). $V_{\rm bg}$ label on top of the plot corresponds to 1D-ABC gate-STS for complementing panel \textbf{e}. When the dip is centered at $E_F$ ($V_{\rm bias}=0$), the tunneling DOS suppression is stronger in 1D-ABC than in 2D-ABC. 
}
    \label{fig3}
\end{figure}
The emergence of ultra-flat bands and the large bandgap provide an opportunity to explore many-body phenomena in a system where electronic interactions are strongly amplified by both strain and quasi-1D confinement. Figure \ref{fig3}a presents a gate-dependent $\mathrm{d}I/\mathrm{d}V$ map, acquired at $T=4\,\mathrm{K}$. In the vicinity of zero backgate, where the CB vHs is pinned to $E_F$, we observe an apparent splitting of $\sim 4\,\mathrm{mV}$ in the $\mathrm{d}I/\mathrm{d}V$ spectrum. The splitting is also evident in Fig. \ref{fig3}b, where the zero-bias intensity in the $\mathrm{d}I/\mathrm{d}V$ spectrum exhibits a dip as a function of $V_\mathrm{bg}$. The zero-bias LDOS remains finite, indicating a correlated metallic phase \cite{zhou-half-and-2021}. Representative spectra in Fig. \ref{fig3}c reveal that the band is initially narrow (FWHM $\approx 5\ \mathrm{mV}$) and nearly empty at $V_{\mathrm{bg}}$ = $-2$~V. As $V_{\mathrm{bg}}$ is increased from $-2$~V to $0$~V, the band broadens and splits, and subsequently recombines into a single band at $V_{\mathrm{bg}} = 1$~V. 

This evolution and spectral-weight distribution are reminiscent of the band splitting associated with partially isospin-polarized states observed in transport measurements of rhombohedral trilayer graphene (RTG) \cite{zhou-half-and-2021}, and the carrier density and effective displacement field probed here, $n\approx0.7\times10^{12}\,\mathrm{cm^{-2}}$, $D_{\mathrm{eff}}\approx0.2\,\mathrm{V/nm}$, is comparable to the regime between the half-metal and symmetric phase \cite{zhou-half-and-2021}. This instability in this 1D-ABC domain emerges near zero $V_{\mathrm{bg}}$, where the DOS of the vHs is enormous and the interlayer potential difference $\Delta_1$ $\sim 20\ \mathrm{meV}$, as estimated from the gap size $\Delta E_{\rm vHs}\sim 35\ \mathrm{meV}$ (also verified by fitting the QPI pattern in Fig. S9 \cite{SM}). As elucidated earlier, we attribute this large $\Delta_1$ at $V_{\mathrm{bg}}$ = $0$~V to the strain-induced pseudo-displacement field in 1D channels. Reaching a comparable $\Delta_1$ and similar isospin polarization in the 2D-ABC domain requires tuning to $V_{\mathrm{bg}}$ = $-9$ V (see Supplementary Sec.~3
and Fig.~S16 \cite{SM}). Although this places the 1D-ABC channels in a spectroscopic regime comparable to 2D RTG under a high displacement field, their strong strain and reduced dimensionality introduce additional ingredients that may modify the underlying phase diagram, thereby providing a complementary RTG platform for exploring correlated phases beyond the 2D limit.

\par
To probe the real-space order associated with this 1D instability, we performed lattice-resolved mappings at $V_\mathrm{bg} = 0$ V. As shown in Fig. \ref{fig3}d, super-resolved lattice maps reveal a clear energy-dependent contrast in lattice order. At energies far from $E_F$ (around $\pm 50$~mV), the lattice retains approximate $C_3$ symmetry. However, close to $E_F$ ($V_{\mathrm{bias}} = 7~\mathrm{mV}$ in Fig. \ref{fig3}d, middle panel) where the flat band splits, the charge density develops a striped modulation oriented along the bond and strain directions. This can be interpreted as a strain-seeded, Stoner-type response of a higher-order vHs \cite{yuan-magic-2019}: strain in the 1D channel breaks the symmetry of the crystal and selects a nematic axis, splitting the otherwise degenerate vHs \cite{mayorov2011interaction, mariani2012fictitious}. Because the LDOS is large near a higher-order vHs, exchange interactions amplify this seed via a Stoner-like factor \cite{yuan-magic-2019}, producing larger splitting and an anisotropic LDOS, observed as stripes aligned with the strain direction. The broken rotational symmetry suggests that the Stoner instability is of the valley isospin, as was observed \cite{morissette2025striped} and calculated \cite{PhysRevLett.133.206604, murshed2025charge} in similar systems under strong displacement fields. Interestingly, we also observe a suppression of quasiparticle coherence near the flat band at $E_F$, suggesting interaction-driven localization tendencies in a flat-band regime where Coulomb interaction dominates (see Fig. S19) \cite{hu-quantum-2024,wu-dichotomy-2025,SM}. 
\par
To further examine the influence of geometrical effects on the tunneling DOS in the correlated regime, we examined the LDOS behavior near $E_F$ \cite{pikus1995coulomb, song2018observation}. Fig. \ref{fig3}e compares $\mathrm{d}I/\mathrm{d}V$ spectra measured at two gate voltages: $V_\mathrm{bg}= -8$ V, where the $35$ meV energy gap lies just above $E_F$, and $V_\mathrm{bg}= -4.7$ V where $E_F$ is centered within the gap. In the latter, the $\mathrm{d}I/\mathrm{d}V$ signal within the gap is strongly suppressed, even though the gap magnitude as a function of $V_\mathrm{bg}$ remains unchanged (Fig. \ref{fig2}d), indicating an enhanced suppression of the tunneling DOS at $E_F$. A similar suppression near $E_F$ appears also in the 2D-ABC region (Fig.~\ref{fig3}f), but is markedly stronger in the 1D-ABC channels (see Fig. S18 for further analysis \cite{SM}). This enhanced suppression reflects the strengthened correlations by confinement \cite{pikus1995coulomb}, which reduces transverse hopping and screening, imposing an additional injection-energy cost, and effectively producing a stronger Coulomb gap. Although a clear power-law decay of the LDOS, a hallmark of Tomonaga–Luttinger liquid behavior, is not observed, the strong near-$E_F$ suppression indicates a crossover toward the 1D limit \cite{voit-one-dimensional-1995,stuhler-tomonagaluttinger-2020}. Applying a perpendicular magnetic field further amplifies this trend by quenching transverse kinetic energy by gapping the bulk; correspondingly, the tunneling DOS develops a power-law suppression, evidencing a magnetic-field-driven crossover toward a purely 1D regime (see Fig. S20 \cite{SM}).
\section{Strain-stabilized topological edge states at ABA-1D-ABC interfaces}
\begin{figure*}
    \centering
   \includegraphics[width=0.95\textwidth]{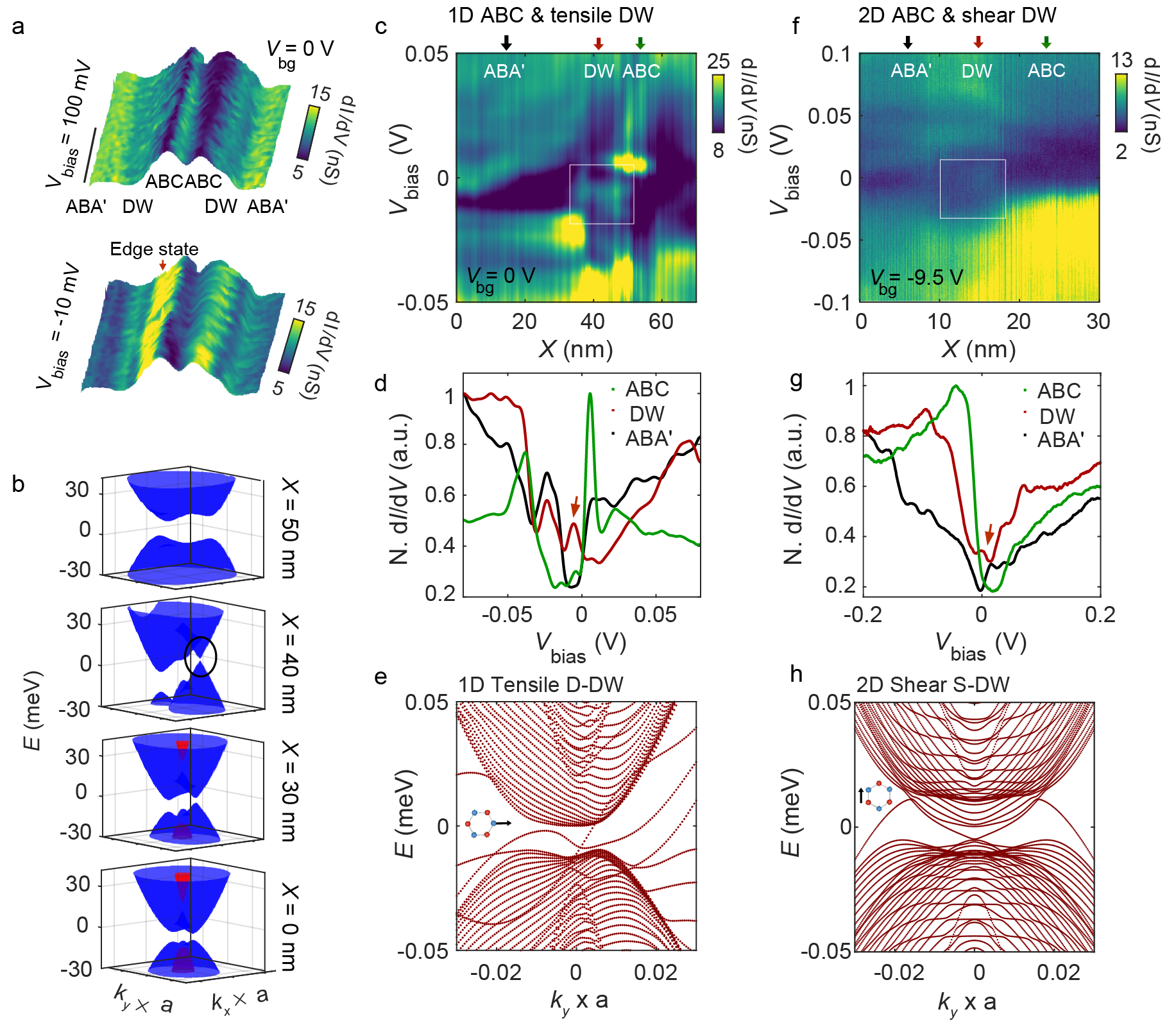}
    \caption{
    \textbf{Topological edge states at the DW of ABA–ABC$_{1D}$ interfaces stabilized by pseudo-electric fields.}  
    \textbf{a}, LDOS map in the vicinity of the twin 1D-ABC channels at $V_{\rm{bias}}$ of 100 mV and -10 mV. Scale bar: 50 nm.
    \textbf{b}, Theoretical local band evolution from a gapped ABA to a gapped ABC region from the Slonczewski-Weiss parametric model. Starting from ABA at $x = 0$, the gap increases when approaching the DW ($x \approx 30$~\AA). At the DW ($x \approx 40$~\AA), the VB and CB exhibit a Dirac band touching, which produces a Dirac point, signaling a band inversion that seeds a topological transition. On the ABC side ($x \approx 50$~\AA), a gap reopens. 
    \textbf{c}, High resolution ${\mathrm{d}I}/{\mathrm{d}V}(V_{\mathrm{bias}}, x)$ line profile. The finite gaps in both the ABA and ABC regions serve as effective confinement barriers for DW states. \(V_{\mathrm{set}}=-300\,\mathrm{mV},\quad
I_{\mathrm{set}}=400\,\mathrm{pA}\).
    \textbf{d}, STS spectra extracted from panel \textbf{c}, showing the ABA', ABC, DW LDOS. The intensities of the CB and VB exhibit slight spatial variations across the ABC channel and this spectrum was acquired on the right-hand side of the channel. The edge state is marked by the red arrow. Spectra in panels \textbf{d} and \textbf{g} are normalized by their maximum value. 
    \textbf{e}, Tensile 1D D-DW band structure with $\Delta_1$ of 30 meV, showing topological edge states. Here, only the $K$-valley is displayed. 
    The inset illustrates the direction of the Burgers vector of the top-layer lattice. 
    \textbf{f}, Shear S-DW edge states (marked by a square) under applied $V_{bg}$ = -9.5 V. \( V_{\mathrm{set}} = 80\,\mathrm{mV} \), \( I_{\mathrm{set}} = 600\,\mathrm{pA} \).
    \textbf{g}, STS spectra extracted from panel \textbf{f}. The edge state is marked by the red arrow.
    \textbf{h}, Shear ABA- S-DW-ABC band structure with $\Delta_1$ of -20 meV, showing topological edge states. In the shear 2D S-DW, both valleys are superposed due to the orientation of the graphene lattice relative to the translationally invariant y-axis. 
    }
    \label{fig4}
\end{figure*}
We now show that the pseudoscalar potential-induced gap discussed above stabilizes topological edge states at zero external $D$. Topological modes at the boundaries between Bernal and rhombohedral graphene have been visualized before, in marginally twisted double bilayer systems~\cite{kerelsky-moireless-2021}. Here, we emphasize the role of the strain-induced pseudo-$D$ in stabilizing them by gapping out their vicinity. Figure~\ref{fig4}a presents real-space $\mathrm{d}I/\mathrm{d}V$ maps across the pair of 1D-ABC channels (shown in Fig. \ref{fig1}c) at two representative energies, both acquired at zero external $D$ ($V_{\text{bg}} = 0$). At $V_{\text{bias}} = 100~\mathrm{mV}$ (top panel), the LDOS is distributed nearly uniformly across the channels. In contrast, at $V_{\text{bias}} = -10~\mathrm{mV}$ (bottom panel), a pronounced state appears localized along the edge of the 1D-ABC channel, with higher intensity on the left-hand channel. A high-resolution $\mathrm{d}I/\mathrm{d}V({V_{\text{bias}}, x})$ line profile further resolves this edge state at $V_{\text{bias}} = -10~\mathrm{mV}$, confined to the DW region between the surrounding ABA and the ABC channel (white square region in Fig. \ref{fig4}c and see Fig. S4 for the mode on the right channel \cite{SM}). Notably, the edge mode lies within the gap of both the ABA and ABC surrounding regions. This is further confirmed by a single-point $\mathrm{d}I/\mathrm{d}V$ spectrum taken in the respective areas (see Fig.~\ref{fig4}d). These edge modes arise from the difference in topological character between the ABA and ABC regions (see Fig.~S22 \cite{SM}), reminiscent of DWs in Bernal multilayer graphene~\cite{martin2008topological,zhang2013valley,vaezi2013topological,yin-direct-2016,kerelsky-moireless-2021,ju-topological-2015,huang-high-temperature-2024, hoke2026imaging}. In the strain-free system, however, an external $D$ is required to open a gap on both sides of the domain boundary. By contrast, in these 1D channels, the tensile-strain–induced gap alone stabilizes the edge modes at zero applied $D$. 
\par
To highlight this distinction, Figs.~\ref{fig4}f and \ref{fig4}g show a shear-strained DW at the edge of the 2D triangular ABC and ABA domain. Because this DW is adjacent to an essentially strain-free bulk region that remains gapless at zero $D$, an external displacement field is required ($V_{\mathrm{bg}} = -9.5$~V in this case) to open a bulk gap upon which the edge state re-emerges (see Fig.~S21 for more supporting data) \cite{yankowitz-electric-2014,SM}. Notably, a comparison between Figs. \ref{fig4}d and \ref{fig4}g indicates that gap opening and the resulting protection of the boundary mode are achieved more effectively when driven by the strain-induced pseudoscalar potential than by a purely electrostatic displacement field. This behavior is consistent with the fact that strain simultaneously reshapes the layer charge distribution and modifies interlayer hopping amplitudes (see Fig.~S13 for details \cite{SM}). Consequently, the enhanced gaps produced by the pseudoscalar potential near DWs can act as effective barriers that stabilize the topological edge modes.
\par
To verify the topological nature of the observed edge state and the spatial evolution of the observed LDOS, we construct a position-dependent Slonczewski–Weiss–McClure (SWMcC) continuum model \cite{garcia2021full}, where the interlayer hopping amplitudes and on-site potentials are continuously parameterized across the ABA–tensile DW–ABC interface (see Supplementary Sec.~4.1 for details \cite{SM}). The resulting band structure evolution (Fig.~\ref{fig4}b) captures the experimentally observed band bending, gap enhancement near the DW, and the emergence of the edge states. In particular, at the DW center ($x \approx 40$~\AA), the VB and CB exhibit a Dirac band-touching point, giving rise to a topological edge state. This continuous evolution from a gapped ABA to a gapped ABC through band inversion captures the phenomenology that we observe experimentally. To rigorously establish the topological character of the DW states, however, it is necessary to go beyond the SWMcC continuum parameterization since it does not capture the electronic properties of the ABA to ABC junction as a whole, but only at a specific position in the junction. To fully capture the formation of topological states, we therefore performed full TB calculations for a tensile 1D junction (half of the dual-ABC structure for simplicity), as shown in Fig.~\ref{fig4}e. These calculations reveal topological edge states, consisting of a pair of counter-propagating valley-polarized modes stemming from different valley Chern numbers protected by an effective U(1) valley-preserving symmetry (also see shear DW in Fig.~\ref{fig4}h) \cite{Wang2024TDBGValleyChern,Li2011Subgap,ju-topological-2015}. Such topological modes can be broken by atomic-scale disorder or interaction-induced valley-mixing order, neither of which is observed in our experiment (also see discussions in Supplementary Sec.~4.2 \cite{SM}).

\section{Discussion}
Our measurements reveal several intriguing aspects of the interplay between the mechanical deformation and electronic structure in graphene-based systems. We observed clear evidence of a strain-induced structural transition accompanied by pronounced modifications in the band structure. While such transitions have been theoretically predicted in bilayer graphene, our data provide the first direct atomic-scale and electronic verification of this phenomenon in TLG. With regard to its impact on electronic properties, we identify the emergence of a strain-induced pseudoscalar potential. In the coupled 1D-ABC channel, this pseudo-displacement field creates flat dispersion and opens an energy gap without an external electromagnetic potential. This behavior is counterintuitive, as uniform strain is conventionally expected to broaden electronic bands and weaken correlation effects \cite{bi-designing-2019}. In contrast, our findings demonstrate that tensile strain can instead be the origin of narrow bands in the multilayer graphene system \cite{bi-designing-2019, checkelsky-flat-2024}. Although the local strain in our experiment ($\sim 1\%$) exceeds the residual strain typically found in uniform 2D devices, comparable magnitudes can occur near wrinkles \cite{strain-banerjee-2020}, DWs, reconstructed moir\'e regions \cite{kazmierczak-strain-2021}, and should therefore be considered in displacement-field-sensitive graphene systems.
\par
In many ways, the pseudo-displacement field can be understood in analogy to a real-displacement field; it generates a flat band that hosts many-body effects, and the associated band gap helps stabilize topological states. While an external displacement field yields a similar outcome, the pseudo-displacement field goes beyond it (also see the comparison between 1D- and 2D-ABC graphene
in Supplementary Sec.~2.2(c) \cite{SM}). For instance, in the single-particle picture, the resulting wavefunction distribution across the layers cannot be reproduced by considering only the real-displacement field. Instead, the strain effect, which modifies the layer polarization and spatial profile of the wavefunction, must be included. Another distinction is that the system under large strain has a different symmetry and interaction strength compared to the unstrained 2D system. Many-body effects in these quasi-1D systems, as observed here, suggest that more novel interaction-driven phenomena under altered symmetry, interaction strength, and modified electronic susceptibility can be realized in these systems. Moreover, the pseudo-displacement field can tune the interlayer potential in unique ways. In TLG, for example, the potential difference between different layers can be parametrized by $\Delta_1$ and $\Delta_2$ \cite{koshino-aba}. Here, $\Delta_1$ denotes the potential difference between the top and bottom layers, whereas $\Delta_2$ represents the potential of the middle layer relative to the average potential of the outer layers. Electrostatic gating modifies only $\Delta_1$, whereas a strain-induced pseudoscalar potential modifies both $\Delta_1$ and $\Delta_2$. This layer-asymmetric potential gaps the bands much more efficiently in ABA TLG (see Fig. S13 \cite{SM}), thereby stabilizing topological edge modes and opening the door for exploring the interplay between the valley Hall effect and quantum Hall edge states under magnetic fields in future experiments \cite{ju-topological-2015,li-gate-controlled-2016,geisenhof2022interplay,deprez-tunable-2021,barrier-one-dimensional-2024,jiang-soliton-dependent-2020}.

\section{Data availability}
The data that support the findings of this study are available from the corresponding authors upon reasonable request.

\section{Code availability}
The code that supports the findings of this study is available from the corresponding authors upon reasonable request.

\section{Acknowledgments}
This work was primarily supported by the BSF (Grant Numbers 2022272). Work at UCSB was primarily supported by the National Science Foundation under award DMR-2226850, with additional support provided by the Gordon and Betty Moore Foundation under award GBMF9471. The work also made use of shared equipment sponsored by the National Science Foundation through Enabling Quantum Leap: Convergent Accelerated Discovery Foundries for Quantum Materials Science, Engineering and Information (Q-AMASE-i) award number DMR-1906325. K.W. and T.T. acknowledge support from the JSPS KAKENHI (Grant Numbers 21H05233 and 23H02052) , the CREST (JPMJCR24A5), JST and World Premier International Research Center Initiative (WPI), MEXT, Japan. A.C, P.A.P, Z.Z and F.G, acknowledge support from the “Severo Ochoa” Programme for Centres of Excellence in R\&D (CEX2020-001039-S/AEI/10.13039/501100011033) financed by MICIU/AEI/10.13039/501100011033 and from NOVMOMAT, Grant PID2022-142162NB-I00 funded by MCIN/AEI/ 10.13039/501100011033 and, by "ERDF A way of making Europe". Z.Z. acknowledges support funding from the European Union's Horizon 2020 research and innovation programme under the Marie Skłodowska-Curie grant agreement No 101034431.  P.A.P acknowledges funding by Grant No.\ JSF-24-05-0002 of the Julian Schwinger Foundation for Physics Research. F.G. acknowledges the support from the  Department of Education of the Basque Government through the project No. \verb|PIBA\2023\1\0007(STRAINER)|.

\section{Author contributions}
N.A, H.B, and A.F.Y. conceived of and directed the project. Y.L. and A.G. performed the STM measurements and analyzed the data. Y.C. and H.S. fabricated the devices. A.C. developed the elastic models to simulate the 1D systems under consideration and performed both the continuum and the tight-binding calculations, supervised by F.G and in collaboration with P.A.P and Z.Z. P.A.P., together with A.C., performed the Berry curvature calculations and analyzed the system's topological properties. Y.L., A.G., A.C., and N.A wrote the paper with input from all the co-authors.

\section{Competing interests}
The authors declare no competing interests.

{\small
\noindent$\dagger$These authors contributed equally.\\
\noindent$^\ast$Corresponding authors: yiwen.liu@weizmann.ac.il;\\ nurit.avraham@weizmann.ac.il; haim.beidenkopf@weizmann.ac.il\\
\noindent$^{\text{\P}}$Present address: Department of Physics, University of Washington, Seattle, WA, USA.\\
\noindent$^{\ddagger}$Present address: Department of Physics, Pohang University of Science and Technology, Pohang 37673, Republic of Korea.\\
\noindent$^{\parallel}$Present address: Center for Theoretical Physics, Hainan University, Haikou 570228, China.
}  

\vspace{0.8em}

\bibliographystyle{apsrev4-2}
\bibliography{reference}

\begin{thebibliography}{98}%
\makeatletter
\providecommand \@ifxundefined [1]{%
 \@ifx{#1\undefined}
}%
\providecommand \@ifnum [1]{%
 \ifnum #1\expandafter \@firstoftwo
 \else \expandafter \@secondoftwo
 \fi
}%
\providecommand \@ifx [1]{%
 \ifx #1\expandafter \@firstoftwo
 \else \expandafter \@secondoftwo
 \fi
}%
\providecommand \natexlab [1]{#1}%
\providecommand \enquote  [1]{``#1''}%
\providecommand \bibnamefont  [1]{#1}%
\providecommand \bibfnamefont [1]{#1}%
\providecommand \citenamefont [1]{#1}%
\providecommand \href@noop [0]{\@secondoftwo}%
\providecommand \href [0]{\begingroup \@sanitize@url \@href}%
\providecommand \@href[1]{\@@startlink{#1}\@@href}%
\providecommand \@@href[1]{\endgroup#1\@@endlink}%
\providecommand \@sanitize@url [0]{\catcode `\\12\catcode `\$12\catcode `\&12\catcode `\#12\catcode `\^12\catcode `\_12\catcode `\%12\relax}%
\providecommand \@@startlink[1]{}%
\providecommand \@@endlink[0]{}%
\providecommand \url  [0]{\begingroup\@sanitize@url \@url }%
\providecommand \@url [1]{\endgroup\@href {#1}{\urlprefix }}%
\providecommand \urlprefix  [0]{URL }%
\providecommand \Eprint [0]{\href }%
\providecommand \doibase [0]{https://doi.org/}%
\providecommand \selectlanguage [0]{\@gobble}%
\providecommand \bibinfo  [0]{\@secondoftwo}%
\providecommand \bibfield  [0]{\@secondoftwo}%
\providecommand \translation [1]{[#1]}%
\providecommand \BibitemOpen [0]{}%
\providecommand \bibitemStop [0]{}%
\providecommand \bibitemNoStop [0]{.\EOS\space}%
\providecommand \EOS [0]{\spacefactor3000\relax}%
\providecommand \BibitemShut  [1]{\csname bibitem#1\endcsname}%
\let\auto@bib@innerbib\@empty
\bibitem [{\citenamefont {Nagaosa}\ and\ \citenamefont {Tokura}(2012)}]{Nagaosa-2012}%
  \BibitemOpen
  \bibfield  {author} {\bibinfo {author} {\bibfnamefont {N.}~\bibnamefont {Nagaosa}}\ and\ \bibinfo {author} {\bibfnamefont {Y.}~\bibnamefont {Tokura}},\ }\href {https://doi.org/10.1088/0031-8949/2012/T146/014020} {\bibfield  {journal} {\bibinfo  {journal} {Phys. Scr.}\ }\textbf {\bibinfo {volume} {T146}},\ \bibinfo {pages} {014020} (\bibinfo {year} {2012})}\BibitemShut {NoStop}%
\bibitem [{\citenamefont {Xiao}\ \emph {et~al.}(2010)\citenamefont {Xiao}, \citenamefont {Chang},\ and\ \citenamefont {Niu}}]{xiao-2010}%
  \BibitemOpen
  \bibfield  {author} {\bibinfo {author} {\bibfnamefont {D.}~\bibnamefont {Xiao}}, \bibinfo {author} {\bibfnamefont {M.-C.}\ \bibnamefont {Chang}},\ and\ \bibinfo {author} {\bibfnamefont {Q.}~\bibnamefont {Niu}},\ }\href {https://doi.org/10.1103/RevModPhys.82.1959} {\bibfield  {journal} {\bibinfo  {journal} {Rev. Mod. Phys.}\ }\textbf {\bibinfo {volume} {82}},\ \bibinfo {pages} {1959} (\bibinfo {year} {2010})}\BibitemShut {NoStop}%
\bibitem [{\citenamefont {Zeljkovic}\ \emph {et~al.}(2015)\citenamefont {Zeljkovic}, \citenamefont {Walkup}, \citenamefont {Assaf}, \citenamefont {Scipioni}, \citenamefont {Sankar}, \citenamefont {Chou},\ and\ \citenamefont {Madhavan}}]{zeljkovic2015strain}%
  \BibitemOpen
  \bibfield  {author} {\bibinfo {author} {\bibfnamefont {I.}~\bibnamefont {Zeljkovic}}, \bibinfo {author} {\bibfnamefont {D.}~\bibnamefont {Walkup}}, \bibinfo {author} {\bibfnamefont {B.~A.}\ \bibnamefont {Assaf}}, \bibinfo {author} {\bibfnamefont {K.~L.}\ \bibnamefont {Scipioni}}, \bibinfo {author} {\bibfnamefont {R.}~\bibnamefont {Sankar}}, \bibinfo {author} {\bibfnamefont {F.}~\bibnamefont {Chou}},\ and\ \bibinfo {author} {\bibfnamefont {V.}~\bibnamefont {Madhavan}},\ }\href {https://www.nature.com/articles/nnano.2015.177} {\bibfield  {journal} {\bibinfo  {journal} {Nat. Nanotechnol.}\ }\textbf {\bibinfo {volume} {10}},\ \bibinfo {pages} {849} (\bibinfo {year} {2015})}\BibitemShut {NoStop}%
\bibitem [{\citenamefont {Liu}\ \emph {et~al.}(2014)\citenamefont {Liu}, \citenamefont {Li}, \citenamefont {Rajput}, \citenamefont {Gilks}, \citenamefont {Lari}, \citenamefont {Galindo}, \citenamefont {Weinert}, \citenamefont {Lazarov},\ and\ \citenamefont {Li}}]{liu2014tuning}%
  \BibitemOpen
  \bibfield  {author} {\bibinfo {author} {\bibfnamefont {Y.}~\bibnamefont {Liu}}, \bibinfo {author} {\bibfnamefont {Y.~Y.}\ \bibnamefont {Li}}, \bibinfo {author} {\bibfnamefont {S.}~\bibnamefont {Rajput}}, \bibinfo {author} {\bibfnamefont {D.}~\bibnamefont {Gilks}}, \bibinfo {author} {\bibfnamefont {L.}~\bibnamefont {Lari}}, \bibinfo {author} {\bibfnamefont {P.~L.}\ \bibnamefont {Galindo}}, \bibinfo {author} {\bibfnamefont {M.}~\bibnamefont {Weinert}}, \bibinfo {author} {\bibfnamefont {V.~K.}\ \bibnamefont {Lazarov}},\ and\ \bibinfo {author} {\bibfnamefont {L.}~\bibnamefont {Li}},\ }\href {https://doi.org/10.1038/nphys2898} {\bibfield  {journal} {\bibinfo  {journal} {Nat. Phys.}\ }\textbf {\bibinfo {volume} {10}},\ \bibinfo {pages} {294} (\bibinfo {year} {2014})}\BibitemShut {NoStop}%
\bibitem [{\citenamefont {Kamboj}\ \emph {et~al.}(2019)\citenamefont {Kamboj}, \citenamefont {Rana}, \citenamefont {Sirohi}, \citenamefont {Vasdev}, \citenamefont {Mandal}, \citenamefont {Marik}, \citenamefont {Singh}, \citenamefont {Das},\ and\ \citenamefont {Sheet}}]{kamboj2019generation}%
  \BibitemOpen
  \bibfield  {author} {\bibinfo {author} {\bibfnamefont {S.}~\bibnamefont {Kamboj}}, \bibinfo {author} {\bibfnamefont {P.~S.}\ \bibnamefont {Rana}}, \bibinfo {author} {\bibfnamefont {A.}~\bibnamefont {Sirohi}}, \bibinfo {author} {\bibfnamefont {A.}~\bibnamefont {Vasdev}}, \bibinfo {author} {\bibfnamefont {M.}~\bibnamefont {Mandal}}, \bibinfo {author} {\bibfnamefont {S.}~\bibnamefont {Marik}}, \bibinfo {author} {\bibfnamefont {R.~P.}\ \bibnamefont {Singh}}, \bibinfo {author} {\bibfnamefont {T.}~\bibnamefont {Das}},\ and\ \bibinfo {author} {\bibfnamefont {G.}~\bibnamefont {Sheet}},\ }\href {https://doi.org/10.1103/PhysRevB.100.115105} {\bibfield  {journal} {\bibinfo  {journal} {Phys. Rev. B}\ }\textbf {\bibinfo {volume} {100}},\ \bibinfo {pages} {115105} (\bibinfo {year} {2019})}\BibitemShut {NoStop}%
\bibitem [{\citenamefont {Cheng}\ \emph {et~al.}(2010)\citenamefont {Cheng}, \citenamefont {Song}, \citenamefont {Zhang}, \citenamefont {Zhang}, \citenamefont {Wang}, \citenamefont {Jia}, \citenamefont {Wang}, \citenamefont {Wang}, \citenamefont {Zhu}, \citenamefont {Chen}, \citenamefont {Ma}, \citenamefont {He}, \citenamefont {Wang}, \citenamefont {Dai}, \citenamefont {Fang}, \citenamefont {Xie}, \citenamefont {Qi}, \citenamefont {Liu}, \citenamefont {Zhang},\ and\ \citenamefont {Xue}}]{cheng2010landau}%
  \BibitemOpen
  \bibfield  {author} {\bibinfo {author} {\bibfnamefont {P.}~\bibnamefont {Cheng}}, \bibinfo {author} {\bibfnamefont {C.}~\bibnamefont {Song}}, \bibinfo {author} {\bibfnamefont {T.}~\bibnamefont {Zhang}}, \bibinfo {author} {\bibfnamefont {Y.}~\bibnamefont {Zhang}}, \bibinfo {author} {\bibfnamefont {Y.}~\bibnamefont {Wang}}, \bibinfo {author} {\bibfnamefont {J.-F.}\ \bibnamefont {Jia}}, \bibinfo {author} {\bibfnamefont {J.}~\bibnamefont {Wang}}, \bibinfo {author} {\bibfnamefont {Y.}~\bibnamefont {Wang}}, \bibinfo {author} {\bibfnamefont {B.-F.}\ \bibnamefont {Zhu}}, \bibinfo {author} {\bibfnamefont {X.}~\bibnamefont {Chen}}, \bibinfo {author} {\bibfnamefont {X.}~\bibnamefont {Ma}}, \bibinfo {author} {\bibfnamefont {K.}~\bibnamefont {He}}, \bibinfo {author} {\bibfnamefont {L.}~\bibnamefont {Wang}}, \bibinfo {author} {\bibfnamefont {X.}~\bibnamefont {Dai}}, \bibinfo {author} {\bibfnamefont {Z.}~\bibnamefont {Fang}}, \bibinfo {author} {\bibfnamefont {X.}~\bibnamefont {Xie}}, \bibinfo {author} {\bibfnamefont
  {X.-L.}\ \bibnamefont {Qi}}, \bibinfo {author} {\bibfnamefont {C.-X.}\ \bibnamefont {Liu}}, \bibinfo {author} {\bibfnamefont {S.-C.}\ \bibnamefont {Zhang}},\ and\ \bibinfo {author} {\bibfnamefont {Q.-K.}\ \bibnamefont {Xue}},\ }\href {https://doi.org/10.1103/PhysRevLett.105.076801} {\bibfield  {journal} {\bibinfo  {journal} {Phys. Rev. Lett.}\ }\textbf {\bibinfo {volume} {105}},\ \bibinfo {pages} {076801} (\bibinfo {year} {2010})}\BibitemShut {NoStop}%
\bibitem [{\citenamefont {Guinea}\ \emph {et~al.}(2010{\natexlab{a}})\citenamefont {Guinea}, \citenamefont {Katsnelson},\ and\ \citenamefont {Geim}}]{guinea-energy-2010}%
  \BibitemOpen
  \bibfield  {author} {\bibinfo {author} {\bibfnamefont {F.}~\bibnamefont {Guinea}}, \bibinfo {author} {\bibfnamefont {M.~I.}\ \bibnamefont {Katsnelson}},\ and\ \bibinfo {author} {\bibfnamefont {A.}~\bibnamefont {Geim}},\ }\href {https://doi.org/10.1038/nphys1420} {\bibfield  {journal} {\bibinfo  {journal} {Nat. Phys.}\ }\textbf {\bibinfo {volume} {6}},\ \bibinfo {pages} {30} (\bibinfo {year} {2010}{\natexlab{a}})}\BibitemShut {NoStop}%
\bibitem [{\citenamefont {Guinea}\ \emph {et~al.}(2010{\natexlab{b}})\citenamefont {Guinea}, \citenamefont {Geim}, \citenamefont {Katsnelson},\ and\ \citenamefont {Novoselov}}]{guinea-generating-2010}%
  \BibitemOpen
  \bibfield  {author} {\bibinfo {author} {\bibfnamefont {F.}~\bibnamefont {Guinea}}, \bibinfo {author} {\bibfnamefont {A.~K.}\ \bibnamefont {Geim}}, \bibinfo {author} {\bibfnamefont {M.~I.}\ \bibnamefont {Katsnelson}},\ and\ \bibinfo {author} {\bibfnamefont {K.~S.}\ \bibnamefont {Novoselov}},\ }\href {https://doi.org/10.1103/PhysRevB.81.035408} {\bibfield  {journal} {\bibinfo  {journal} {Phys. Rev. B}\ }\textbf {\bibinfo {volume} {81}},\ \bibinfo {pages} {035408} (\bibinfo {year} {2010}{\natexlab{b}})}\BibitemShut {NoStop}%
\bibitem [{\citenamefont {Mann}\ \emph {et~al.}(2020)\citenamefont {Mann}, \citenamefont {Horsley},\ and\ \citenamefont {Mariani}}]{mann-tunable-2020}%
  \BibitemOpen
  \bibfield  {author} {\bibinfo {author} {\bibfnamefont {C.-R.}\ \bibnamefont {Mann}}, \bibinfo {author} {\bibfnamefont {S.~A.~R.}\ \bibnamefont {Horsley}},\ and\ \bibinfo {author} {\bibfnamefont {E.}~\bibnamefont {Mariani}},\ }\href {https://doi.org/10.1038/s41566-020-0688-8} {\bibfield  {journal} {\bibinfo  {journal} {Nat. Photonics}\ }\textbf {\bibinfo {volume} {14}},\ \bibinfo {pages} {669} (\bibinfo {year} {2020})}\BibitemShut {NoStop}%
\bibitem [{\citenamefont {Vozmediano}\ \emph {et~al.}(2010)\citenamefont {Vozmediano}, \citenamefont {Katsnelson},\ and\ \citenamefont {Guinea}}]{vozmediano-gauge-2010}%
  \BibitemOpen
  \bibfield  {author} {\bibinfo {author} {\bibfnamefont {M.~A.~H.}\ \bibnamefont {Vozmediano}}, \bibinfo {author} {\bibfnamefont {M.~I.}\ \bibnamefont {Katsnelson}},\ and\ \bibinfo {author} {\bibfnamefont {F.}~\bibnamefont {Guinea}},\ }\href {https://doi.org/10.1016/j.physrep.2010.07.003} {\bibfield  {journal} {\bibinfo  {journal} {Phys. Rep.}\ }\textbf {\bibinfo {volume} {496}},\ \bibinfo {pages} {109} (\bibinfo {year} {2010})}\BibitemShut {NoStop}%
\bibitem [{\citenamefont {Ceferino}\ and\ \citenamefont {Guinea}(2024)}]{ceferino-pseudomagnetic-2024}%
  \BibitemOpen
  \bibfield  {author} {\bibinfo {author} {\bibfnamefont {A.}~\bibnamefont {Ceferino}}\ and\ \bibinfo {author} {\bibfnamefont {F.}~\bibnamefont {Guinea}},\ }\href {https://doi.org/10.1088/2053-1583/ad3b0e} {\bibfield  {journal} {\bibinfo  {journal} {2D Mater.}\ }\textbf {\bibinfo {volume} {11}},\ \bibinfo {pages} {035015} (\bibinfo {year} {2024})}\BibitemShut {NoStop}%
\bibitem [{\citenamefont {Levy}\ \emph {et~al.}(2010)\citenamefont {Levy}, \citenamefont {Burke}, \citenamefont {Meaker}, \citenamefont {Panlasigui}, \citenamefont {Zettl}, \citenamefont {Guinea}, \citenamefont {Castro~Neto},\ and\ \citenamefont {Crommie}}]{levy-strain-induced-2010}%
  \BibitemOpen
  \bibfield  {author} {\bibinfo {author} {\bibfnamefont {N.}~\bibnamefont {Levy}}, \bibinfo {author} {\bibfnamefont {S.~A.}\ \bibnamefont {Burke}}, \bibinfo {author} {\bibfnamefont {K.~L.}\ \bibnamefont {Meaker}}, \bibinfo {author} {\bibfnamefont {M.}~\bibnamefont {Panlasigui}}, \bibinfo {author} {\bibfnamefont {A.}~\bibnamefont {Zettl}}, \bibinfo {author} {\bibfnamefont {F.}~\bibnamefont {Guinea}}, \bibinfo {author} {\bibfnamefont {A.~H.}\ \bibnamefont {Castro~Neto}},\ and\ \bibinfo {author} {\bibfnamefont {M.~F.}\ \bibnamefont {Crommie}},\ }\href {https://doi.org/10.1126/science.1191700} {\bibfield  {journal} {\bibinfo  {journal} {Science}\ }\textbf {\bibinfo {volume} {329}},\ \bibinfo {pages} {544} (\bibinfo {year} {2010})}\BibitemShut {NoStop}%
\bibitem [{\citenamefont {Zhou}\ \emph {et~al.}(2023)\citenamefont {Zhou}, \citenamefont {Auerbach}, \citenamefont {Uzan}, \citenamefont {Zhou}, \citenamefont {Banu}, \citenamefont {Zhi}, \citenamefont {Huber}, \citenamefont {Watanabe}, \citenamefont {Taniguchi}, \citenamefont {Myasoedov}, \citenamefont {Yan},\ and\ \citenamefont {Zeldov}}]{zhou-imaging-2023}%
  \BibitemOpen
  \bibfield  {author} {\bibinfo {author} {\bibfnamefont {H.}~\bibnamefont {Zhou}}, \bibinfo {author} {\bibfnamefont {N.}~\bibnamefont {Auerbach}}, \bibinfo {author} {\bibfnamefont {M.}~\bibnamefont {Uzan}}, \bibinfo {author} {\bibfnamefont {Y.}~\bibnamefont {Zhou}}, \bibinfo {author} {\bibfnamefont {N.}~\bibnamefont {Banu}}, \bibinfo {author} {\bibfnamefont {W.}~\bibnamefont {Zhi}}, \bibinfo {author} {\bibfnamefont {M.~E.}\ \bibnamefont {Huber}}, \bibinfo {author} {\bibfnamefont {K.}~\bibnamefont {Watanabe}}, \bibinfo {author} {\bibfnamefont {T.}~\bibnamefont {Taniguchi}}, \bibinfo {author} {\bibfnamefont {Y.}~\bibnamefont {Myasoedov}}, \bibinfo {author} {\bibfnamefont {B.}~\bibnamefont {Yan}},\ and\ \bibinfo {author} {\bibfnamefont {E.}~\bibnamefont {Zeldov}},\ }\href {https://doi.org/10.1038/s41586-023-06763-5} {\bibfield  {journal} {\bibinfo  {journal} {Nature}\ }\textbf {\bibinfo {volume} {624}},\ \bibinfo {pages} {275} (\bibinfo {year} {2023})}\BibitemShut {NoStop}%
\bibitem [{\citenamefont {Nigge}\ \emph {et~al.}(2019)\citenamefont {Nigge}, \citenamefont {Qu}, \citenamefont {Lantagne-Hurtubise}, \citenamefont {Marsell}, \citenamefont {Link}, \citenamefont {Tom}, \citenamefont {Zonno}, \citenamefont {Michiardi}, \citenamefont {Schneider}, \citenamefont {Zhdanovich}, \citenamefont {Levy}, \citenamefont {Starke}, \citenamefont {Gutierrez}, \citenamefont {Bonn}, \citenamefont {Burke}, \citenamefont {Franz},\ and\ \citenamefont {Damascelli}}]{nigge-room-2019}%
  \BibitemOpen
  \bibfield  {author} {\bibinfo {author} {\bibfnamefont {P.}~\bibnamefont {Nigge}}, \bibinfo {author} {\bibfnamefont {A.~C.}\ \bibnamefont {Qu}}, \bibinfo {author} {\bibfnamefont {E.}~\bibnamefont {Lantagne-Hurtubise}}, \bibinfo {author} {\bibfnamefont {E.}~\bibnamefont {Marsell}}, \bibinfo {author} {\bibfnamefont {S.}~\bibnamefont {Link}}, \bibinfo {author} {\bibfnamefont {G.}~\bibnamefont {Tom}}, \bibinfo {author} {\bibfnamefont {M.}~\bibnamefont {Zonno}}, \bibinfo {author} {\bibfnamefont {M.}~\bibnamefont {Michiardi}}, \bibinfo {author} {\bibfnamefont {M.}~\bibnamefont {Schneider}}, \bibinfo {author} {\bibfnamefont {S.}~\bibnamefont {Zhdanovich}}, \bibinfo {author} {\bibfnamefont {G.}~\bibnamefont {Levy}}, \bibinfo {author} {\bibfnamefont {U.}~\bibnamefont {Starke}}, \bibinfo {author} {\bibfnamefont {C.}~\bibnamefont {Gutierrez}}, \bibinfo {author} {\bibfnamefont {D.}~\bibnamefont {Bonn}}, \bibinfo {author} {\bibfnamefont {S.~A.}\ \bibnamefont {Burke}}, \bibinfo {author} {\bibfnamefont {M.}~\bibnamefont
  {Franz}},\ and\ \bibinfo {author} {\bibfnamefont {A.}~\bibnamefont {Damascelli}},\ }\href {https://doi.org/10.1126/sciadv.aaw5593} {\bibfield  {journal} {\bibinfo  {journal} {Sci. Adv.}\ }\textbf {\bibinfo {volume} {5}},\ \bibinfo {pages} {eaaw5593} (\bibinfo {year} {2019})}\BibitemShut {NoStop}%
\bibitem [{\citenamefont {Georgi}\ \emph {et~al.}(2017)\citenamefont {Georgi}, \citenamefont {Nemes-Incze}, \citenamefont {Carrillo-Bastos}, \citenamefont {Faria}, \citenamefont {Viola~Kusminskiy}, \citenamefont {Zhai}, \citenamefont {Schneider}, \citenamefont {Subramaniam}, \citenamefont {Mashoff}, \citenamefont {Freitag}, \citenamefont {Liebmann}, \citenamefont {Pratzer}, \citenamefont {Wirtz}, \citenamefont {Woods}, \citenamefont {Gorbachev}, \citenamefont {Cao}, \citenamefont {Novoselov}, \citenamefont {Sandler},\ and\ \citenamefont {Morgenstern}}]{georgi-tuning-2017}%
  \BibitemOpen
  \bibfield  {author} {\bibinfo {author} {\bibfnamefont {A.}~\bibnamefont {Georgi}}, \bibinfo {author} {\bibfnamefont {P.}~\bibnamefont {Nemes-Incze}}, \bibinfo {author} {\bibfnamefont {R.}~\bibnamefont {Carrillo-Bastos}}, \bibinfo {author} {\bibfnamefont {D.}~\bibnamefont {Faria}}, \bibinfo {author} {\bibfnamefont {S.}~\bibnamefont {Viola~Kusminskiy}}, \bibinfo {author} {\bibfnamefont {D.}~\bibnamefont {Zhai}}, \bibinfo {author} {\bibfnamefont {M.}~\bibnamefont {Schneider}}, \bibinfo {author} {\bibfnamefont {D.}~\bibnamefont {Subramaniam}}, \bibinfo {author} {\bibfnamefont {T.}~\bibnamefont {Mashoff}}, \bibinfo {author} {\bibfnamefont {N.~M.}\ \bibnamefont {Freitag}}, \bibinfo {author} {\bibfnamefont {M.}~\bibnamefont {Liebmann}}, \bibinfo {author} {\bibfnamefont {M.}~\bibnamefont {Pratzer}}, \bibinfo {author} {\bibfnamefont {L.}~\bibnamefont {Wirtz}}, \bibinfo {author} {\bibfnamefont {C.~R.}\ \bibnamefont {Woods}}, \bibinfo {author} {\bibfnamefont {R.~V.}\ \bibnamefont {Gorbachev}}, \bibinfo {author}
  {\bibfnamefont {Y.}~\bibnamefont {Cao}}, \bibinfo {author} {\bibfnamefont {K.~S.}\ \bibnamefont {Novoselov}}, \bibinfo {author} {\bibfnamefont {N.}~\bibnamefont {Sandler}},\ and\ \bibinfo {author} {\bibfnamefont {M.}~\bibnamefont {Morgenstern}},\ }\href {https://doi.org/10.1021/acs.nanolett.6b04870} {\bibfield  {journal} {\bibinfo  {journal} {Nano Lett.}\ }\textbf {\bibinfo {volume} {17}},\ \bibinfo {pages} {2240} (\bibinfo {year} {2017})}\BibitemShut {NoStop}%
\bibitem [{\citenamefont {Liu}\ \emph {et~al.}(2022)\citenamefont {Liu}, \citenamefont {Zhan}, \citenamefont {Wu}, \citenamefont {Yan}, \citenamefont {Yuan},\ and\ \citenamefont {He}}]{liu-realizing-2022}%
  \BibitemOpen
  \bibfield  {author} {\bibinfo {author} {\bibfnamefont {Y.-W.}\ \bibnamefont {Liu}}, \bibinfo {author} {\bibfnamefont {Z.}~\bibnamefont {Zhan}}, \bibinfo {author} {\bibfnamefont {Z.}~\bibnamefont {Wu}}, \bibinfo {author} {\bibfnamefont {C.}~\bibnamefont {Yan}}, \bibinfo {author} {\bibfnamefont {S.}~\bibnamefont {Yuan}},\ and\ \bibinfo {author} {\bibfnamefont {L.}~\bibnamefont {He}},\ }\href {https://doi.org/10.1103/PhysRevLett.129.056803} {\bibfield  {journal} {\bibinfo  {journal} {Phys. Rev. Lett.}\ }\textbf {\bibinfo {volume} {129}},\ \bibinfo {pages} {056803} (\bibinfo {year} {2022})}\BibitemShut {NoStop}%
\bibitem [{\citenamefont {Liu}\ \emph {et~al.}(2018)\citenamefont {Liu}, \citenamefont {Rodrigues}, \citenamefont {Luo}, \citenamefont {Li}, \citenamefont {Carvalho}, \citenamefont {Yang}, \citenamefont {Laksono}, \citenamefont {Lu}, \citenamefont {Bao}, \citenamefont {Xu}, \citenamefont {Tan}, \citenamefont {Qiu}, \citenamefont {Sow}, \citenamefont {Feng}, \citenamefont {Neto}, \citenamefont {Adam}, \citenamefont {Lu},\ and\ \citenamefont {Loh}}]{liu-tailoring-2018}%
  \BibitemOpen
  \bibfield  {author} {\bibinfo {author} {\bibfnamefont {Y.}~\bibnamefont {Liu}}, \bibinfo {author} {\bibfnamefont {J.~N.~B.}\ \bibnamefont {Rodrigues}}, \bibinfo {author} {\bibfnamefont {Y.~Z.}\ \bibnamefont {Luo}}, \bibinfo {author} {\bibfnamefont {L.}~\bibnamefont {Li}}, \bibinfo {author} {\bibfnamefont {A.}~\bibnamefont {Carvalho}}, \bibinfo {author} {\bibfnamefont {M.}~\bibnamefont {Yang}}, \bibinfo {author} {\bibfnamefont {E.}~\bibnamefont {Laksono}}, \bibinfo {author} {\bibfnamefont {J.}~\bibnamefont {Lu}}, \bibinfo {author} {\bibfnamefont {Y.}~\bibnamefont {Bao}}, \bibinfo {author} {\bibfnamefont {H.}~\bibnamefont {Xu}}, \bibinfo {author} {\bibfnamefont {S.~J.~R.}\ \bibnamefont {Tan}}, \bibinfo {author} {\bibfnamefont {Z.}~\bibnamefont {Qiu}}, \bibinfo {author} {\bibfnamefont {C.~H.}\ \bibnamefont {Sow}}, \bibinfo {author} {\bibfnamefont {Y.~P.}\ \bibnamefont {Feng}}, \bibinfo {author} {\bibfnamefont {A.~H.~C.}\ \bibnamefont {Neto}}, \bibinfo {author} {\bibfnamefont {S.}~\bibnamefont {Adam}}, \bibinfo
  {author} {\bibfnamefont {J.}~\bibnamefont {Lu}},\ and\ \bibinfo {author} {\bibfnamefont {K.~P.}\ \bibnamefont {Loh}},\ }\href {https://doi.org/10.1038/s41565-018-0178-z} {\bibfield  {journal} {\bibinfo  {journal} {Nat. Nanotechnol.}\ }\textbf {\bibinfo {volume} {13}},\ \bibinfo {pages} {828} (\bibinfo {year} {2018})}\BibitemShut {NoStop}%
\bibitem [{\citenamefont {Ho}\ \emph {et~al.}(2021)\citenamefont {Ho}, \citenamefont {Chang}, \citenamefont {Hsieh}, \citenamefont {Lo}, \citenamefont {Huang}, \citenamefont {Vu}, \citenamefont {Ortix},\ and\ \citenamefont {Chen}}]{ho2021hall}%
  \BibitemOpen
  \bibfield  {author} {\bibinfo {author} {\bibfnamefont {S.-C.}\ \bibnamefont {Ho}}, \bibinfo {author} {\bibfnamefont {C.-H.}\ \bibnamefont {Chang}}, \bibinfo {author} {\bibfnamefont {Y.-C.}\ \bibnamefont {Hsieh}}, \bibinfo {author} {\bibfnamefont {S.-T.}\ \bibnamefont {Lo}}, \bibinfo {author} {\bibfnamefont {B.}~\bibnamefont {Huang}}, \bibinfo {author} {\bibfnamefont {T.-H.-Y.}\ \bibnamefont {Vu}}, \bibinfo {author} {\bibfnamefont {C.}~\bibnamefont {Ortix}},\ and\ \bibinfo {author} {\bibfnamefont {T.-M.}\ \bibnamefont {Chen}},\ }\href {https://doi.org/10.1038/s41928-021-00537-5} {\bibfield  {journal} {\bibinfo  {journal} {Nat. Electron.}\ }\textbf {\bibinfo {volume} {4}},\ \bibinfo {pages} {116} (\bibinfo {year} {2021})}\BibitemShut {NoStop}%
\bibitem [{\citenamefont {Hsu}\ \emph {et~al.}(2020)\citenamefont {Hsu}, \citenamefont {Teague}, \citenamefont {Wang},\ and\ \citenamefont {Yeh}}]{hsu-nanoscale-2020}%
  \BibitemOpen
  \bibfield  {author} {\bibinfo {author} {\bibfnamefont {C.-C.}\ \bibnamefont {Hsu}}, \bibinfo {author} {\bibfnamefont {M.~L.}\ \bibnamefont {Teague}}, \bibinfo {author} {\bibfnamefont {J.-Q.}\ \bibnamefont {Wang}},\ and\ \bibinfo {author} {\bibfnamefont {N.-C.}\ \bibnamefont {Yeh}},\ }\href {https://doi.org/10.1126/sciadv.aat9488} {\bibfield  {journal} {\bibinfo  {journal} {Sci. Adv.}\ }\textbf {\bibinfo {volume} {6}},\ \bibinfo {pages} {eaat9488} (\bibinfo {year} {2020})}\BibitemShut {NoStop}%
\bibitem [{\citenamefont {Kang}\ \emph {et~al.}(2021)\citenamefont {Kang}, \citenamefont {Sun}, \citenamefont {Luo}, \citenamefont {Lu}, \citenamefont {Chen}, \citenamefont {Kim}, \citenamefont {Jung}, \citenamefont {Gao}, \citenamefont {Parluhutan}, \citenamefont {Ge}, \citenamefont {Koh}, \citenamefont {Giovanni}, \citenamefont {Sum}, \citenamefont {Wang}, \citenamefont {Li},\ and\ \citenamefont {Nam}}]{kang-pseudo-magnetic-2021}%
  \BibitemOpen
  \bibfield  {author} {\bibinfo {author} {\bibfnamefont {D.-H.}\ \bibnamefont {Kang}}, \bibinfo {author} {\bibfnamefont {H.}~\bibnamefont {Sun}}, \bibinfo {author} {\bibfnamefont {M.}~\bibnamefont {Luo}}, \bibinfo {author} {\bibfnamefont {K.}~\bibnamefont {Lu}}, \bibinfo {author} {\bibfnamefont {M.}~\bibnamefont {Chen}}, \bibinfo {author} {\bibfnamefont {Y.}~\bibnamefont {Kim}}, \bibinfo {author} {\bibfnamefont {Y.}~\bibnamefont {Jung}}, \bibinfo {author} {\bibfnamefont {X.}~\bibnamefont {Gao}}, \bibinfo {author} {\bibfnamefont {S.~J.}\ \bibnamefont {Parluhutan}}, \bibinfo {author} {\bibfnamefont {J.}~\bibnamefont {Ge}}, \bibinfo {author} {\bibfnamefont {S.~W.}\ \bibnamefont {Koh}}, \bibinfo {author} {\bibfnamefont {D.}~\bibnamefont {Giovanni}}, \bibinfo {author} {\bibfnamefont {T.~C.}\ \bibnamefont {Sum}}, \bibinfo {author} {\bibfnamefont {Q.~J.}\ \bibnamefont {Wang}}, \bibinfo {author} {\bibfnamefont {H.}~\bibnamefont {Li}},\ and\ \bibinfo {author} {\bibfnamefont {D.}~\bibnamefont {Nam}},\ }\href
  {https://doi.org/10.1038/s41467-021-25304-0} {\bibfield  {journal} {\bibinfo  {journal} {Nat. Commun.}\ }\textbf {\bibinfo {volume} {12}},\ \bibinfo {pages} {5087} (\bibinfo {year} {2021})}\BibitemShut {NoStop}%
\bibitem [{\citenamefont {Shi}\ \emph {et~al.}(2020)\citenamefont {Shi}, \citenamefont {Zhan}, \citenamefont {Qi}, \citenamefont {Huang}, \citenamefont {Veen}, \citenamefont {Silva-Guillén}, \citenamefont {Zhang}, \citenamefont {Li}, \citenamefont {Xie}, \citenamefont {Ji}, \citenamefont {Katsnelson}, \citenamefont {Yuan}, \citenamefont {Qin},\ and\ \citenamefont {Zhang}}]{shi-large-area-2020}%
  \BibitemOpen
  \bibfield  {author} {\bibinfo {author} {\bibfnamefont {H.}~\bibnamefont {Shi}}, \bibinfo {author} {\bibfnamefont {Z.}~\bibnamefont {Zhan}}, \bibinfo {author} {\bibfnamefont {Z.}~\bibnamefont {Qi}}, \bibinfo {author} {\bibfnamefont {K.}~\bibnamefont {Huang}}, \bibinfo {author} {\bibfnamefont {E.~v.}\ \bibnamefont {Veen}}, \bibinfo {author} {\bibfnamefont {J.~A.}\ \bibnamefont {Silva-Guillén}}, \bibinfo {author} {\bibfnamefont {R.}~\bibnamefont {Zhang}}, \bibinfo {author} {\bibfnamefont {P.}~\bibnamefont {Li}}, \bibinfo {author} {\bibfnamefont {K.}~\bibnamefont {Xie}}, \bibinfo {author} {\bibfnamefont {H.}~\bibnamefont {Ji}}, \bibinfo {author} {\bibfnamefont {M.~I.}\ \bibnamefont {Katsnelson}}, \bibinfo {author} {\bibfnamefont {S.}~\bibnamefont {Yuan}}, \bibinfo {author} {\bibfnamefont {S.}~\bibnamefont {Qin}},\ and\ \bibinfo {author} {\bibfnamefont {Z.}~\bibnamefont {Zhang}},\ }\href {https://doi.org/10.1038/s41467-019-14207-w} {\bibfield  {journal} {\bibinfo  {journal} {Nat. Commun.}\ }\textbf {\bibinfo
  {volume} {11}},\ \bibinfo {pages} {371} (\bibinfo {year} {2020})}\BibitemShut {NoStop}%
\bibitem [{\citenamefont {Mao}\ \emph {et~al.}(2020)\citenamefont {Mao}, \citenamefont {Milovanovi{\'c}}, \citenamefont {An{\dj}elkovi{\'c}}, \citenamefont {Lai}, \citenamefont {Cao}, \citenamefont {Watanabe}, \citenamefont {Taniguchi}, \citenamefont {Covaci}, \citenamefont {Peeters}, \citenamefont {Geim}, \citenamefont {Jiang},\ and\ \citenamefont {Andrei}}]{mao2020evidence}%
  \BibitemOpen
  \bibfield  {author} {\bibinfo {author} {\bibfnamefont {J.}~\bibnamefont {Mao}}, \bibinfo {author} {\bibfnamefont {S.~P.}\ \bibnamefont {Milovanovi{\'c}}}, \bibinfo {author} {\bibfnamefont {M.}~\bibnamefont {An{\dj}elkovi{\'c}}}, \bibinfo {author} {\bibfnamefont {X.}~\bibnamefont {Lai}}, \bibinfo {author} {\bibfnamefont {Y.}~\bibnamefont {Cao}}, \bibinfo {author} {\bibfnamefont {K.}~\bibnamefont {Watanabe}}, \bibinfo {author} {\bibfnamefont {T.}~\bibnamefont {Taniguchi}}, \bibinfo {author} {\bibfnamefont {L.}~\bibnamefont {Covaci}}, \bibinfo {author} {\bibfnamefont {F.~M.}\ \bibnamefont {Peeters}}, \bibinfo {author} {\bibfnamefont {A.~K.}\ \bibnamefont {Geim}}, \bibinfo {author} {\bibfnamefont {Y.}~\bibnamefont {Jiang}},\ and\ \bibinfo {author} {\bibfnamefont {E.~Y.}\ \bibnamefont {Andrei}},\ }\href {https://www.nature.com/articles/s41586-020-2567-3} {\bibfield  {journal} {\bibinfo  {journal} {Nature}\ }\textbf {\bibinfo {volume} {584}},\ \bibinfo {pages} {215} (\bibinfo {year} {2020})}\BibitemShut {NoStop}%
\bibitem [{\citenamefont {Suzuura}\ and\ \citenamefont {Ando}(2002)}]{suzuura2002phonons}%
  \BibitemOpen
  \bibfield  {author} {\bibinfo {author} {\bibfnamefont {H.}~\bibnamefont {Suzuura}}\ and\ \bibinfo {author} {\bibfnamefont {T.}~\bibnamefont {Ando}},\ }\href {https://doi.org/10.1103/PhysRevB.65.235412} {\bibfield  {journal} {\bibinfo  {journal} {Phys. Rev. B}\ }\textbf {\bibinfo {volume} {65}},\ \bibinfo {pages} {235412} (\bibinfo {year} {2002})}\BibitemShut {NoStop}%
\bibitem [{\citenamefont {Aharonov}\ and\ \citenamefont {Bohm}(1959)}]{aharonov-significance-1959}%
  \BibitemOpen
  \bibfield  {author} {\bibinfo {author} {\bibfnamefont {Y.}~\bibnamefont {Aharonov}}\ and\ \bibinfo {author} {\bibfnamefont {D.}~\bibnamefont {Bohm}},\ }\href {https://doi.org/10.1103/PhysRev.115.485} {\bibfield  {journal} {\bibinfo  {journal} {Phys. Rev.}\ }\textbf {\bibinfo {volume} {115}},\ \bibinfo {pages} {485} (\bibinfo {year} {1959})}\BibitemShut {NoStop}%
\bibitem [{\citenamefont {Ilan}\ \emph {et~al.}(2020)\citenamefont {Ilan}, \citenamefont {Grushin},\ and\ \citenamefont {Pikulin}}]{ilan-pseudo-electromagnetic-2020}%
  \BibitemOpen
  \bibfield  {author} {\bibinfo {author} {\bibfnamefont {R.}~\bibnamefont {Ilan}}, \bibinfo {author} {\bibfnamefont {A.~G.}\ \bibnamefont {Grushin}},\ and\ \bibinfo {author} {\bibfnamefont {D.~I.}\ \bibnamefont {Pikulin}},\ }\href {https://doi.org/10.1038/s42254-019-0121-8} {\bibfield  {journal} {\bibinfo  {journal} {Nat. Rev. Phys.}\ }\textbf {\bibinfo {volume} {2}},\ \bibinfo {pages} {29} (\bibinfo {year} {2020})}\BibitemShut {NoStop}%
\bibitem [{\citenamefont {Choi}\ \emph {et~al.}(2010{\natexlab{a}})\citenamefont {Choi}, \citenamefont {Jhi},\ and\ \citenamefont {Son}}]{choi-controlling-2010}%
  \BibitemOpen
  \bibfield  {author} {\bibinfo {author} {\bibfnamefont {S.-M.}\ \bibnamefont {Choi}}, \bibinfo {author} {\bibfnamefont {S.-H.}\ \bibnamefont {Jhi}},\ and\ \bibinfo {author} {\bibfnamefont {Y.-W.}\ \bibnamefont {Son}},\ }\href {https://pubs.acs.org/doi/10.1021/nl101617x} {\bibfield  {journal} {\bibinfo  {journal} {Nano Lett.}\ }\textbf {\bibinfo {volume} {10}},\ \bibinfo {pages} {3486} (\bibinfo {year} {2010}{\natexlab{a}})}\BibitemShut {NoStop}%
\bibitem [{\citenamefont {Choi}\ \emph {et~al.}(2010{\natexlab{b}})\citenamefont {Choi}, \citenamefont {Jhi},\ and\ \citenamefont {Son}}]{strain-choi-2010}%
  \BibitemOpen
  \bibfield  {author} {\bibinfo {author} {\bibfnamefont {S.-M.}\ \bibnamefont {Choi}}, \bibinfo {author} {\bibfnamefont {S.-H.}\ \bibnamefont {Jhi}},\ and\ \bibinfo {author} {\bibfnamefont {Y.-W.}\ \bibnamefont {Son}},\ }\href {https://doi.org/10.1103/PhysRevB.81.081407} {\bibfield  {journal} {\bibinfo  {journal} {Phys. Rev. B}\ }\textbf {\bibinfo {volume} {81}},\ \bibinfo {pages} {081407} (\bibinfo {year} {2010}{\natexlab{b}})}\BibitemShut {NoStop}%
\bibitem [{\citenamefont {Kerjouan}\ \emph {et~al.}(2024)\citenamefont {Kerjouan}, \citenamefont {Rosticher}, \citenamefont {Pierret}, \citenamefont {Watanabe}, \citenamefont {Taniguchi}, \citenamefont {Dhillon}, \citenamefont {Ferreira}, \citenamefont {Dolfi}, \citenamefont {Goerbig}, \citenamefont {Plaçais},\ and\ \citenamefont {Mangeney}}]{kerjouan-quantum-2024}%
  \BibitemOpen
  \bibfield  {author} {\bibinfo {author} {\bibfnamefont {R.}~\bibnamefont {Kerjouan}}, \bibinfo {author} {\bibfnamefont {M.}~\bibnamefont {Rosticher}}, \bibinfo {author} {\bibfnamefont {A.}~\bibnamefont {Pierret}}, \bibinfo {author} {\bibfnamefont {K.}~\bibnamefont {Watanabe}}, \bibinfo {author} {\bibfnamefont {T.}~\bibnamefont {Taniguchi}}, \bibinfo {author} {\bibfnamefont {S.}~\bibnamefont {Dhillon}}, \bibinfo {author} {\bibfnamefont {R.}~\bibnamefont {Ferreira}}, \bibinfo {author} {\bibfnamefont {D.}~\bibnamefont {Dolfi}}, \bibinfo {author} {\bibfnamefont {M.}~\bibnamefont {Goerbig}}, \bibinfo {author} {\bibfnamefont {B.}~\bibnamefont {Plaçais}},\ and\ \bibinfo {author} {\bibfnamefont {J.}~\bibnamefont {Mangeney}},\ }\href {https://doi.org/10.1103/PhysRevApplied.22.024076} {\bibfield  {journal} {\bibinfo  {journal} {Phys. Rev. Applied}\ }\textbf {\bibinfo {volume} {22}},\ \bibinfo {pages} {024076} (\bibinfo {year} {2024})}\BibitemShut {NoStop}%
\bibitem [{\citenamefont {Shioya}\ \emph {et~al.}(2015)\citenamefont {Shioya}, \citenamefont {Russo}, \citenamefont {Yamamoto}, \citenamefont {Craciun},\ and\ \citenamefont {Tarucha}}]{shioya-electron-2015}%
  \BibitemOpen
  \bibfield  {author} {\bibinfo {author} {\bibfnamefont {H.}~\bibnamefont {Shioya}}, \bibinfo {author} {\bibfnamefont {S.}~\bibnamefont {Russo}}, \bibinfo {author} {\bibfnamefont {M.}~\bibnamefont {Yamamoto}}, \bibinfo {author} {\bibfnamefont {M.~F.}\ \bibnamefont {Craciun}},\ and\ \bibinfo {author} {\bibfnamefont {S.}~\bibnamefont {Tarucha}},\ }\href {https://doi.org/10.1021/acs.nanolett.5b03027} {\bibfield  {journal} {\bibinfo  {journal} {Nano Lett.}\ }\textbf {\bibinfo {volume} {15}},\ \bibinfo {pages} {7943} (\bibinfo {year} {2015})}\BibitemShut {NoStop}%
\bibitem [{\citenamefont {Wang}\ \emph {et~al.}(2021)\citenamefont {Wang}, \citenamefont {Baumgartner}, \citenamefont {Makk}, \citenamefont {Zihlmann}, \citenamefont {Varghese}, \citenamefont {Indolese}, \citenamefont {Watanabe}, \citenamefont {Taniguchi},\ and\ \citenamefont {Schönenberger}}]{wang-global-2021}%
  \BibitemOpen
  \bibfield  {author} {\bibinfo {author} {\bibfnamefont {L.}~\bibnamefont {Wang}}, \bibinfo {author} {\bibfnamefont {A.}~\bibnamefont {Baumgartner}}, \bibinfo {author} {\bibfnamefont {P.}~\bibnamefont {Makk}}, \bibinfo {author} {\bibfnamefont {S.}~\bibnamefont {Zihlmann}}, \bibinfo {author} {\bibfnamefont {B.~S.}\ \bibnamefont {Varghese}}, \bibinfo {author} {\bibfnamefont {D.~I.}\ \bibnamefont {Indolese}}, \bibinfo {author} {\bibfnamefont {K.}~\bibnamefont {Watanabe}}, \bibinfo {author} {\bibfnamefont {T.}~\bibnamefont {Taniguchi}},\ and\ \bibinfo {author} {\bibfnamefont {C.}~\bibnamefont {Schönenberger}},\ }\href {https://doi.org/10.1038/s42005-021-00651-y} {\bibfield  {journal} {\bibinfo  {journal} {Commun. Phys.}\ }\textbf {\bibinfo {volume} {4}},\ \bibinfo {pages} {147} (\bibinfo {year} {2021})}\BibitemShut {NoStop}%
\bibitem [{\citenamefont {Layek}\ \emph {et~al.}(2026)\citenamefont {Layek}, \citenamefont {Hingankar}, \citenamefont {Mukherjee}, \citenamefont {Chakraborty}, \citenamefont {Jangade}, \citenamefont {Kumar}, \citenamefont {Sangani}, \citenamefont {Basu}, \citenamefont {Bhuvaneswari}, \citenamefont {Watanabe}, \citenamefont {Taniguchi}, \citenamefont {Agarwal}, \citenamefont {Waghmare},\ and\ \citenamefont {Deshmukh}}]{layek2025modulation}%
  \BibitemOpen
  \bibfield  {author} {\bibinfo {author} {\bibfnamefont {S.}~\bibnamefont {Layek}}, \bibinfo {author} {\bibfnamefont {M.~A.}\ \bibnamefont {Hingankar}}, \bibinfo {author} {\bibfnamefont {A.}~\bibnamefont {Mukherjee}}, \bibinfo {author} {\bibfnamefont {A.}~\bibnamefont {Chakraborty}}, \bibinfo {author} {\bibfnamefont {D.~A.}\ \bibnamefont {Jangade}}, \bibinfo {author} {\bibfnamefont {A.}~\bibnamefont {Kumar}}, \bibinfo {author} {\bibfnamefont {L.~D.~V.}\ \bibnamefont {Sangani}}, \bibinfo {author} {\bibfnamefont {A.}~\bibnamefont {Basu}}, \bibinfo {author} {\bibfnamefont {R.}~\bibnamefont {Bhuvaneswari}}, \bibinfo {author} {\bibfnamefont {K.}~\bibnamefont {Watanabe}}, \bibinfo {author} {\bibfnamefont {T.}~\bibnamefont {Taniguchi}}, \bibinfo {author} {\bibfnamefont {A.}~\bibnamefont {Agarwal}}, \bibinfo {author} {\bibfnamefont {U.~V.}\ \bibnamefont {Waghmare}},\ and\ \bibinfo {author} {\bibfnamefont {M.~M.}\ \bibnamefont {Deshmukh}},\ }\href {https://doi.org/10.1038/s41467-026-70893-3} {\bibfield  {journal}
  {\bibinfo  {journal} {Nat. Commun.}\ }\textbf {\bibinfo {volume} {17}},\ \bibinfo {pages} {4366} (\bibinfo {year} {2026})}\BibitemShut {NoStop}%
\bibitem [{\citenamefont {Zhou}\ \emph {et~al.}(2021{\natexlab{a}})\citenamefont {Zhou}, \citenamefont {Xie}, \citenamefont {Ghazaryan}, \citenamefont {Holder}, \citenamefont {Ehrets}, \citenamefont {Spanton}, \citenamefont {Taniguchi}, \citenamefont {Watanabe}, \citenamefont {Berg}, \citenamefont {Serbyn},\ and\ \citenamefont {Young}}]{zhou-half-and-2021}%
  \BibitemOpen
  \bibfield  {author} {\bibinfo {author} {\bibfnamefont {H.}~\bibnamefont {Zhou}}, \bibinfo {author} {\bibfnamefont {T.}~\bibnamefont {Xie}}, \bibinfo {author} {\bibfnamefont {A.}~\bibnamefont {Ghazaryan}}, \bibinfo {author} {\bibfnamefont {T.}~\bibnamefont {Holder}}, \bibinfo {author} {\bibfnamefont {J.~R.}\ \bibnamefont {Ehrets}}, \bibinfo {author} {\bibfnamefont {E.~M.}\ \bibnamefont {Spanton}}, \bibinfo {author} {\bibfnamefont {T.}~\bibnamefont {Taniguchi}}, \bibinfo {author} {\bibfnamefont {K.}~\bibnamefont {Watanabe}}, \bibinfo {author} {\bibfnamefont {E.}~\bibnamefont {Berg}}, \bibinfo {author} {\bibfnamefont {M.}~\bibnamefont {Serbyn}},\ and\ \bibinfo {author} {\bibfnamefont {A.~F.}\ \bibnamefont {Young}},\ }\href {https://www.nature.com/articles/s41586-021-03938-w} {\bibfield  {journal} {\bibinfo  {journal} {Nature}\ }\textbf {\bibinfo {volume} {598}},\ \bibinfo {pages} {429} (\bibinfo {year} {2021}{\natexlab{a}})}\BibitemShut {NoStop}%
\bibitem [{\citenamefont {Chen}(2022)}]{chen-correlated-2022}%
  \BibitemOpen
  \bibfield  {author} {\bibinfo {author} {\bibfnamefont {G.}~\bibnamefont {Chen}},\ }\href {https://doi.org/10.1007/s44214-022-00010-0} {\bibfield  {journal} {\bibinfo  {journal} {Quantum Front.}\ }\textbf {\bibinfo {volume} {1}},\ \bibinfo {pages} {8} (\bibinfo {year} {2022})}\BibitemShut {NoStop}%
\bibitem [{\citenamefont {Liu}\ \emph {et~al.}(2024{\natexlab{a}})\citenamefont {Liu}, \citenamefont {Zheng}, \citenamefont {Sha}, \citenamefont {Lyu}, \citenamefont {Li}, \citenamefont {Park}, \citenamefont {Ren}, \citenamefont {Watanabe}, \citenamefont {Taniguchi}, \citenamefont {Jia}, \citenamefont {Luo}, \citenamefont {Shi}, \citenamefont {Jung},\ and\ \citenamefont {Chen}}]{liu2024spontaneous}%
  \BibitemOpen
  \bibfield  {author} {\bibinfo {author} {\bibfnamefont {K.}~\bibnamefont {Liu}}, \bibinfo {author} {\bibfnamefont {J.}~\bibnamefont {Zheng}}, \bibinfo {author} {\bibfnamefont {Y.}~\bibnamefont {Sha}}, \bibinfo {author} {\bibfnamefont {B.}~\bibnamefont {Lyu}}, \bibinfo {author} {\bibfnamefont {F.}~\bibnamefont {Li}}, \bibinfo {author} {\bibfnamefont {Y.}~\bibnamefont {Park}}, \bibinfo {author} {\bibfnamefont {Y.}~\bibnamefont {Ren}}, \bibinfo {author} {\bibfnamefont {K.}~\bibnamefont {Watanabe}}, \bibinfo {author} {\bibfnamefont {T.}~\bibnamefont {Taniguchi}}, \bibinfo {author} {\bibfnamefont {J.}~\bibnamefont {Jia}}, \bibinfo {author} {\bibfnamefont {W.}~\bibnamefont {Luo}}, \bibinfo {author} {\bibfnamefont {Z.}~\bibnamefont {Shi}}, \bibinfo {author} {\bibfnamefont {J.}~\bibnamefont {Jung}},\ and\ \bibinfo {author} {\bibfnamefont {G.}~\bibnamefont {Chen}},\ }\href {https://doi.org/10.1038/s41565-023-01558-1} {\bibfield  {journal} {\bibinfo  {journal} {Nat. Nanotechnol.}\ }\textbf {\bibinfo {volume} {19}},\
  \bibinfo {pages} {188} (\bibinfo {year} {2024}{\natexlab{a}})}\BibitemShut {NoStop}%
\bibitem [{\citenamefont {Lu}\ \emph {et~al.}(2024)\citenamefont {Lu}, \citenamefont {Han}, \citenamefont {Yao}, \citenamefont {Reddy}, \citenamefont {Yang}, \citenamefont {Seo}, \citenamefont {Watanabe}, \citenamefont {Taniguchi}, \citenamefont {Fu},\ and\ \citenamefont {Ju}}]{lu2024fractional}%
  \BibitemOpen
  \bibfield  {author} {\bibinfo {author} {\bibfnamefont {Z.}~\bibnamefont {Lu}}, \bibinfo {author} {\bibfnamefont {T.}~\bibnamefont {Han}}, \bibinfo {author} {\bibfnamefont {Y.}~\bibnamefont {Yao}}, \bibinfo {author} {\bibfnamefont {A.~P.}\ \bibnamefont {Reddy}}, \bibinfo {author} {\bibfnamefont {J.}~\bibnamefont {Yang}}, \bibinfo {author} {\bibfnamefont {J.}~\bibnamefont {Seo}}, \bibinfo {author} {\bibfnamefont {K.}~\bibnamefont {Watanabe}}, \bibinfo {author} {\bibfnamefont {T.}~\bibnamefont {Taniguchi}}, \bibinfo {author} {\bibfnamefont {L.}~\bibnamefont {Fu}},\ and\ \bibinfo {author} {\bibfnamefont {L.}~\bibnamefont {Ju}},\ }\href {https://www.nature.com/articles/s41586-023-07010-7} {\bibfield  {journal} {\bibinfo  {journal} {Nature}\ }\textbf {\bibinfo {volume} {626}},\ \bibinfo {pages} {759} (\bibinfo {year} {2024})}\BibitemShut {NoStop}%
\bibitem [{\citenamefont {Xie}\ \emph {et~al.}(2025)\citenamefont {Xie}, \citenamefont {Huo}, \citenamefont {Lu}, \citenamefont {Feng}, \citenamefont {Zhang}, \citenamefont {Wang}, \citenamefont {Yang}, \citenamefont {Watanabe}, \citenamefont {Taniguchi}, \citenamefont {Liu}, \citenamefont {Song}, \citenamefont {Xie}, \citenamefont {Liu},\ and\ \citenamefont {Lu}}]{xie2025tunable}%
  \BibitemOpen
  \bibfield  {author} {\bibinfo {author} {\bibfnamefont {J.}~\bibnamefont {Xie}}, \bibinfo {author} {\bibfnamefont {Z.}~\bibnamefont {Huo}}, \bibinfo {author} {\bibfnamefont {X.}~\bibnamefont {Lu}}, \bibinfo {author} {\bibfnamefont {Z.}~\bibnamefont {Feng}}, \bibinfo {author} {\bibfnamefont {Z.}~\bibnamefont {Zhang}}, \bibinfo {author} {\bibfnamefont {W.}~\bibnamefont {Wang}}, \bibinfo {author} {\bibfnamefont {Q.}~\bibnamefont {Yang}}, \bibinfo {author} {\bibfnamefont {K.}~\bibnamefont {Watanabe}}, \bibinfo {author} {\bibfnamefont {T.}~\bibnamefont {Taniguchi}}, \bibinfo {author} {\bibfnamefont {K.}~\bibnamefont {Liu}}, \bibinfo {author} {\bibfnamefont {Z.}~\bibnamefont {Song}}, \bibinfo {author} {\bibfnamefont {X.~C.}\ \bibnamefont {Xie}}, \bibinfo {author} {\bibfnamefont {J.}~\bibnamefont {Liu}},\ and\ \bibinfo {author} {\bibfnamefont {X.}~\bibnamefont {Lu}},\ }\href {https://www.nature.com/articles/s41563-025-02225-7} {\bibfield  {journal} {\bibinfo  {journal} {Nat. Mater.}\ }\textbf {\bibinfo {volume}
  {24}},\ \bibinfo {pages} {1042} (\bibinfo {year} {2025})}\BibitemShut {NoStop}%
\bibitem [{\citenamefont {Zhou}\ \emph {et~al.}(2024)\citenamefont {Zhou}, \citenamefont {Ding}, \citenamefont {Hua}, \citenamefont {Zhang}, \citenamefont {Watanabe}, \citenamefont {Taniguchi}, \citenamefont {Zhu},\ and\ \citenamefont {Xu}}]{zhou2024layer}%
  \BibitemOpen
  \bibfield  {author} {\bibinfo {author} {\bibfnamefont {W.}~\bibnamefont {Zhou}}, \bibinfo {author} {\bibfnamefont {J.}~\bibnamefont {Ding}}, \bibinfo {author} {\bibfnamefont {J.}~\bibnamefont {Hua}}, \bibinfo {author} {\bibfnamefont {L.}~\bibnamefont {Zhang}}, \bibinfo {author} {\bibfnamefont {K.}~\bibnamefont {Watanabe}}, \bibinfo {author} {\bibfnamefont {T.}~\bibnamefont {Taniguchi}}, \bibinfo {author} {\bibfnamefont {W.}~\bibnamefont {Zhu}},\ and\ \bibinfo {author} {\bibfnamefont {S.}~\bibnamefont {Xu}},\ }\href {https://www.nature.com/articles/s41467-024-46913-5} {\bibfield  {journal} {\bibinfo  {journal} {Nat. Commun.}\ }\textbf {\bibinfo {volume} {15}},\ \bibinfo {pages} {2597} (\bibinfo {year} {2024})}\BibitemShut {NoStop}%
\bibitem [{\citenamefont {Kumar}\ \emph {et~al.}(2026)\citenamefont {Kumar}, \citenamefont {Waleffe}, \citenamefont {Okounkova}, \citenamefont {Tejani}, \citenamefont {Phong}, \citenamefont {Watanabe}, \citenamefont {Taniguchi}, \citenamefont {Lewandowski}, \citenamefont {Folk},\ and\ \citenamefont {Yankowitz}}]{kumar2025superconductivity}%
  \BibitemOpen
  \bibfield  {author} {\bibinfo {author} {\bibfnamefont {M.}~\bibnamefont {Kumar}}, \bibinfo {author} {\bibfnamefont {D.}~\bibnamefont {Waleffe}}, \bibinfo {author} {\bibfnamefont {A.}~\bibnamefont {Okounkova}}, \bibinfo {author} {\bibfnamefont {R.}~\bibnamefont {Tejani}}, \bibinfo {author} {\bibfnamefont {V.~T.}\ \bibnamefont {Phong}}, \bibinfo {author} {\bibfnamefont {K.}~\bibnamefont {Watanabe}}, \bibinfo {author} {\bibfnamefont {T.}~\bibnamefont {Taniguchi}}, \bibinfo {author} {\bibfnamefont {C.}~\bibnamefont {Lewandowski}}, \bibinfo {author} {\bibfnamefont {J.}~\bibnamefont {Folk}},\ and\ \bibinfo {author} {\bibfnamefont {M.}~\bibnamefont {Yankowitz}},\ }\href {https://doi.org/10.1038/s41567-026-03277-5} {\bibfield  {journal} {\bibinfo  {journal} {Nat. Phys.}\ }\textbf {\bibinfo {volume} {22}},\ \bibinfo {pages} {862} (\bibinfo {year} {2026})}\BibitemShut {NoStop}%
\bibitem [{\citenamefont {Redekop}\ \emph {et~al.}(2024)\citenamefont {Redekop}, \citenamefont {Zhang}, \citenamefont {Park}, \citenamefont {Cai}, \citenamefont {Anderson}, \citenamefont {Sheekey}, \citenamefont {Arp}, \citenamefont {Babikyan}, \citenamefont {Salters}, \citenamefont {Watanabe}, \citenamefont {Taniguchi}, \citenamefont {Huber}, \citenamefont {Xu},\ and\ \citenamefont {Young}}]{redekop-direct-2024}%
  \BibitemOpen
  \bibfield  {author} {\bibinfo {author} {\bibfnamefont {E.}~\bibnamefont {Redekop}}, \bibinfo {author} {\bibfnamefont {C.}~\bibnamefont {Zhang}}, \bibinfo {author} {\bibfnamefont {H.}~\bibnamefont {Park}}, \bibinfo {author} {\bibfnamefont {J.}~\bibnamefont {Cai}}, \bibinfo {author} {\bibfnamefont {E.}~\bibnamefont {Anderson}}, \bibinfo {author} {\bibfnamefont {O.}~\bibnamefont {Sheekey}}, \bibinfo {author} {\bibfnamefont {T.}~\bibnamefont {Arp}}, \bibinfo {author} {\bibfnamefont {G.}~\bibnamefont {Babikyan}}, \bibinfo {author} {\bibfnamefont {S.}~\bibnamefont {Salters}}, \bibinfo {author} {\bibfnamefont {K.}~\bibnamefont {Watanabe}}, \bibinfo {author} {\bibfnamefont {T.}~\bibnamefont {Taniguchi}}, \bibinfo {author} {\bibfnamefont {M.~E.}\ \bibnamefont {Huber}}, \bibinfo {author} {\bibfnamefont {X.}~\bibnamefont {Xu}},\ and\ \bibinfo {author} {\bibfnamefont {A.~F.}\ \bibnamefont {Young}},\ }\href {https://doi.org/10.1038/s41586-024-08153-x} {\bibfield  {journal} {\bibinfo  {journal} {Nature}\ }\textbf
  {\bibinfo {volume} {635}},\ \bibinfo {pages} {584} (\bibinfo {year} {2024})}\BibitemShut {NoStop}%
\bibitem [{\citenamefont {Zhou}\ \emph {et~al.}(2021{\natexlab{b}})\citenamefont {Zhou}, \citenamefont {Xie}, \citenamefont {Taniguchi}, \citenamefont {Watanabe},\ and\ \citenamefont {Young}}]{zhou-superconductivity-2021}%
  \BibitemOpen
  \bibfield  {author} {\bibinfo {author} {\bibfnamefont {H.}~\bibnamefont {Zhou}}, \bibinfo {author} {\bibfnamefont {T.}~\bibnamefont {Xie}}, \bibinfo {author} {\bibfnamefont {T.}~\bibnamefont {Taniguchi}}, \bibinfo {author} {\bibfnamefont {K.}~\bibnamefont {Watanabe}},\ and\ \bibinfo {author} {\bibfnamefont {A.~F.}\ \bibnamefont {Young}},\ }\href {https://doi.org/10.1038/s41586-021-03926-0} {\bibfield  {journal} {\bibinfo  {journal} {Nature}\ }\textbf {\bibinfo {volume} {598}},\ \bibinfo {pages} {434} (\bibinfo {year} {2021}{\natexlab{b}})}\BibitemShut {NoStop}%
\bibitem [{\citenamefont {Han}\ \emph {et~al.}(2025)\citenamefont {Han}, \citenamefont {Lu}, \citenamefont {Hadjri}, \citenamefont {Shi}, \citenamefont {Wu}, \citenamefont {Xu}, \citenamefont {Yao}, \citenamefont {Cotten}, \citenamefont {Sharifi~Sedeh}, \citenamefont {Weldeyesus}, \citenamefont {Yang}, \citenamefont {Seo}, \citenamefont {Ye}, \citenamefont {Zhou}, \citenamefont {Liu}, \citenamefont {Shi}, \citenamefont {Hua}, \citenamefont {Watanabe}, \citenamefont {Taniguchi}, \citenamefont {Xiong}, \citenamefont {Zumb{\"u}hl}, \citenamefont {Fu},\ and\ \citenamefont {Ju}}]{han2025signatures}%
  \BibitemOpen
  \bibfield  {author} {\bibinfo {author} {\bibfnamefont {T.}~\bibnamefont {Han}}, \bibinfo {author} {\bibfnamefont {Z.}~\bibnamefont {Lu}}, \bibinfo {author} {\bibfnamefont {Z.}~\bibnamefont {Hadjri}}, \bibinfo {author} {\bibfnamefont {L.}~\bibnamefont {Shi}}, \bibinfo {author} {\bibfnamefont {Z.}~\bibnamefont {Wu}}, \bibinfo {author} {\bibfnamefont {W.}~\bibnamefont {Xu}}, \bibinfo {author} {\bibfnamefont {Y.}~\bibnamefont {Yao}}, \bibinfo {author} {\bibfnamefont {A.~A.}\ \bibnamefont {Cotten}}, \bibinfo {author} {\bibfnamefont {O.}~\bibnamefont {Sharifi~Sedeh}}, \bibinfo {author} {\bibfnamefont {H.}~\bibnamefont {Weldeyesus}}, \bibinfo {author} {\bibfnamefont {J.}~\bibnamefont {Yang}}, \bibinfo {author} {\bibfnamefont {J.}~\bibnamefont {Seo}}, \bibinfo {author} {\bibfnamefont {S.}~\bibnamefont {Ye}}, \bibinfo {author} {\bibfnamefont {M.}~\bibnamefont {Zhou}}, \bibinfo {author} {\bibfnamefont {H.}~\bibnamefont {Liu}}, \bibinfo {author} {\bibfnamefont {G.}~\bibnamefont {Shi}}, \bibinfo {author} {\bibfnamefont
  {Z.}~\bibnamefont {Hua}}, \bibinfo {author} {\bibfnamefont {K.}~\bibnamefont {Watanabe}}, \bibinfo {author} {\bibfnamefont {T.}~\bibnamefont {Taniguchi}}, \bibinfo {author} {\bibfnamefont {P.}~\bibnamefont {Xiong}}, \bibinfo {author} {\bibfnamefont {D.~M.}\ \bibnamefont {Zumb{\"u}hl}}, \bibinfo {author} {\bibfnamefont {L.}~\bibnamefont {Fu}},\ and\ \bibinfo {author} {\bibfnamefont {L.}~\bibnamefont {Ju}},\ }\href {https://www.nature.com/articles/s41586-025-09169-7} {\bibfield  {journal} {\bibinfo  {journal} {Nature}\ }\textbf {\bibinfo {volume} {643}},\ \bibinfo {pages} {654} (\bibinfo {year} {2025})}\BibitemShut {NoStop}%
\bibitem [{\citenamefont {Cai}\ \emph {et~al.}(2023)\citenamefont {Cai}, \citenamefont {Anderson}, \citenamefont {Wang}, \citenamefont {Zhang}, \citenamefont {Liu}, \citenamefont {Holtzmann}, \citenamefont {Zhang}, \citenamefont {Fan}, \citenamefont {Taniguchi}, \citenamefont {Watanabe}, \citenamefont {Ran}, \citenamefont {Cao}, \citenamefont {Fu}, \citenamefont {Xiao}, \citenamefont {Yao},\ and\ \citenamefont {Xu}}]{cai2023signatures}%
  \BibitemOpen
  \bibfield  {author} {\bibinfo {author} {\bibfnamefont {J.}~\bibnamefont {Cai}}, \bibinfo {author} {\bibfnamefont {E.}~\bibnamefont {Anderson}}, \bibinfo {author} {\bibfnamefont {C.}~\bibnamefont {Wang}}, \bibinfo {author} {\bibfnamefont {X.}~\bibnamefont {Zhang}}, \bibinfo {author} {\bibfnamefont {X.}~\bibnamefont {Liu}}, \bibinfo {author} {\bibfnamefont {W.}~\bibnamefont {Holtzmann}}, \bibinfo {author} {\bibfnamefont {Y.}~\bibnamefont {Zhang}}, \bibinfo {author} {\bibfnamefont {F.}~\bibnamefont {Fan}}, \bibinfo {author} {\bibfnamefont {T.}~\bibnamefont {Taniguchi}}, \bibinfo {author} {\bibfnamefont {K.}~\bibnamefont {Watanabe}}, \bibinfo {author} {\bibfnamefont {Y.}~\bibnamefont {Ran}}, \bibinfo {author} {\bibfnamefont {T.}~\bibnamefont {Cao}}, \bibinfo {author} {\bibfnamefont {L.}~\bibnamefont {Fu}}, \bibinfo {author} {\bibfnamefont {D.}~\bibnamefont {Xiao}}, \bibinfo {author} {\bibfnamefont {W.}~\bibnamefont {Yao}},\ and\ \bibinfo {author} {\bibfnamefont {X.}~\bibnamefont {Xu}},\ }\href
  {https://doi.org/10.1038/s41586-023-06289-w} {\bibfield  {journal} {\bibinfo  {journal} {Nature}\ }\textbf {\bibinfo {volume} {622}},\ \bibinfo {pages} {63} (\bibinfo {year} {2023})}\BibitemShut {NoStop}%
\bibitem [{\citenamefont {Song}\ \emph {et~al.}(2025)\citenamefont {Song}, \citenamefont {Yang}, \citenamefont {Lin}, \citenamefont {Liu}, \citenamefont {Wu}, \citenamefont {Chen}, \citenamefont {Ke}, \citenamefont {Lee}, \citenamefont {Liu}, \citenamefont {Zhu}, \citenamefont {Kivshar}, \citenamefont {Li},\ and\ \citenamefont {Zhang}}]{song2025artificial}%
  \BibitemOpen
  \bibfield  {author} {\bibinfo {author} {\bibfnamefont {W.}~\bibnamefont {Song}}, \bibinfo {author} {\bibfnamefont {Y.}~\bibnamefont {Yang}}, \bibinfo {author} {\bibfnamefont {Z.}~\bibnamefont {Lin}}, \bibinfo {author} {\bibfnamefont {X.}~\bibnamefont {Liu}}, \bibinfo {author} {\bibfnamefont {S.}~\bibnamefont {Wu}}, \bibinfo {author} {\bibfnamefont {C.}~\bibnamefont {Chen}}, \bibinfo {author} {\bibfnamefont {Y.}~\bibnamefont {Ke}}, \bibinfo {author} {\bibfnamefont {C.}~\bibnamefont {Lee}}, \bibinfo {author} {\bibfnamefont {W.}~\bibnamefont {Liu}}, \bibinfo {author} {\bibfnamefont {S.}~\bibnamefont {Zhu}}, \bibinfo {author} {\bibfnamefont {Y.~S.}\ \bibnamefont {Kivshar}}, \bibinfo {author} {\bibfnamefont {T.}~\bibnamefont {Li}},\ and\ \bibinfo {author} {\bibfnamefont {S.}~\bibnamefont {Zhang}},\ }\href {https://doi.org/10.1038/s42254-025-00869-7} {\bibfield  {journal} {\bibinfo  {journal} {Nat. Rev. Phys.}\ }\textbf {\bibinfo {volume} {7}},\ \bibinfo {pages} {606} (\bibinfo {year} {2025})}\BibitemShut
  {NoStop}%
\bibitem [{\citenamefont {Ma}\ \emph {et~al.}(2025)\citenamefont {Ma}, \citenamefont {Liu}, \citenamefont {Cai}, \citenamefont {Watanabe}, \citenamefont {Taniguchi}, \citenamefont {Xu}, \citenamefont {Chu},\ and\ \citenamefont {Yankowitz}}]{ma2025}%
  \BibitemOpen
  \bibfield  {author} {\bibinfo {author} {\bibfnamefont {X.}~\bibnamefont {Ma}}, \bibinfo {author} {\bibfnamefont {Z.}~\bibnamefont {Liu}}, \bibinfo {author} {\bibfnamefont {J.}~\bibnamefont {Cai}}, \bibinfo {author} {\bibfnamefont {K.}~\bibnamefont {Watanabe}}, \bibinfo {author} {\bibfnamefont {T.}~\bibnamefont {Taniguchi}}, \bibinfo {author} {\bibfnamefont {X.}~\bibnamefont {Xu}}, \bibinfo {author} {\bibfnamefont {J.-H.}\ \bibnamefont {Chu}},\ and\ \bibinfo {author} {\bibfnamefont {M.}~\bibnamefont {Yankowitz}},\ }\href {https://arxiv.org/abs/2505.10506} {\bibfield  {journal} {\bibinfo  {journal} {arXiv:2505.10506}\ } (\bibinfo {year} {2025})}\BibitemShut {NoStop}%
\bibitem [{\citenamefont {Couto}\ \emph {et~al.}(2014)\citenamefont {Couto}, \citenamefont {Costanzo}, \citenamefont {Engels}, \citenamefont {Ki}, \citenamefont {Watanabe}, \citenamefont {Taniguchi}, \citenamefont {Stampfer}, \citenamefont {Guinea},\ and\ \citenamefont {Morpurgo}}]{couto2014random}%
  \BibitemOpen
  \bibfield  {author} {\bibinfo {author} {\bibfnamefont {N.~J.~G.}\ \bibnamefont {Couto}}, \bibinfo {author} {\bibfnamefont {D.}~\bibnamefont {Costanzo}}, \bibinfo {author} {\bibfnamefont {S.}~\bibnamefont {Engels}}, \bibinfo {author} {\bibfnamefont {D.-K.}\ \bibnamefont {Ki}}, \bibinfo {author} {\bibfnamefont {K.}~\bibnamefont {Watanabe}}, \bibinfo {author} {\bibfnamefont {T.}~\bibnamefont {Taniguchi}}, \bibinfo {author} {\bibfnamefont {C.}~\bibnamefont {Stampfer}}, \bibinfo {author} {\bibfnamefont {F.}~\bibnamefont {Guinea}},\ and\ \bibinfo {author} {\bibfnamefont {A.~F.}\ \bibnamefont {Morpurgo}},\ }\href {https://doi.org/10.1103/PhysRevX.4.041019} {\bibfield  {journal} {\bibinfo  {journal} {Phys. Rev. X}\ }\textbf {\bibinfo {volume} {4}},\ \bibinfo {pages} {041019} (\bibinfo {year} {2014})}\BibitemShut {NoStop}%
\bibitem [{\citenamefont {Hoke}\ \emph {et~al.}(2026)\citenamefont {Hoke}, \citenamefont {Li}, \citenamefont {Hu}, \citenamefont {May-Mann}, \citenamefont {Watanabe}, \citenamefont {Taniguchi}, \citenamefont {Devakul},\ and\ \citenamefont {Feldman}}]{hoke2026imaging}%
  \BibitemOpen
  \bibfield  {author} {\bibinfo {author} {\bibfnamefont {J.~C.}\ \bibnamefont {Hoke}}, \bibinfo {author} {\bibfnamefont {Y.}~\bibnamefont {Li}}, \bibinfo {author} {\bibfnamefont {Y.}~\bibnamefont {Hu}}, \bibinfo {author} {\bibfnamefont {J.}~\bibnamefont {May-Mann}}, \bibinfo {author} {\bibfnamefont {K.}~\bibnamefont {Watanabe}}, \bibinfo {author} {\bibfnamefont {T.}~\bibnamefont {Taniguchi}}, \bibinfo {author} {\bibfnamefont {T.}~\bibnamefont {Devakul}},\ and\ \bibinfo {author} {\bibfnamefont {B.~E.}\ \bibnamefont {Feldman}},\ }\href {https://doi.org/10.1038/s41563-025-02423-3} {\bibfield  {journal} {\bibinfo  {journal} {Nat. Mater.}\ }\textbf {\bibinfo {volume} {25}},\ \bibinfo {pages} {775} (\bibinfo {year} {2026})}\BibitemShut {NoStop}%
\bibitem [{\citenamefont {Li}\ \emph {et~al.}(2020)\citenamefont {Li}, \citenamefont {Utama}, \citenamefont {Wang}, \citenamefont {Zhao}, \citenamefont {Zhao}, \citenamefont {Xiao}, \citenamefont {Jiang}, \citenamefont {Jiang}, \citenamefont {Taniguchi}, \citenamefont {Watanabe}, \citenamefont {Weber-Bargioni}, \citenamefont {Zettl},\ and\ \citenamefont {Wang}}]{li2020global}%
  \BibitemOpen
  \bibfield  {author} {\bibinfo {author} {\bibfnamefont {H.}~\bibnamefont {Li}}, \bibinfo {author} {\bibfnamefont {M.~B.}\ \bibnamefont {Utama}}, \bibinfo {author} {\bibfnamefont {S.}~\bibnamefont {Wang}}, \bibinfo {author} {\bibfnamefont {W.}~\bibnamefont {Zhao}}, \bibinfo {author} {\bibfnamefont {S.}~\bibnamefont {Zhao}}, \bibinfo {author} {\bibfnamefont {X.}~\bibnamefont {Xiao}}, \bibinfo {author} {\bibfnamefont {Y.}~\bibnamefont {Jiang}}, \bibinfo {author} {\bibfnamefont {L.}~\bibnamefont {Jiang}}, \bibinfo {author} {\bibfnamefont {T.}~\bibnamefont {Taniguchi}}, \bibinfo {author} {\bibfnamefont {K.}~\bibnamefont {Watanabe}}, \bibinfo {author} {\bibfnamefont {A.}~\bibnamefont {Weber-Bargioni}}, \bibinfo {author} {\bibfnamefont {A.}~\bibnamefont {Zettl}},\ and\ \bibinfo {author} {\bibfnamefont {F.}~\bibnamefont {Wang}},\ }\href {https://pubs.acs.org/doi/full/10.1021/acs.nanolett.9b05092} {\bibfield  {journal} {\bibinfo  {journal} {Nano Lett.}\ }\textbf {\bibinfo {volume} {20}},\ \bibinfo {pages} {3106}
  (\bibinfo {year} {2020})}\BibitemShut {NoStop}%
\bibitem [{\citenamefont {Kerelsky}\ \emph {et~al.}(2021)\citenamefont {Kerelsky}, \citenamefont {Rubio-Verdú}, \citenamefont {Xian}, \citenamefont {Kennes}, \citenamefont {Halbertal}, \citenamefont {Finney}, \citenamefont {Song}, \citenamefont {Turkel}, \citenamefont {Wang}, \citenamefont {Watanabe}, \citenamefont {Taniguchi}, \citenamefont {Hone}, \citenamefont {Dean}, \citenamefont {Basov}, \citenamefont {Rubio},\ and\ \citenamefont {Pasupathy}}]{kerelsky-moireless-2021}%
  \BibitemOpen
  \bibfield  {author} {\bibinfo {author} {\bibfnamefont {A.}~\bibnamefont {Kerelsky}}, \bibinfo {author} {\bibfnamefont {C.}~\bibnamefont {Rubio-Verdú}}, \bibinfo {author} {\bibfnamefont {L.}~\bibnamefont {Xian}}, \bibinfo {author} {\bibfnamefont {D.~M.}\ \bibnamefont {Kennes}}, \bibinfo {author} {\bibfnamefont {D.}~\bibnamefont {Halbertal}}, \bibinfo {author} {\bibfnamefont {N.}~\bibnamefont {Finney}}, \bibinfo {author} {\bibfnamefont {L.}~\bibnamefont {Song}}, \bibinfo {author} {\bibfnamefont {S.}~\bibnamefont {Turkel}}, \bibinfo {author} {\bibfnamefont {L.}~\bibnamefont {Wang}}, \bibinfo {author} {\bibfnamefont {K.}~\bibnamefont {Watanabe}}, \bibinfo {author} {\bibfnamefont {T.}~\bibnamefont {Taniguchi}}, \bibinfo {author} {\bibfnamefont {J.}~\bibnamefont {Hone}}, \bibinfo {author} {\bibfnamefont {C.}~\bibnamefont {Dean}}, \bibinfo {author} {\bibfnamefont {D.~N.}\ \bibnamefont {Basov}}, \bibinfo {author} {\bibfnamefont {A.}~\bibnamefont {Rubio}},\ and\ \bibinfo {author} {\bibfnamefont {A.~N.}\
  \bibnamefont {Pasupathy}},\ }\href {https://www.pnas.org/doi/full/10.1073/pnas.2017366118} {\bibfield  {journal} {\bibinfo  {journal} {Proc. Natl. Acad. Sci. U.S.A.}\ }\textbf {\bibinfo {volume} {118}},\ \bibinfo {pages} {e2017366118} (\bibinfo {year} {2021})}\BibitemShut {NoStop}%
\bibitem [{\citenamefont {Halbertal}\ \emph {et~al.}(2021)\citenamefont {Halbertal}, \citenamefont {Finney}, \citenamefont {Sunku}, \citenamefont {Kerelsky}, \citenamefont {Rubio-Verd{\'u}}, \citenamefont {Shabani}, \citenamefont {Xian}, \citenamefont {Carr}, \citenamefont {Chen}, \citenamefont {Zhang}, \citenamefont {Wang}, \citenamefont {Gonzalez-Acevedo}, \citenamefont {McLeod}, \citenamefont {Rhodes}, \citenamefont {Watanabe}, \citenamefont {Taniguchi}, \citenamefont {Kaxiras}, \citenamefont {Dean}, \citenamefont {Hone}, \citenamefont {Pasupathy}, \citenamefont {Kennes}, \citenamefont {Rubio},\ and\ \citenamefont {Basov}}]{halbertal-moire-2021}%
  \BibitemOpen
  \bibfield  {author} {\bibinfo {author} {\bibfnamefont {D.}~\bibnamefont {Halbertal}}, \bibinfo {author} {\bibfnamefont {N.~R.}\ \bibnamefont {Finney}}, \bibinfo {author} {\bibfnamefont {S.~S.}\ \bibnamefont {Sunku}}, \bibinfo {author} {\bibfnamefont {A.}~\bibnamefont {Kerelsky}}, \bibinfo {author} {\bibfnamefont {C.}~\bibnamefont {Rubio-Verd{\'u}}}, \bibinfo {author} {\bibfnamefont {S.}~\bibnamefont {Shabani}}, \bibinfo {author} {\bibfnamefont {L.}~\bibnamefont {Xian}}, \bibinfo {author} {\bibfnamefont {S.}~\bibnamefont {Carr}}, \bibinfo {author} {\bibfnamefont {S.}~\bibnamefont {Chen}}, \bibinfo {author} {\bibfnamefont {C.}~\bibnamefont {Zhang}}, \bibinfo {author} {\bibfnamefont {L.}~\bibnamefont {Wang}}, \bibinfo {author} {\bibfnamefont {D.}~\bibnamefont {Gonzalez-Acevedo}}, \bibinfo {author} {\bibfnamefont {A.~S.}\ \bibnamefont {McLeod}}, \bibinfo {author} {\bibfnamefont {D.}~\bibnamefont {Rhodes}}, \bibinfo {author} {\bibfnamefont {K.}~\bibnamefont {Watanabe}}, \bibinfo {author} {\bibfnamefont
  {T.}~\bibnamefont {Taniguchi}}, \bibinfo {author} {\bibfnamefont {E.}~\bibnamefont {Kaxiras}}, \bibinfo {author} {\bibfnamefont {C.~R.}\ \bibnamefont {Dean}}, \bibinfo {author} {\bibfnamefont {J.~C.}\ \bibnamefont {Hone}}, \bibinfo {author} {\bibfnamefont {A.~N.}\ \bibnamefont {Pasupathy}}, \bibinfo {author} {\bibfnamefont {D.~M.}\ \bibnamefont {Kennes}}, \bibinfo {author} {\bibfnamefont {A.}~\bibnamefont {Rubio}},\ and\ \bibinfo {author} {\bibfnamefont {D.~N.}\ \bibnamefont {Basov}},\ }\href {https://doi.org/10.1038/s41467-020-20486-3} {\bibfield  {journal} {\bibinfo  {journal} {Nat. Commun.}\ }\textbf {\bibinfo {volume} {12}},\ \bibinfo {pages} {242} (\bibinfo {year} {2021})}\BibitemShut {NoStop}%
\bibitem [{\citenamefont {Park}\ \emph {et~al.}(2025)\citenamefont {Park}, \citenamefont {Park}, \citenamefont {Yananose}, \citenamefont {Ko}, \citenamefont {Kim}, \citenamefont {Engelke}, \citenamefont {Zhang}, \citenamefont {Davydov}, \citenamefont {Green}, \citenamefont {Kim}, \citenamefont {Park}, \citenamefont {Lee}, \citenamefont {Kim}, \citenamefont {Kim}, \citenamefont {Watanabe}, \citenamefont {Taniguchi}, \citenamefont {Yang}, \citenamefont {Wang}, \citenamefont {Kim}, \citenamefont {Son},\ and\ \citenamefont {Yoo}}]{park2025unconventional}%
  \BibitemOpen
  \bibfield  {author} {\bibinfo {author} {\bibfnamefont {D.}~\bibnamefont {Park}}, \bibinfo {author} {\bibfnamefont {C.}~\bibnamefont {Park}}, \bibinfo {author} {\bibfnamefont {K.}~\bibnamefont {Yananose}}, \bibinfo {author} {\bibfnamefont {E.}~\bibnamefont {Ko}}, \bibinfo {author} {\bibfnamefont {B.}~\bibnamefont {Kim}}, \bibinfo {author} {\bibfnamefont {R.}~\bibnamefont {Engelke}}, \bibinfo {author} {\bibfnamefont {X.}~\bibnamefont {Zhang}}, \bibinfo {author} {\bibfnamefont {K.}~\bibnamefont {Davydov}}, \bibinfo {author} {\bibfnamefont {M.}~\bibnamefont {Green}}, \bibinfo {author} {\bibfnamefont {H.-M.}\ \bibnamefont {Kim}}, \bibinfo {author} {\bibfnamefont {S.~H.}\ \bibnamefont {Park}}, \bibinfo {author} {\bibfnamefont {J.~H.}\ \bibnamefont {Lee}}, \bibinfo {author} {\bibfnamefont {S.-G.}\ \bibnamefont {Kim}}, \bibinfo {author} {\bibfnamefont {H.}~\bibnamefont {Kim}}, \bibinfo {author} {\bibfnamefont {K.}~\bibnamefont {Watanabe}}, \bibinfo {author} {\bibfnamefont {T.}~\bibnamefont {Taniguchi}}, \bibinfo
  {author} {\bibfnamefont {S.~M.}\ \bibnamefont {Yang}}, \bibinfo {author} {\bibfnamefont {K.}~\bibnamefont {Wang}}, \bibinfo {author} {\bibfnamefont {P.}~\bibnamefont {Kim}}, \bibinfo {author} {\bibfnamefont {Y.-W.}\ \bibnamefont {Son}},\ and\ \bibinfo {author} {\bibfnamefont {H.}~\bibnamefont {Yoo}},\ }\href {https://www.nature.com/articles/s41586-025-08932-0} {\bibfield  {journal} {\bibinfo  {journal} {Nature}\ }\textbf {\bibinfo {volume} {641}},\ \bibinfo {pages} {896} (\bibinfo {year} {2025})}\BibitemShut {NoStop}%
\bibitem [{\citenamefont {Alden}\ \emph {et~al.}(2013)\citenamefont {Alden}, \citenamefont {Tsen}, \citenamefont {Huang}, \citenamefont {Hovden}, \citenamefont {Brown}, \citenamefont {Park}, \citenamefont {Muller},\ and\ \citenamefont {McEuen}}]{alden-strain-2013}%
  \BibitemOpen
  \bibfield  {author} {\bibinfo {author} {\bibfnamefont {J.~S.}\ \bibnamefont {Alden}}, \bibinfo {author} {\bibfnamefont {A.~W.}\ \bibnamefont {Tsen}}, \bibinfo {author} {\bibfnamefont {P.~Y.}\ \bibnamefont {Huang}}, \bibinfo {author} {\bibfnamefont {R.}~\bibnamefont {Hovden}}, \bibinfo {author} {\bibfnamefont {L.}~\bibnamefont {Brown}}, \bibinfo {author} {\bibfnamefont {J.}~\bibnamefont {Park}}, \bibinfo {author} {\bibfnamefont {D.~A.}\ \bibnamefont {Muller}},\ and\ \bibinfo {author} {\bibfnamefont {P.~L.}\ \bibnamefont {McEuen}},\ }\href {https://doi.org/10.1073/pnas.1309394110} {\bibfield  {journal} {\bibinfo  {journal} {Proc. Natl. Acad. Sci. U.S.A.}\ }\textbf {\bibinfo {volume} {110}},\ \bibinfo {pages} {11256} (\bibinfo {year} {2013})}\BibitemShut {NoStop}%
\bibitem [{\citenamefont {Kazmierczak}\ \emph {et~al.}(2021)\citenamefont {Kazmierczak}, \citenamefont {Van~Winkle}, \citenamefont {Ophus}, \citenamefont {Bustillo}, \citenamefont {Carr}, \citenamefont {Brown}, \citenamefont {Ciston}, \citenamefont {Taniguchi}, \citenamefont {Watanabe},\ and\ \citenamefont {Bediako}}]{kazmierczak-strain-2021}%
  \BibitemOpen
  \bibfield  {author} {\bibinfo {author} {\bibfnamefont {N.~P.}\ \bibnamefont {Kazmierczak}}, \bibinfo {author} {\bibfnamefont {M.}~\bibnamefont {Van~Winkle}}, \bibinfo {author} {\bibfnamefont {C.}~\bibnamefont {Ophus}}, \bibinfo {author} {\bibfnamefont {K.~C.}\ \bibnamefont {Bustillo}}, \bibinfo {author} {\bibfnamefont {S.}~\bibnamefont {Carr}}, \bibinfo {author} {\bibfnamefont {H.~G.}\ \bibnamefont {Brown}}, \bibinfo {author} {\bibfnamefont {J.}~\bibnamefont {Ciston}}, \bibinfo {author} {\bibfnamefont {T.}~\bibnamefont {Taniguchi}}, \bibinfo {author} {\bibfnamefont {K.}~\bibnamefont {Watanabe}},\ and\ \bibinfo {author} {\bibfnamefont {D.~K.}\ \bibnamefont {Bediako}},\ }\href {https://doi.org/10.1038/s41563-021-00973-w} {\bibfield  {journal} {\bibinfo  {journal} {Nat. Mater.}\ }\textbf {\bibinfo {volume} {20}},\ \bibinfo {pages} {956} (\bibinfo {year} {2021})}\BibitemShut {NoStop}%
\bibitem [{SM()}]{SM}%
  \BibitemOpen
  \href@noop {} {}\bibinfo {note} {See the Supplementary Material for additional information on the following topics: I.Device fabrication and characterization. II.Additional STM and QPI measurements. III. Additional theoretical calculations for the elastic and electronic properties.}\BibitemShut {Stop}%
\bibitem [{\citenamefont {Lebedeva}\ and\ \citenamefont {Popov}(2020{\natexlab{a}})}]{lebedeva-energetics-2019}%
  \BibitemOpen
  \bibfield  {author} {\bibinfo {author} {\bibfnamefont {I.~V.}\ \bibnamefont {Lebedeva}}\ and\ \bibinfo {author} {\bibfnamefont {A.~M.}\ \bibnamefont {Popov}},\ }\href {https://doi.org/10.1021/acs.jpcc.9b08306} {\bibfield  {journal} {\bibinfo  {journal} {J. Phys. Chem. C}\ }\textbf {\bibinfo {volume} {124}},\ \bibinfo {pages} {2120} (\bibinfo {year} {2020}{\natexlab{a}})}\BibitemShut {NoStop}%
\bibitem [{\citenamefont {Lebedeva}\ and\ \citenamefont {Popov}(2020{\natexlab{b}})}]{lebedeva-two-2020}%
  \BibitemOpen
  \bibfield  {author} {\bibinfo {author} {\bibfnamefont {I.~V.}\ \bibnamefont {Lebedeva}}\ and\ \bibinfo {author} {\bibfnamefont {A.~M.}\ \bibnamefont {Popov}},\ }\href {https://doi.org/10.1103/PhysRevLett.124.116101} {\bibfield  {journal} {\bibinfo  {journal} {Phys. Rev. Lett.}\ }\textbf {\bibinfo {volume} {124}},\ \bibinfo {pages} {116101} (\bibinfo {year} {2020}{\natexlab{b}})}\BibitemShut {NoStop}%
\bibitem [{\citenamefont {Escudero}\ \emph {et~al.}(2024)\citenamefont {Escudero}, \citenamefont {Sinner}, \citenamefont {Zhan}, \citenamefont {Pantaleón},\ and\ \citenamefont {Guinea}}]{escudero-designing-2024}%
  \BibitemOpen
  \bibfield  {author} {\bibinfo {author} {\bibfnamefont {F.}~\bibnamefont {Escudero}}, \bibinfo {author} {\bibfnamefont {A.}~\bibnamefont {Sinner}}, \bibinfo {author} {\bibfnamefont {Z.}~\bibnamefont {Zhan}}, \bibinfo {author} {\bibfnamefont {P.~A.}\ \bibnamefont {Pantaleón}},\ and\ \bibinfo {author} {\bibfnamefont {F.}~\bibnamefont {Guinea}},\ }\href {https://doi.org/10.1103/PhysRevResearch.6.023203} {\bibfield  {journal} {\bibinfo  {journal} {Phys. Rev. Research}\ }\textbf {\bibinfo {volume} {6}},\ \bibinfo {pages} {023203} (\bibinfo {year} {2024})}\BibitemShut {NoStop}%
\bibitem [{\citenamefont {Ouyang}\ \emph {et~al.}(2026)\citenamefont {Ouyang}, \citenamefont {Yu}, \citenamefont {Li}, \citenamefont {Jia}, \citenamefont {Wang}, \citenamefont {Xiao}, \citenamefont {Zhang}, \citenamefont {Hu}, \citenamefont {Pantale{\'o}n}, \citenamefont {Zhan}, \citenamefont {Zhou}, \citenamefont {Guinea}, \citenamefont {Xue},\ and\ \citenamefont {Li}}]{ouyang2025structural}%
  \BibitemOpen
  \bibfield  {author} {\bibinfo {author} {\bibfnamefont {P.}~\bibnamefont {Ouyang}}, \bibinfo {author} {\bibfnamefont {J.}~\bibnamefont {Yu}}, \bibinfo {author} {\bibfnamefont {Q.}~\bibnamefont {Li}}, \bibinfo {author} {\bibfnamefont {G.}~\bibnamefont {Jia}}, \bibinfo {author} {\bibfnamefont {Y.}~\bibnamefont {Wang}}, \bibinfo {author} {\bibfnamefont {K.}~\bibnamefont {Xiao}}, \bibinfo {author} {\bibfnamefont {H.}~\bibnamefont {Zhang}}, \bibinfo {author} {\bibfnamefont {Z.}~\bibnamefont {Hu}}, \bibinfo {author} {\bibfnamefont {P.~A.}\ \bibnamefont {Pantale{\'o}n}}, \bibinfo {author} {\bibfnamefont {Z.}~\bibnamefont {Zhan}}, \bibinfo {author} {\bibfnamefont {S.}~\bibnamefont {Zhou}}, \bibinfo {author} {\bibfnamefont {F.}~\bibnamefont {Guinea}}, \bibinfo {author} {\bibfnamefont {Q.-K.}\ \bibnamefont {Xue}},\ and\ \bibinfo {author} {\bibfnamefont {W.}~\bibnamefont {Li}},\ }\href {https://doi.org/10.1093/nsr/nwaf568} {\bibfield  {journal} {\bibinfo  {journal} {Natl. Sci. Rev.}\ }\textbf {\bibinfo {volume} {13}},\
  \bibinfo {pages} {nwaf568} (\bibinfo {year} {2026})}\BibitemShut {NoStop}%
\bibitem [{\citenamefont {Koshino}(2013)}]{koshino-electronic-2013}%
  \BibitemOpen
  \bibfield  {author} {\bibinfo {author} {\bibfnamefont {M.}~\bibnamefont {Koshino}},\ }\href {https://doi.org/10.1103/PhysRevB.88.115409} {\bibfield  {journal} {\bibinfo  {journal} {Phys. Rev. B}\ }\textbf {\bibinfo {volume} {88}},\ \bibinfo {pages} {115409} (\bibinfo {year} {2013})}\BibitemShut {NoStop}%
\bibitem [{\citenamefont {Jiang}\ \emph {et~al.}(2016)\citenamefont {Jiang}, \citenamefont {Shi}, \citenamefont {Zeng}, \citenamefont {Wang}, \citenamefont {Kang}, \citenamefont {Joshi}, \citenamefont {Jin}, \citenamefont {Ju}, \citenamefont {Kim}, \citenamefont {Lyu}, \citenamefont {Shen}, \citenamefont {Crommie}, \citenamefont {Gao},\ and\ \citenamefont {Wang}}]{jiang-soliton-dependent-2016}%
  \BibitemOpen
  \bibfield  {author} {\bibinfo {author} {\bibfnamefont {L.}~\bibnamefont {Jiang}}, \bibinfo {author} {\bibfnamefont {Z.}~\bibnamefont {Shi}}, \bibinfo {author} {\bibfnamefont {B.}~\bibnamefont {Zeng}}, \bibinfo {author} {\bibfnamefont {S.}~\bibnamefont {Wang}}, \bibinfo {author} {\bibfnamefont {J.-H.}\ \bibnamefont {Kang}}, \bibinfo {author} {\bibfnamefont {T.}~\bibnamefont {Joshi}}, \bibinfo {author} {\bibfnamefont {C.}~\bibnamefont {Jin}}, \bibinfo {author} {\bibfnamefont {L.}~\bibnamefont {Ju}}, \bibinfo {author} {\bibfnamefont {J.}~\bibnamefont {Kim}}, \bibinfo {author} {\bibfnamefont {T.}~\bibnamefont {Lyu}}, \bibinfo {author} {\bibfnamefont {Y.-R.}\ \bibnamefont {Shen}}, \bibinfo {author} {\bibfnamefont {M.}~\bibnamefont {Crommie}}, \bibinfo {author} {\bibfnamefont {H.-J.}\ \bibnamefont {Gao}},\ and\ \bibinfo {author} {\bibfnamefont {F.}~\bibnamefont {Wang}},\ }\href {https://doi.org/10.1038/nmat4653} {\bibfield  {journal} {\bibinfo  {journal} {Nat. Mater.}\ }\textbf {\bibinfo {volume} {15}},\ \bibinfo
  {pages} {840} (\bibinfo {year} {2016})}\BibitemShut {NoStop}%
\bibitem [{\citenamefont {Mesple}\ \emph {et~al.}(2023)\citenamefont {Mesple}, \citenamefont {Walet}, \citenamefont {Trambly~de Laissardi{\`e}re}, \citenamefont {Guinea}, \citenamefont {Do{\v{s}}enovi{\'c}}, \citenamefont {Okuno}, \citenamefont {Paillet}, \citenamefont {Michon}, \citenamefont {Chapelier},\ and\ \citenamefont {Renard}}]{mesple-giant-2023}%
  \BibitemOpen
  \bibfield  {author} {\bibinfo {author} {\bibfnamefont {F.}~\bibnamefont {Mesple}}, \bibinfo {author} {\bibfnamefont {N.~R.}\ \bibnamefont {Walet}}, \bibinfo {author} {\bibfnamefont {G.}~\bibnamefont {Trambly~de Laissardi{\`e}re}}, \bibinfo {author} {\bibfnamefont {F.}~\bibnamefont {Guinea}}, \bibinfo {author} {\bibfnamefont {D.}~\bibnamefont {Do{\v{s}}enovi{\'c}}}, \bibinfo {author} {\bibfnamefont {H.}~\bibnamefont {Okuno}}, \bibinfo {author} {\bibfnamefont {C.}~\bibnamefont {Paillet}}, \bibinfo {author} {\bibfnamefont {A.}~\bibnamefont {Michon}}, \bibinfo {author} {\bibfnamefont {C.}~\bibnamefont {Chapelier}},\ and\ \bibinfo {author} {\bibfnamefont {V.~T.}\ \bibnamefont {Renard}},\ }\href {https://doi.org/10.1002/adma.202306312} {\bibfield  {journal} {\bibinfo  {journal} {Adv. Mater.}\ }\textbf {\bibinfo {volume} {35}},\ \bibinfo {pages} {2306312} (\bibinfo {year} {2023})}\BibitemShut {NoStop}%
\bibitem [{\citenamefont {Jiang}\ \emph {et~al.}(2018)\citenamefont {Jiang}, \citenamefont {Wang}, \citenamefont {Shi}, \citenamefont {Jin}, \citenamefont {Utama}, \citenamefont {Zhao}, \citenamefont {Shen}, \citenamefont {Gao}, \citenamefont {Zhang},\ and\ \citenamefont {Wang}}]{jiang-manipulation-2018}%
  \BibitemOpen
  \bibfield  {author} {\bibinfo {author} {\bibfnamefont {L.}~\bibnamefont {Jiang}}, \bibinfo {author} {\bibfnamefont {S.}~\bibnamefont {Wang}}, \bibinfo {author} {\bibfnamefont {Z.}~\bibnamefont {Shi}}, \bibinfo {author} {\bibfnamefont {C.}~\bibnamefont {Jin}}, \bibinfo {author} {\bibfnamefont {M.~I.~B.}\ \bibnamefont {Utama}}, \bibinfo {author} {\bibfnamefont {S.}~\bibnamefont {Zhao}}, \bibinfo {author} {\bibfnamefont {Y.-R.}\ \bibnamefont {Shen}}, \bibinfo {author} {\bibfnamefont {H.-J.}\ \bibnamefont {Gao}}, \bibinfo {author} {\bibfnamefont {G.}~\bibnamefont {Zhang}},\ and\ \bibinfo {author} {\bibfnamefont {F.}~\bibnamefont {Wang}},\ }\href {https://doi.org/10.1038/s41565-017-0042-6} {\bibfield  {journal} {\bibinfo  {journal} {Nat. Nanotechnol.}\ }\textbf {\bibinfo {volume} {13}},\ \bibinfo {pages} {204} (\bibinfo {year} {2018})}\BibitemShut {NoStop}%
\bibitem [{\citenamefont {Hasdeo}\ and\ \citenamefont {Song}(2017)}]{hasdeo-long-lived-2017}%
  \BibitemOpen
  \bibfield  {author} {\bibinfo {author} {\bibfnamefont {E.~H.}\ \bibnamefont {Hasdeo}}\ and\ \bibinfo {author} {\bibfnamefont {J.~C.}\ \bibnamefont {Song}},\ }\href {https://doi.org/10.1021/acs.nanolett.7b02584} {\bibfield  {journal} {\bibinfo  {journal} {Nano Lett.}\ }\textbf {\bibinfo {volume} {17}},\ \bibinfo {pages} {7252} (\bibinfo {year} {2017})}\BibitemShut {NoStop}%
\bibitem [{\citenamefont {Liu}\ \emph {et~al.}(2024{\natexlab{b}})\citenamefont {Liu}, \citenamefont {Gupta}, \citenamefont {Choi}, \citenamefont {Vituri}, \citenamefont {Stoyanov}, \citenamefont {Xiao}, \citenamefont {Wang}, \citenamefont {Zhou}, \citenamefont {Barick}, \citenamefont {Taniguchi}, \citenamefont {Watanabe}, \citenamefont {Yan}, \citenamefont {Berg}, \citenamefont {Young}, \citenamefont {Beidenkopf},\ and\ \citenamefont {Avraham}}]{liu-visualizing-2024}%
  \BibitemOpen
  \bibfield  {author} {\bibinfo {author} {\bibfnamefont {Y.}~\bibnamefont {Liu}}, \bibinfo {author} {\bibfnamefont {A.}~\bibnamefont {Gupta}}, \bibinfo {author} {\bibfnamefont {Y.}~\bibnamefont {Choi}}, \bibinfo {author} {\bibfnamefont {Y.}~\bibnamefont {Vituri}}, \bibinfo {author} {\bibfnamefont {H.}~\bibnamefont {Stoyanov}}, \bibinfo {author} {\bibfnamefont {J.}~\bibnamefont {Xiao}}, \bibinfo {author} {\bibfnamefont {Y.}~\bibnamefont {Wang}}, \bibinfo {author} {\bibfnamefont {H.}~\bibnamefont {Zhou}}, \bibinfo {author} {\bibfnamefont {B.}~\bibnamefont {Barick}}, \bibinfo {author} {\bibfnamefont {T.}~\bibnamefont {Taniguchi}}, \bibinfo {author} {\bibfnamefont {K.}~\bibnamefont {Watanabe}}, \bibinfo {author} {\bibfnamefont {B.}~\bibnamefont {Yan}}, \bibinfo {author} {\bibfnamefont {E.}~\bibnamefont {Berg}}, \bibinfo {author} {\bibfnamefont {A.~F.}\ \bibnamefont {Young}}, \bibinfo {author} {\bibfnamefont {H.}~\bibnamefont {Beidenkopf}},\ and\ \bibinfo {author} {\bibfnamefont {N.}~\bibnamefont {Avraham}},\
  }\href {https://arxiv.org/abs/2411.11163} {\bibfield  {journal} {\bibinfo  {journal} {arXiv:2411.11163}\ } (\bibinfo {year} {2024}{\natexlab{b}})}\BibitemShut {NoStop}%
\bibitem [{\citenamefont {Yuan}\ \emph {et~al.}(2019)\citenamefont {Yuan}, \citenamefont {Isobe},\ and\ \citenamefont {Fu}}]{yuan-magic-2019}%
  \BibitemOpen
  \bibfield  {author} {\bibinfo {author} {\bibfnamefont {N.~F.~Q.}\ \bibnamefont {Yuan}}, \bibinfo {author} {\bibfnamefont {H.}~\bibnamefont {Isobe}},\ and\ \bibinfo {author} {\bibfnamefont {L.}~\bibnamefont {Fu}},\ }\href {https://doi.org/10.1038/s41467-019-13670-9} {\bibfield  {journal} {\bibinfo  {journal} {Nat. Commun.}\ }\textbf {\bibinfo {volume} {10}},\ \bibinfo {pages} {5769} (\bibinfo {year} {2019})}\BibitemShut {NoStop}%
\bibitem [{\citenamefont {Jiang}\ \emph {et~al.}(2017)\citenamefont {Jiang}, \citenamefont {Mao}, \citenamefont {Duan}, \citenamefont {Lai}, \citenamefont {Watanabe}, \citenamefont {Taniguchi},\ and\ \citenamefont {Andrei}}]{jiang-visualizing-2017}%
  \BibitemOpen
  \bibfield  {author} {\bibinfo {author} {\bibfnamefont {Y.}~\bibnamefont {Jiang}}, \bibinfo {author} {\bibfnamefont {J.}~\bibnamefont {Mao}}, \bibinfo {author} {\bibfnamefont {J.}~\bibnamefont {Duan}}, \bibinfo {author} {\bibfnamefont {X.}~\bibnamefont {Lai}}, \bibinfo {author} {\bibfnamefont {K.}~\bibnamefont {Watanabe}}, \bibinfo {author} {\bibfnamefont {T.}~\bibnamefont {Taniguchi}},\ and\ \bibinfo {author} {\bibfnamefont {E.~Y.}\ \bibnamefont {Andrei}},\ }\href {https://doi.org/10.1021/acs.nanolett.6b05228} {\bibfield  {journal} {\bibinfo  {journal} {Nano Lett.}\ }\textbf {\bibinfo {volume} {17}},\ \bibinfo {pages} {2839} (\bibinfo {year} {2017})}\BibitemShut {NoStop}%
\bibitem [{\citenamefont {Sequeira}\ \emph {et~al.}(2024)\citenamefont {Sequeira}, \citenamefont {Barabas}, \citenamefont {Barajas-Aguilar}, \citenamefont {Bacani}, \citenamefont {Nakatsuji}, \citenamefont {Koshino}, \citenamefont {Taniguchi}, \citenamefont {Watanabe},\ and\ \citenamefont {Sanchez-Yamagishi}}]{sequeira2024manipulating}%
  \BibitemOpen
  \bibfield  {author} {\bibinfo {author} {\bibfnamefont {I.}~\bibnamefont {Sequeira}}, \bibinfo {author} {\bibfnamefont {A.~Z.}\ \bibnamefont {Barabas}}, \bibinfo {author} {\bibfnamefont {A.~H.}\ \bibnamefont {Barajas-Aguilar}}, \bibinfo {author} {\bibfnamefont {M.~G.}\ \bibnamefont {Bacani}}, \bibinfo {author} {\bibfnamefont {N.}~\bibnamefont {Nakatsuji}}, \bibinfo {author} {\bibfnamefont {M.}~\bibnamefont {Koshino}}, \bibinfo {author} {\bibfnamefont {T.}~\bibnamefont {Taniguchi}}, \bibinfo {author} {\bibfnamefont {K.}~\bibnamefont {Watanabe}},\ and\ \bibinfo {author} {\bibfnamefont {J.~D.}\ \bibnamefont {Sanchez-Yamagishi}},\ }\href {https://doi.org/10.1021/acs.nanolett.4c04201} {\bibfield  {journal} {\bibinfo  {journal} {Nano Lett.}\ }\textbf {\bibinfo {volume} {24}},\ \bibinfo {pages} {15662} (\bibinfo {year} {2024})}\BibitemShut {NoStop}%
\bibitem [{\citenamefont {Kapfer}\ \emph {et~al.}(2023)\citenamefont {Kapfer}, \citenamefont {Jessen}, \citenamefont {Eisele}, \citenamefont {Fu}, \citenamefont {Danielsen}, \citenamefont {Darlington}, \citenamefont {Moore}, \citenamefont {Finney}, \citenamefont {Marchese}, \citenamefont {Hsieh}, \citenamefont {Majchrzak}, \citenamefont {Jiang}, \citenamefont {Biswas}, \citenamefont {Dudin}, \citenamefont {Avila}, \citenamefont {Watanabe}, \citenamefont {Taniguchi}, \citenamefont {Ulstrup}, \citenamefont {Bøggild}, \citenamefont {Schuck}, \citenamefont {Basov}, \citenamefont {Hone},\ and\ \citenamefont {Dean}}]{kapfer2023programming}%
  \BibitemOpen
  \bibfield  {author} {\bibinfo {author} {\bibfnamefont {M.}~\bibnamefont {Kapfer}}, \bibinfo {author} {\bibfnamefont {B.~S.}\ \bibnamefont {Jessen}}, \bibinfo {author} {\bibfnamefont {M.~E.}\ \bibnamefont {Eisele}}, \bibinfo {author} {\bibfnamefont {M.}~\bibnamefont {Fu}}, \bibinfo {author} {\bibfnamefont {D.~R.}\ \bibnamefont {Danielsen}}, \bibinfo {author} {\bibfnamefont {T.~P.}\ \bibnamefont {Darlington}}, \bibinfo {author} {\bibfnamefont {S.~L.}\ \bibnamefont {Moore}}, \bibinfo {author} {\bibfnamefont {N.~R.}\ \bibnamefont {Finney}}, \bibinfo {author} {\bibfnamefont {A.}~\bibnamefont {Marchese}}, \bibinfo {author} {\bibfnamefont {V.}~\bibnamefont {Hsieh}}, \bibinfo {author} {\bibfnamefont {P.}~\bibnamefont {Majchrzak}}, \bibinfo {author} {\bibfnamefont {Z.}~\bibnamefont {Jiang}}, \bibinfo {author} {\bibfnamefont {D.}~\bibnamefont {Biswas}}, \bibinfo {author} {\bibfnamefont {P.}~\bibnamefont {Dudin}}, \bibinfo {author} {\bibfnamefont {J.}~\bibnamefont {Avila}}, \bibinfo {author} {\bibfnamefont
  {K.}~\bibnamefont {Watanabe}}, \bibinfo {author} {\bibfnamefont {T.}~\bibnamefont {Taniguchi}}, \bibinfo {author} {\bibfnamefont {S.}~\bibnamefont {Ulstrup}}, \bibinfo {author} {\bibfnamefont {P.}~\bibnamefont {Bøggild}}, \bibinfo {author} {\bibfnamefont {P.~J.}\ \bibnamefont {Schuck}}, \bibinfo {author} {\bibfnamefont {D.~N.}\ \bibnamefont {Basov}}, \bibinfo {author} {\bibfnamefont {J.}~\bibnamefont {Hone}},\ and\ \bibinfo {author} {\bibfnamefont {C.~R.}\ \bibnamefont {Dean}},\ }\href {https://doi.org/10.1126/science.ade9995} {\bibfield  {journal} {\bibinfo  {journal} {Science}\ }\textbf {\bibinfo {volume} {381}},\ \bibinfo {pages} {677} (\bibinfo {year} {2023})}\BibitemShut {NoStop}%
\bibitem [{\citenamefont {Li}\ \emph {et~al.}(2024)\citenamefont {Li}, \citenamefont {Xiang}, \citenamefont {Wang}, \citenamefont {Naik}, \citenamefont {Kim}, \citenamefont {Nie}, \citenamefont {Li}, \citenamefont {Ge}, \citenamefont {He}, \citenamefont {Ou}, \citenamefont {Banerjee}, \citenamefont {Taniguchi}, \citenamefont {Watanabe}, \citenamefont {Tongay}, \citenamefont {Zettl}, \citenamefont {Louie}, \citenamefont {Zaletel}, \citenamefont {Crommie},\ and\ \citenamefont {Wang}}]{li-imaging-2024}%
  \BibitemOpen
  \bibfield  {author} {\bibinfo {author} {\bibfnamefont {H.}~\bibnamefont {Li}}, \bibinfo {author} {\bibfnamefont {Z.}~\bibnamefont {Xiang}}, \bibinfo {author} {\bibfnamefont {T.}~\bibnamefont {Wang}}, \bibinfo {author} {\bibfnamefont {M.~H.}\ \bibnamefont {Naik}}, \bibinfo {author} {\bibfnamefont {W.}~\bibnamefont {Kim}}, \bibinfo {author} {\bibfnamefont {J.}~\bibnamefont {Nie}}, \bibinfo {author} {\bibfnamefont {S.}~\bibnamefont {Li}}, \bibinfo {author} {\bibfnamefont {Z.}~\bibnamefont {Ge}}, \bibinfo {author} {\bibfnamefont {Z.}~\bibnamefont {He}}, \bibinfo {author} {\bibfnamefont {Y.}~\bibnamefont {Ou}}, \bibinfo {author} {\bibfnamefont {R.}~\bibnamefont {Banerjee}}, \bibinfo {author} {\bibfnamefont {T.}~\bibnamefont {Taniguchi}}, \bibinfo {author} {\bibfnamefont {K.}~\bibnamefont {Watanabe}}, \bibinfo {author} {\bibfnamefont {S.}~\bibnamefont {Tongay}}, \bibinfo {author} {\bibfnamefont {A.}~\bibnamefont {Zettl}}, \bibinfo {author} {\bibfnamefont {S.~G.}\ \bibnamefont {Louie}}, \bibinfo {author}
  {\bibfnamefont {M.~P.}\ \bibnamefont {Zaletel}}, \bibinfo {author} {\bibfnamefont {M.~F.}\ \bibnamefont {Crommie}},\ and\ \bibinfo {author} {\bibfnamefont {F.}~\bibnamefont {Wang}},\ }\href {https://doi.org/10.1038/s41586-024-07596-6} {\bibfield  {journal} {\bibinfo  {journal} {Nature}\ }\textbf {\bibinfo {volume} {631}},\ \bibinfo {pages} {765} (\bibinfo {year} {2024})}\BibitemShut {NoStop}%
\bibitem [{\citenamefont {Bi}\ \emph {et~al.}(2019)\citenamefont {Bi}, \citenamefont {Yuan},\ and\ \citenamefont {Fu}}]{bi-designing-2019}%
  \BibitemOpen
  \bibfield  {author} {\bibinfo {author} {\bibfnamefont {Z.}~\bibnamefont {Bi}}, \bibinfo {author} {\bibfnamefont {N.~F.~Q.}\ \bibnamefont {Yuan}},\ and\ \bibinfo {author} {\bibfnamefont {L.}~\bibnamefont {Fu}},\ }\href {https://doi.org/10.1103/PhysRevB.100.035448} {\bibfield  {journal} {\bibinfo  {journal} {Phys. Rev. B}\ }\textbf {\bibinfo {volume} {100}},\ \bibinfo {pages} {035448} (\bibinfo {year} {2019})}\BibitemShut {NoStop}%
\bibitem [{\citenamefont {Mayorov}\ \emph {et~al.}(2011)\citenamefont {Mayorov}, \citenamefont {Elias}, \citenamefont {Mucha-Kruczy{\'n}ski}, \citenamefont {Gorbachev}, \citenamefont {Tudorovskiy}, \citenamefont {Zhukov}, \citenamefont {Morozov}, \citenamefont {Katsnelson}, \citenamefont {Fal'ko}, \citenamefont {Geim},\ and\ \citenamefont {Novoselov}}]{mayorov2011interaction}%
  \BibitemOpen
  \bibfield  {author} {\bibinfo {author} {\bibfnamefont {A.~S.}\ \bibnamefont {Mayorov}}, \bibinfo {author} {\bibfnamefont {D.~C.}\ \bibnamefont {Elias}}, \bibinfo {author} {\bibfnamefont {M.}~\bibnamefont {Mucha-Kruczy{\'n}ski}}, \bibinfo {author} {\bibfnamefont {R.~V.}\ \bibnamefont {Gorbachev}}, \bibinfo {author} {\bibfnamefont {T.}~\bibnamefont {Tudorovskiy}}, \bibinfo {author} {\bibfnamefont {A.}~\bibnamefont {Zhukov}}, \bibinfo {author} {\bibfnamefont {S.~V.}\ \bibnamefont {Morozov}}, \bibinfo {author} {\bibfnamefont {M.~I.}\ \bibnamefont {Katsnelson}}, \bibinfo {author} {\bibfnamefont {V.~I.}\ \bibnamefont {Fal'ko}}, \bibinfo {author} {\bibfnamefont {A.~K.}\ \bibnamefont {Geim}},\ and\ \bibinfo {author} {\bibfnamefont {K.~S.}\ \bibnamefont {Novoselov}},\ }\href {https://www.science.org/doi/10.1126/science.1208683} {\bibfield  {journal} {\bibinfo  {journal} {Science}\ }\textbf {\bibinfo {volume} {333}},\ \bibinfo {pages} {860} (\bibinfo {year} {2011})}\BibitemShut {NoStop}%
\bibitem [{\citenamefont {Mariani}\ \emph {et~al.}(2012)\citenamefont {Mariani}, \citenamefont {Pearce},\ and\ \citenamefont {von Oppen}}]{mariani2012fictitious}%
  \BibitemOpen
  \bibfield  {author} {\bibinfo {author} {\bibfnamefont {E.}~\bibnamefont {Mariani}}, \bibinfo {author} {\bibfnamefont {A.~J.}\ \bibnamefont {Pearce}},\ and\ \bibinfo {author} {\bibfnamefont {F.}~\bibnamefont {von Oppen}},\ }\href {https://journals.aps.org/prb/abstract/10.1103/PhysRevB.86.165448} {\bibfield  {journal} {\bibinfo  {journal} {Phys. Rev. B}\ }\textbf {\bibinfo {volume} {86}},\ \bibinfo {pages} {165448} (\bibinfo {year} {2012})}\BibitemShut {NoStop}%
\bibitem [{\citenamefont {Qin}\ \emph {et~al.}(2025)\citenamefont {Qin}, \citenamefont {Wu}, \citenamefont {Nguyen}, \citenamefont {Morissette}, \citenamefont {Zhang}, \citenamefont {Watanabe}, \citenamefont {Taniguchi},\ and\ \citenamefont {Li}}]{morissette2025striped}%
  \BibitemOpen
  \bibfield  {author} {\bibinfo {author} {\bibfnamefont {P.}~\bibnamefont {Qin}}, \bibinfo {author} {\bibfnamefont {H.-T.}\ \bibnamefont {Wu}}, \bibinfo {author} {\bibfnamefont {R.~Q.}\ \bibnamefont {Nguyen}}, \bibinfo {author} {\bibfnamefont {E.}~\bibnamefont {Morissette}}, \bibinfo {author} {\bibfnamefont {N.~J.}\ \bibnamefont {Zhang}}, \bibinfo {author} {\bibfnamefont {K.}~\bibnamefont {Watanabe}}, \bibinfo {author} {\bibfnamefont {T.}~\bibnamefont {Taniguchi}},\ and\ \bibinfo {author} {\bibfnamefont {J.~I.~A.}\ \bibnamefont {Li}},\ }\href {https://arxiv.org/abs/2504.05129} {\bibfield  {journal} {\bibinfo  {journal} {arXiv:2504.05129}\ } (\bibinfo {year} {2025})}\BibitemShut {NoStop}%
\bibitem [{\citenamefont {Chakraborty}\ \emph {et~al.}(2024)\citenamefont {Chakraborty}, \citenamefont {Moessner},\ and\ \citenamefont {Doucot}}]{PhysRevLett.133.206604}%
  \BibitemOpen
  \bibfield  {author} {\bibinfo {author} {\bibfnamefont {N.}~\bibnamefont {Chakraborty}}, \bibinfo {author} {\bibfnamefont {R.}~\bibnamefont {Moessner}},\ and\ \bibinfo {author} {\bibfnamefont {B.}~\bibnamefont {Doucot}},\ }\href {https://doi.org/10.1103/PhysRevLett.133.206604} {\bibfield  {journal} {\bibinfo  {journal} {Phys. Rev. Lett.}\ }\textbf {\bibinfo {volume} {133}},\ \bibinfo {pages} {206604} (\bibinfo {year} {2024})}\BibitemShut {NoStop}%
\bibitem [{\citenamefont {Murshed}\ and\ \citenamefont {Roy}(2026)}]{murshed2025charge}%
  \BibitemOpen
  \bibfield  {author} {\bibinfo {author} {\bibfnamefont {S.~A.}\ \bibnamefont {Murshed}}\ and\ \bibinfo {author} {\bibfnamefont {B.}~\bibnamefont {Roy}},\ }\href {https://doi.org/10.1103/png3-tym5} {\bibfield  {journal} {\bibinfo  {journal} {Phys. Rev. B}\ }\textbf {\bibinfo {volume} {114}},\ \bibinfo {pages} {125412} (\bibinfo {year} {2026})}\BibitemShut {NoStop}%
\bibitem [{\citenamefont {Hu}\ \emph {et~al.}(2024)\citenamefont {Hu}, \citenamefont {Chen},\ and\ \citenamefont {Si}}]{hu-quantum-2024}%
  \BibitemOpen
  \bibfield  {author} {\bibinfo {author} {\bibfnamefont {H.}~\bibnamefont {Hu}}, \bibinfo {author} {\bibfnamefont {L.}~\bibnamefont {Chen}},\ and\ \bibinfo {author} {\bibfnamefont {Q.}~\bibnamefont {Si}},\ }\href {https://doi.org/10.1038/s41567-024-02679-7} {\bibfield  {journal} {\bibinfo  {journal} {Nat. Phys.}\ }\textbf {\bibinfo {volume} {20}},\ \bibinfo {pages} {1863} (\bibinfo {year} {2024})}\BibitemShut {NoStop}%
\bibitem [{\citenamefont {Wu}\ \emph {et~al.}(2026)\citenamefont {Wu}, \citenamefont {Huang}, \citenamefont {Hu}, \citenamefont {Chen}, \citenamefont {Hao}, \citenamefont {Shi}, \citenamefont {Malinowski}, \citenamefont {Guo}, \citenamefont {Jang}, \citenamefont {Zhu}, \citenamefont {May}, \citenamefont {Werner}, \citenamefont {Wang}, \citenamefont {Chen}, \citenamefont {Xie}, \citenamefont {Gao}, \citenamefont {Zhang}, \citenamefont {Yue}, \citenamefont {Ren}, \citenamefont {Hashimoto}, \citenamefont {Lu}, \citenamefont {Fedorov}, \citenamefont {Mo}, \citenamefont {Kono}, \citenamefont {He}, \citenamefont {Birgeneau}, \citenamefont {Dai}, \citenamefont {Xu}, \citenamefont {Cao}, \citenamefont {Si}, \citenamefont {Chu},\ and\ \citenamefont {Yi}}]{wu-dichotomy-2025}%
  \BibitemOpen
  \bibfield  {author} {\bibinfo {author} {\bibfnamefont {H.}~\bibnamefont {Wu}}, \bibinfo {author} {\bibfnamefont {J.}~\bibnamefont {Huang}}, \bibinfo {author} {\bibfnamefont {C.}~\bibnamefont {Hu}}, \bibinfo {author} {\bibfnamefont {L.}~\bibnamefont {Chen}}, \bibinfo {author} {\bibfnamefont {Y.}~\bibnamefont {Hao}}, \bibinfo {author} {\bibfnamefont {Y.}~\bibnamefont {Shi}}, \bibinfo {author} {\bibfnamefont {P.}~\bibnamefont {Malinowski}}, \bibinfo {author} {\bibfnamefont {Y.}~\bibnamefont {Guo}}, \bibinfo {author} {\bibfnamefont {B.~G.}\ \bibnamefont {Jang}}, \bibinfo {author} {\bibfnamefont {J.-X.}\ \bibnamefont {Zhu}}, \bibinfo {author} {\bibfnamefont {A.~F.}\ \bibnamefont {May}}, \bibinfo {author} {\bibfnamefont {T.}~\bibnamefont {Werner}}, \bibinfo {author} {\bibfnamefont {S.}~\bibnamefont {Wang}}, \bibinfo {author} {\bibfnamefont {X.}~\bibnamefont {Chen}}, \bibinfo {author} {\bibfnamefont {Y.}~\bibnamefont {Xie}}, \bibinfo {author} {\bibfnamefont {B.}~\bibnamefont {Gao}}, \bibinfo {author}
  {\bibfnamefont {Y.}~\bibnamefont {Zhang}}, \bibinfo {author} {\bibfnamefont {Z.}~\bibnamefont {Yue}}, \bibinfo {author} {\bibfnamefont {Z.}~\bibnamefont {Ren}}, \bibinfo {author} {\bibfnamefont {M.}~\bibnamefont {Hashimoto}}, \bibinfo {author} {\bibfnamefont {D.}~\bibnamefont {Lu}}, \bibinfo {author} {\bibfnamefont {A.}~\bibnamefont {Fedorov}}, \bibinfo {author} {\bibfnamefont {S.-K.}\ \bibnamefont {Mo}}, \bibinfo {author} {\bibfnamefont {J.}~\bibnamefont {Kono}}, \bibinfo {author} {\bibfnamefont {Y.}~\bibnamefont {He}}, \bibinfo {author} {\bibfnamefont {R.~J.}\ \bibnamefont {Birgeneau}}, \bibinfo {author} {\bibfnamefont {P.}~\bibnamefont {Dai}}, \bibinfo {author} {\bibfnamefont {X.}~\bibnamefont {Xu}}, \bibinfo {author} {\bibfnamefont {H.}~\bibnamefont {Cao}}, \bibinfo {author} {\bibfnamefont {Q.}~\bibnamefont {Si}}, \bibinfo {author} {\bibfnamefont {J.-H.}\ \bibnamefont {Chu}},\ and\ \bibinfo {author} {\bibfnamefont {M.}~\bibnamefont {Yi}},\ }\href {https://doi.org/10.1103/tbk8-zg11} {\bibfield  {journal}
  {\bibinfo  {journal} {Phys. Rev. B}\ }\textbf {\bibinfo {volume} {113}},\ \bibinfo {pages} {125120} (\bibinfo {year} {2026})}\BibitemShut {NoStop}%
\bibitem [{\citenamefont {Pikus}\ and\ \citenamefont {Efros}(1995)}]{pikus1995coulomb}%
  \BibitemOpen
  \bibfield  {author} {\bibinfo {author} {\bibfnamefont {F.~G.}\ \bibnamefont {Pikus}}\ and\ \bibinfo {author} {\bibfnamefont {A.~L.}\ \bibnamefont {Efros}},\ }\href {https://doi.org/10.1103/PhysRevB.51.16871} {\bibfield  {journal} {\bibinfo  {journal} {Phys. Rev. B}\ }\textbf {\bibinfo {volume} {51}},\ \bibinfo {pages} {16871} (\bibinfo {year} {1995})}\BibitemShut {NoStop}%
\bibitem [{\citenamefont {Song}\ \emph {et~al.}(2018)\citenamefont {Song}, \citenamefont {Jia}, \citenamefont {Zhang}, \citenamefont {Zhu}, \citenamefont {Shi}, \citenamefont {Wang}, \citenamefont {Zhu}, \citenamefont {Yuan}, \citenamefont {Zhang}, \citenamefont {Xing},\ and\ \citenamefont {Li}}]{song2018observation}%
  \BibitemOpen
  \bibfield  {author} {\bibinfo {author} {\bibfnamefont {Y.-H.}\ \bibnamefont {Song}}, \bibinfo {author} {\bibfnamefont {Z.-Y.}\ \bibnamefont {Jia}}, \bibinfo {author} {\bibfnamefont {D.}~\bibnamefont {Zhang}}, \bibinfo {author} {\bibfnamefont {X.-Y.}\ \bibnamefont {Zhu}}, \bibinfo {author} {\bibfnamefont {Z.-Q.}\ \bibnamefont {Shi}}, \bibinfo {author} {\bibfnamefont {H.}~\bibnamefont {Wang}}, \bibinfo {author} {\bibfnamefont {L.}~\bibnamefont {Zhu}}, \bibinfo {author} {\bibfnamefont {Q.-Q.}\ \bibnamefont {Yuan}}, \bibinfo {author} {\bibfnamefont {H.}~\bibnamefont {Zhang}}, \bibinfo {author} {\bibfnamefont {D.-Y.}\ \bibnamefont {Xing}},\ and\ \bibinfo {author} {\bibfnamefont {S.-C.}\ \bibnamefont {Li}},\ }\href {https://doi.org/10.1038/s41467-018-06635-x} {\bibfield  {journal} {\bibinfo  {journal} {Nat. Commun.}\ }\textbf {\bibinfo {volume} {9}},\ \bibinfo {pages} {4071} (\bibinfo {year} {2018})}\BibitemShut {NoStop}%
\bibitem [{\citenamefont {Voit}(1995)}]{voit-one-dimensional-1995}%
  \BibitemOpen
  \bibfield  {author} {\bibinfo {author} {\bibfnamefont {J.}~\bibnamefont {Voit}},\ }\href {https://doi.org/10.1088/0034-4885/58/9/002} {\bibfield  {journal} {\bibinfo  {journal} {Rep. Prog. Phys.}\ }\textbf {\bibinfo {volume} {58}},\ \bibinfo {pages} {977} (\bibinfo {year} {1995})}\BibitemShut {NoStop}%
\bibitem [{\citenamefont {Stühler}\ \emph {et~al.}(2020)\citenamefont {Stühler}, \citenamefont {Reis}, \citenamefont {Müller}, \citenamefont {Helbig}, \citenamefont {Schwemmer}, \citenamefont {Thomale}, \citenamefont {Schäfer},\ and\ \citenamefont {Claessen}}]{stuhler-tomonagaluttinger-2020}%
  \BibitemOpen
  \bibfield  {author} {\bibinfo {author} {\bibfnamefont {R.}~\bibnamefont {Stühler}}, \bibinfo {author} {\bibfnamefont {F.}~\bibnamefont {Reis}}, \bibinfo {author} {\bibfnamefont {T.}~\bibnamefont {Müller}}, \bibinfo {author} {\bibfnamefont {T.}~\bibnamefont {Helbig}}, \bibinfo {author} {\bibfnamefont {T.}~\bibnamefont {Schwemmer}}, \bibinfo {author} {\bibfnamefont {R.}~\bibnamefont {Thomale}}, \bibinfo {author} {\bibfnamefont {J.}~\bibnamefont {Schäfer}},\ and\ \bibinfo {author} {\bibfnamefont {R.}~\bibnamefont {Claessen}},\ }\href {https://doi.org/10.1038/s41567-019-0697-z} {\bibfield  {journal} {\bibinfo  {journal} {Nat. Phys.}\ }\textbf {\bibinfo {volume} {16}},\ \bibinfo {pages} {47} (\bibinfo {year} {2020})}\BibitemShut {NoStop}%
\bibitem [{\citenamefont {Martin}\ \emph {et~al.}(2008)\citenamefont {Martin}, \citenamefont {Blanter},\ and\ \citenamefont {Morpurgo}}]{martin2008topological}%
  \BibitemOpen
  \bibfield  {author} {\bibinfo {author} {\bibfnamefont {I.}~\bibnamefont {Martin}}, \bibinfo {author} {\bibfnamefont {Y.~M.}\ \bibnamefont {Blanter}},\ and\ \bibinfo {author} {\bibfnamefont {A.}~\bibnamefont {Morpurgo}},\ }\href {https://journals.aps.org/prl/abstract/10.1103/PhysRevLett.100.036804} {\bibfield  {journal} {\bibinfo  {journal} {Phys. Rev. Lett.}\ }\textbf {\bibinfo {volume} {100}},\ \bibinfo {pages} {036804} (\bibinfo {year} {2008})}\BibitemShut {NoStop}%
\bibitem [{\citenamefont {Zhang}\ \emph {et~al.}(2013)\citenamefont {Zhang}, \citenamefont {MacDonald},\ and\ \citenamefont {Mele}}]{zhang2013valley}%
  \BibitemOpen
  \bibfield  {author} {\bibinfo {author} {\bibfnamefont {F.}~\bibnamefont {Zhang}}, \bibinfo {author} {\bibfnamefont {A.~H.}\ \bibnamefont {MacDonald}},\ and\ \bibinfo {author} {\bibfnamefont {E.~J.}\ \bibnamefont {Mele}},\ }\href {https://www.pnas.org/doi/10.1073/pnas.1308853110} {\bibfield  {journal} {\bibinfo  {journal} {Proc. Natl. Acad. Sci. U.S.A.}\ }\textbf {\bibinfo {volume} {110}},\ \bibinfo {pages} {10546} (\bibinfo {year} {2013})}\BibitemShut {NoStop}%
\bibitem [{\citenamefont {Vaezi}\ \emph {et~al.}(2013)\citenamefont {Vaezi}, \citenamefont {Liang}, \citenamefont {Ngai}, \citenamefont {Yang},\ and\ \citenamefont {Kim}}]{vaezi2013topological}%
  \BibitemOpen
  \bibfield  {author} {\bibinfo {author} {\bibfnamefont {A.}~\bibnamefont {Vaezi}}, \bibinfo {author} {\bibfnamefont {Y.}~\bibnamefont {Liang}}, \bibinfo {author} {\bibfnamefont {D.~H.}\ \bibnamefont {Ngai}}, \bibinfo {author} {\bibfnamefont {L.}~\bibnamefont {Yang}},\ and\ \bibinfo {author} {\bibfnamefont {E.-A.}\ \bibnamefont {Kim}},\ }\href {https://journals.aps.org/prx/abstract/10.1103/PhysRevX.3.021018} {\bibfield  {journal} {\bibinfo  {journal} {Phys. Rev. X}\ }\textbf {\bibinfo {volume} {3}},\ \bibinfo {pages} {021018} (\bibinfo {year} {2013})}\BibitemShut {NoStop}%
\bibitem [{\citenamefont {Yin}\ \emph {et~al.}(2016)\citenamefont {Yin}, \citenamefont {Jiang}, \citenamefont {Qiao},\ and\ \citenamefont {He}}]{yin-direct-2016}%
  \BibitemOpen
  \bibfield  {author} {\bibinfo {author} {\bibfnamefont {L.-J.}\ \bibnamefont {Yin}}, \bibinfo {author} {\bibfnamefont {H.}~\bibnamefont {Jiang}}, \bibinfo {author} {\bibfnamefont {J.-B.}\ \bibnamefont {Qiao}},\ and\ \bibinfo {author} {\bibfnamefont {L.}~\bibnamefont {He}},\ }\href {https://www.nature.com/articles/ncomms11760} {\bibfield  {journal} {\bibinfo  {journal} {Nat. Commun.}\ }\textbf {\bibinfo {volume} {7}},\ \bibinfo {pages} {11760} (\bibinfo {year} {2016})}\BibitemShut {NoStop}%
\bibitem [{\citenamefont {Ju}\ \emph {et~al.}(2015)\citenamefont {Ju}, \citenamefont {Shi}, \citenamefont {Nair}, \citenamefont {Lv}, \citenamefont {Jin}, \citenamefont {Velasco}, \citenamefont {Ojeda-Aristizabal}, \citenamefont {Bechtel}, \citenamefont {Martin}, \citenamefont {Zettl}, \citenamefont {Analytis},\ and\ \citenamefont {Wang}}]{ju-topological-2015}%
  \BibitemOpen
  \bibfield  {author} {\bibinfo {author} {\bibfnamefont {L.}~\bibnamefont {Ju}}, \bibinfo {author} {\bibfnamefont {Z.}~\bibnamefont {Shi}}, \bibinfo {author} {\bibfnamefont {N.}~\bibnamefont {Nair}}, \bibinfo {author} {\bibfnamefont {Y.}~\bibnamefont {Lv}}, \bibinfo {author} {\bibfnamefont {C.}~\bibnamefont {Jin}}, \bibinfo {author} {\bibfnamefont {J.}~\bibnamefont {Velasco}}, \bibinfo {author} {\bibfnamefont {C.}~\bibnamefont {Ojeda-Aristizabal}}, \bibinfo {author} {\bibfnamefont {H.~A.}\ \bibnamefont {Bechtel}}, \bibinfo {author} {\bibfnamefont {M.~C.}\ \bibnamefont {Martin}}, \bibinfo {author} {\bibfnamefont {A.}~\bibnamefont {Zettl}}, \bibinfo {author} {\bibfnamefont {J.}~\bibnamefont {Analytis}},\ and\ \bibinfo {author} {\bibfnamefont {F.}~\bibnamefont {Wang}},\ }\href {https://doi.org/10.1038/nature14364} {\bibfield  {journal} {\bibinfo  {journal} {Nature}\ }\textbf {\bibinfo {volume} {520}},\ \bibinfo {pages} {650} (\bibinfo {year} {2015})}\BibitemShut {NoStop}%
\bibitem [{\citenamefont {Huang}\ \emph {et~al.}(2024)\citenamefont {Huang}, \citenamefont {Fu}, \citenamefont {Watanabe}, \citenamefont {Taniguchi},\ and\ \citenamefont {Zhu}}]{huang-high-temperature-2024}%
  \BibitemOpen
  \bibfield  {author} {\bibinfo {author} {\bibfnamefont {K.}~\bibnamefont {Huang}}, \bibinfo {author} {\bibfnamefont {H.}~\bibnamefont {Fu}}, \bibinfo {author} {\bibfnamefont {K.}~\bibnamefont {Watanabe}}, \bibinfo {author} {\bibfnamefont {T.}~\bibnamefont {Taniguchi}},\ and\ \bibinfo {author} {\bibfnamefont {J.}~\bibnamefont {Zhu}},\ }\href {https://doi.org/10.1126/science.adj3742} {\bibfield  {journal} {\bibinfo  {journal} {Science}\ }\textbf {\bibinfo {volume} {385}},\ \bibinfo {pages} {657} (\bibinfo {year} {2024})}\BibitemShut {NoStop}%
\bibitem [{\citenamefont {Yankowitz}\ \emph {et~al.}(2014)\citenamefont {Yankowitz}, \citenamefont {Wang}, \citenamefont {Birdwell}, \citenamefont {Chen}, \citenamefont {Watanabe}, \citenamefont {Taniguchi}, \citenamefont {Jacquod}, \citenamefont {San-Jose}, \citenamefont {Jarillo-Herrero},\ and\ \citenamefont {LeRoy}}]{yankowitz-electric-2014}%
  \BibitemOpen
  \bibfield  {author} {\bibinfo {author} {\bibfnamefont {M.}~\bibnamefont {Yankowitz}}, \bibinfo {author} {\bibfnamefont {J.~I.-J.}\ \bibnamefont {Wang}}, \bibinfo {author} {\bibfnamefont {A.~G.}\ \bibnamefont {Birdwell}}, \bibinfo {author} {\bibfnamefont {Y.-A.}\ \bibnamefont {Chen}}, \bibinfo {author} {\bibfnamefont {K.}~\bibnamefont {Watanabe}}, \bibinfo {author} {\bibfnamefont {T.}~\bibnamefont {Taniguchi}}, \bibinfo {author} {\bibfnamefont {P.}~\bibnamefont {Jacquod}}, \bibinfo {author} {\bibfnamefont {P.}~\bibnamefont {San-Jose}}, \bibinfo {author} {\bibfnamefont {P.}~\bibnamefont {Jarillo-Herrero}},\ and\ \bibinfo {author} {\bibfnamefont {B.~J.}\ \bibnamefont {LeRoy}},\ }\href {https://doi.org/10.1038/nmat3965} {\bibfield  {journal} {\bibinfo  {journal} {Nat. Mater.}\ }\textbf {\bibinfo {volume} {13}},\ \bibinfo {pages} {786} (\bibinfo {year} {2014})}\BibitemShut {NoStop}%
\bibitem [{\citenamefont {Garcia-Ruiz}\ \emph {et~al.}(2021)\citenamefont {Garcia-Ruiz}, \citenamefont {Deng}, \citenamefont {Enaldiev},\ and\ \citenamefont {Fal'ko}}]{garcia2021full}%
  \BibitemOpen
  \bibfield  {author} {\bibinfo {author} {\bibfnamefont {A.}~\bibnamefont {Garcia-Ruiz}}, \bibinfo {author} {\bibfnamefont {H.-Y.}\ \bibnamefont {Deng}}, \bibinfo {author} {\bibfnamefont {V.~V.}\ \bibnamefont {Enaldiev}},\ and\ \bibinfo {author} {\bibfnamefont {V.~I.}\ \bibnamefont {Fal'ko}},\ }\href {https://doi.org/10.1103/PhysRevB.104.085402} {\bibfield  {journal} {\bibinfo  {journal} {Phys. Rev. B}\ }\textbf {\bibinfo {volume} {104}},\ \bibinfo {pages} {085402} (\bibinfo {year} {2021})}\BibitemShut {NoStop}%
\bibitem [{\citenamefont {Wang}\ \emph {et~al.}(2024)\citenamefont {Wang}, \citenamefont {Burg}, \citenamefont {Lian}, \citenamefont {Watanabe}, \citenamefont {Taniguchi}, \citenamefont {Bernevig},\ and\ \citenamefont {Tutuc}}]{Wang2024TDBGValleyChern}%
  \BibitemOpen
  \bibfield  {author} {\bibinfo {author} {\bibfnamefont {Y.}~\bibnamefont {Wang}}, \bibinfo {author} {\bibfnamefont {G.~W.}\ \bibnamefont {Burg}}, \bibinfo {author} {\bibfnamefont {B.}~\bibnamefont {Lian}}, \bibinfo {author} {\bibfnamefont {K.}~\bibnamefont {Watanabe}}, \bibinfo {author} {\bibfnamefont {T.}~\bibnamefont {Taniguchi}}, \bibinfo {author} {\bibfnamefont {B.~A.}\ \bibnamefont {Bernevig}},\ and\ \bibinfo {author} {\bibfnamefont {E.}~\bibnamefont {Tutuc}},\ }\href {https://doi.org/10.1103/PhysRevLett.133.246401} {\bibfield  {journal} {\bibinfo  {journal} {Phys. Rev. Lett.}\ }\textbf {\bibinfo {volume} {133}},\ \bibinfo {pages} {246401} (\bibinfo {year} {2024})}\BibitemShut {NoStop}%
\bibitem [{\citenamefont {Li}\ \emph {et~al.}(2011)\citenamefont {Li}, \citenamefont {Martin}, \citenamefont {B{\"u}ttiker},\ and\ \citenamefont {Morpurgo}}]{Li2011Subgap}%
  \BibitemOpen
  \bibfield  {author} {\bibinfo {author} {\bibfnamefont {J.}~\bibnamefont {Li}}, \bibinfo {author} {\bibfnamefont {I.}~\bibnamefont {Martin}}, \bibinfo {author} {\bibfnamefont {M.}~\bibnamefont {B{\"u}ttiker}},\ and\ \bibinfo {author} {\bibfnamefont {A.~F.}\ \bibnamefont {Morpurgo}},\ }\href {https://doi.org/10.1038/nphys1822} {\bibfield  {journal} {\bibinfo  {journal} {Nat. Phys.}\ }\textbf {\bibinfo {volume} {7}},\ \bibinfo {pages} {38} (\bibinfo {year} {2011})}\BibitemShut {NoStop}%
\bibitem [{\citenamefont {Checkelsky}\ \emph {et~al.}(2024)\citenamefont {Checkelsky}, \citenamefont {Bernevig}, \citenamefont {Coleman}, \citenamefont {Si},\ and\ \citenamefont {Paschen}}]{checkelsky-flat-2024}%
  \BibitemOpen
  \bibfield  {author} {\bibinfo {author} {\bibfnamefont {J.~G.}\ \bibnamefont {Checkelsky}}, \bibinfo {author} {\bibfnamefont {B.~A.}\ \bibnamefont {Bernevig}}, \bibinfo {author} {\bibfnamefont {P.}~\bibnamefont {Coleman}}, \bibinfo {author} {\bibfnamefont {Q.}~\bibnamefont {Si}},\ and\ \bibinfo {author} {\bibfnamefont {S.}~\bibnamefont {Paschen}},\ }\href {https://doi.org/10.1038/s41578-023-00644-z} {\bibfield  {journal} {\bibinfo  {journal} {Nat. Rev. Mater.}\ }\textbf {\bibinfo {volume} {9}},\ \bibinfo {pages} {509} (\bibinfo {year} {2024})}\BibitemShut {NoStop}%
\bibitem [{\citenamefont {Banerjee}\ \emph {et~al.}(2020)\citenamefont {Banerjee}, \citenamefont {Nguyen}, \citenamefont {Granzier-Nakajima}, \citenamefont {Pabbi}, \citenamefont {Lherbier}, \citenamefont {Binion}, \citenamefont {Charlier}, \citenamefont {Terrones},\ and\ \citenamefont {Hudson}}]{strain-banerjee-2020}%
  \BibitemOpen
  \bibfield  {author} {\bibinfo {author} {\bibfnamefont {R.}~\bibnamefont {Banerjee}}, \bibinfo {author} {\bibfnamefont {V.-H.}\ \bibnamefont {Nguyen}}, \bibinfo {author} {\bibfnamefont {T.}~\bibnamefont {Granzier-Nakajima}}, \bibinfo {author} {\bibfnamefont {L.}~\bibnamefont {Pabbi}}, \bibinfo {author} {\bibfnamefont {A.}~\bibnamefont {Lherbier}}, \bibinfo {author} {\bibfnamefont {A.~R.}\ \bibnamefont {Binion}}, \bibinfo {author} {\bibfnamefont {J.-C.}\ \bibnamefont {Charlier}}, \bibinfo {author} {\bibfnamefont {M.}~\bibnamefont {Terrones}},\ and\ \bibinfo {author} {\bibfnamefont {E.~W.}\ \bibnamefont {Hudson}},\ }\href {https://doi.org/10.1021/acs.nanolett.9b05108} {\bibfield  {journal} {\bibinfo  {journal} {Nano Lett.}\ }\textbf {\bibinfo {volume} {20}},\ \bibinfo {pages} {3113} (\bibinfo {year} {2020})}\BibitemShut {NoStop}%
\bibitem [{\citenamefont {Koshino}\ and\ \citenamefont {McCann}(2009)}]{koshino-aba}%
  \BibitemOpen
  \bibfield  {author} {\bibinfo {author} {\bibfnamefont {M.}~\bibnamefont {Koshino}}\ and\ \bibinfo {author} {\bibfnamefont {E.}~\bibnamefont {McCann}},\ }\href {https://doi.org/10.1103/PhysRevB.79.125443} {\bibfield  {journal} {\bibinfo  {journal} {Phys. Rev. B}\ }\textbf {\bibinfo {volume} {79}},\ \bibinfo {pages} {125443} (\bibinfo {year} {2009})}\BibitemShut {NoStop}%
\bibitem [{\citenamefont {Li}\ \emph {et~al.}(2016)\citenamefont {Li}, \citenamefont {Wang}, \citenamefont {McFaul}, \citenamefont {Zern}, \citenamefont {Ren}, \citenamefont {Watanabe}, \citenamefont {Taniguchi}, \citenamefont {Qiao},\ and\ \citenamefont {Zhu}}]{li-gate-controlled-2016}%
  \BibitemOpen
  \bibfield  {author} {\bibinfo {author} {\bibfnamefont {J.}~\bibnamefont {Li}}, \bibinfo {author} {\bibfnamefont {K.}~\bibnamefont {Wang}}, \bibinfo {author} {\bibfnamefont {K.~J.}\ \bibnamefont {McFaul}}, \bibinfo {author} {\bibfnamefont {Z.}~\bibnamefont {Zern}}, \bibinfo {author} {\bibfnamefont {Y.}~\bibnamefont {Ren}}, \bibinfo {author} {\bibfnamefont {K.}~\bibnamefont {Watanabe}}, \bibinfo {author} {\bibfnamefont {T.}~\bibnamefont {Taniguchi}}, \bibinfo {author} {\bibfnamefont {Z.}~\bibnamefont {Qiao}},\ and\ \bibinfo {author} {\bibfnamefont {J.}~\bibnamefont {Zhu}},\ }\href {https://doi.org/10.1038/nnano.2016.158} {\bibfield  {journal} {\bibinfo  {journal} {Nat. Nanotechnol.}\ }\textbf {\bibinfo {volume} {11}},\ \bibinfo {pages} {1060} (\bibinfo {year} {2016})}\BibitemShut {NoStop}%
\bibitem [{\citenamefont {Geisenhof}\ \emph {et~al.}(2022)\citenamefont {Geisenhof}, \citenamefont {Winterer}, \citenamefont {Seiler}, \citenamefont {Lenz}, \citenamefont {Martin},\ and\ \citenamefont {Weitz}}]{geisenhof2022interplay}%
  \BibitemOpen
  \bibfield  {author} {\bibinfo {author} {\bibfnamefont {F.~R.}\ \bibnamefont {Geisenhof}}, \bibinfo {author} {\bibfnamefont {F.}~\bibnamefont {Winterer}}, \bibinfo {author} {\bibfnamefont {A.~M.}\ \bibnamefont {Seiler}}, \bibinfo {author} {\bibfnamefont {J.}~\bibnamefont {Lenz}}, \bibinfo {author} {\bibfnamefont {I.}~\bibnamefont {Martin}},\ and\ \bibinfo {author} {\bibfnamefont {R.~T.}\ \bibnamefont {Weitz}},\ }\href {https://doi.org/10.1038/s41467-022-31680-y} {\bibfield  {journal} {\bibinfo  {journal} {Nat. Commun.}\ }\textbf {\bibinfo {volume} {13}},\ \bibinfo {pages} {4187} (\bibinfo {year} {2022})}\BibitemShut {NoStop}%
\bibitem [{\citenamefont {Déprez}\ \emph {et~al.}(2021)\citenamefont {Déprez}, \citenamefont {Veyrat}, \citenamefont {Vignaud}, \citenamefont {Nayak}, \citenamefont {Watanabe}, \citenamefont {Taniguchi}, \citenamefont {Gay}, \citenamefont {Sellier},\ and\ \citenamefont {Sacépé}}]{deprez-tunable-2021}%
  \BibitemOpen
  \bibfield  {author} {\bibinfo {author} {\bibfnamefont {C.}~\bibnamefont {Déprez}}, \bibinfo {author} {\bibfnamefont {L.}~\bibnamefont {Veyrat}}, \bibinfo {author} {\bibfnamefont {H.}~\bibnamefont {Vignaud}}, \bibinfo {author} {\bibfnamefont {G.}~\bibnamefont {Nayak}}, \bibinfo {author} {\bibfnamefont {K.}~\bibnamefont {Watanabe}}, \bibinfo {author} {\bibfnamefont {T.}~\bibnamefont {Taniguchi}}, \bibinfo {author} {\bibfnamefont {F.}~\bibnamefont {Gay}}, \bibinfo {author} {\bibfnamefont {H.}~\bibnamefont {Sellier}},\ and\ \bibinfo {author} {\bibfnamefont {B.}~\bibnamefont {Sacépé}},\ }\href {https://doi.org/10.1038/s41565-021-00847-x} {\bibfield  {journal} {\bibinfo  {journal} {Nat. Nanotechnol.}\ }\textbf {\bibinfo {volume} {16}},\ \bibinfo {pages} {555} (\bibinfo {year} {2021})}\BibitemShut {NoStop}%
\bibitem [{\citenamefont {Barrier}\ \emph {et~al.}(2024)\citenamefont {Barrier}, \citenamefont {Kim}, \citenamefont {Kumar}, \citenamefont {Xin}, \citenamefont {Kumaravadivel}, \citenamefont {Hague}, \citenamefont {Nguyen}, \citenamefont {Berdyugin}, \citenamefont {Moulsdale}, \citenamefont {Enaldiev}, \citenamefont {Prance}, \citenamefont {Koppens}, \citenamefont {Gorbachev}, \citenamefont {Watanabe}, \citenamefont {Taniguchi}, \citenamefont {Glazman}, \citenamefont {Grigorieva}, \citenamefont {Fal'ko},\ and\ \citenamefont {Geim}}]{barrier-one-dimensional-2024}%
  \BibitemOpen
  \bibfield  {author} {\bibinfo {author} {\bibfnamefont {J.}~\bibnamefont {Barrier}}, \bibinfo {author} {\bibfnamefont {M.}~\bibnamefont {Kim}}, \bibinfo {author} {\bibfnamefont {R.~K.}\ \bibnamefont {Kumar}}, \bibinfo {author} {\bibfnamefont {N.}~\bibnamefont {Xin}}, \bibinfo {author} {\bibfnamefont {P.}~\bibnamefont {Kumaravadivel}}, \bibinfo {author} {\bibfnamefont {L.}~\bibnamefont {Hague}}, \bibinfo {author} {\bibfnamefont {E.}~\bibnamefont {Nguyen}}, \bibinfo {author} {\bibfnamefont {A.~I.}\ \bibnamefont {Berdyugin}}, \bibinfo {author} {\bibfnamefont {C.}~\bibnamefont {Moulsdale}}, \bibinfo {author} {\bibfnamefont {V.~V.}\ \bibnamefont {Enaldiev}}, \bibinfo {author} {\bibfnamefont {J.~R.}\ \bibnamefont {Prance}}, \bibinfo {author} {\bibfnamefont {F.~H.~L.}\ \bibnamefont {Koppens}}, \bibinfo {author} {\bibfnamefont {R.~V.}\ \bibnamefont {Gorbachev}}, \bibinfo {author} {\bibfnamefont {K.}~\bibnamefont {Watanabe}}, \bibinfo {author} {\bibfnamefont {T.}~\bibnamefont {Taniguchi}}, \bibinfo {author}
  {\bibfnamefont {L.~I.}\ \bibnamefont {Glazman}}, \bibinfo {author} {\bibfnamefont {I.~V.}\ \bibnamefont {Grigorieva}}, \bibinfo {author} {\bibfnamefont {V.~I.}\ \bibnamefont {Fal'ko}},\ and\ \bibinfo {author} {\bibfnamefont {A.~K.}\ \bibnamefont {Geim}},\ }\href {https://doi.org/10.1038/s41586-024-07271-w} {\bibfield  {journal} {\bibinfo  {journal} {Nature}\ }\textbf {\bibinfo {volume} {628}},\ \bibinfo {pages} {741} (\bibinfo {year} {2024})}\BibitemShut {NoStop}%
\bibitem [{\citenamefont {Jiang}\ \emph {et~al.}(2020)\citenamefont {Jiang}, \citenamefont {Wang}, \citenamefont {Zhao}, \citenamefont {Crommie},\ and\ \citenamefont {Wang}}]{jiang-soliton-dependent-2020}%
  \BibitemOpen
  \bibfield  {author} {\bibinfo {author} {\bibfnamefont {L.}~\bibnamefont {Jiang}}, \bibinfo {author} {\bibfnamefont {S.}~\bibnamefont {Wang}}, \bibinfo {author} {\bibfnamefont {S.}~\bibnamefont {Zhao}}, \bibinfo {author} {\bibfnamefont {M.}~\bibnamefont {Crommie}},\ and\ \bibinfo {author} {\bibfnamefont {F.}~\bibnamefont {Wang}},\ }\href {https://doi.org/10.1021/acs.nanolett.0c01911} {\bibfield  {journal} {\bibinfo  {journal} {Nano Lett.}\ }\textbf {\bibinfo {volume} {20}},\ \bibinfo {pages} {5936} (\bibinfo {year} {2020})}\BibitemShut {NoStop}%
\end{thebibliography}%

\end{document}